\documentclass[superscriptaddress,twocolumn,aps,preprintnumbers,notitlepage]{revtex4-1}
\usepackage{amsmath,amssymb,esint}
\usepackage{lipsum} 
\usepackage[latin9]{inputenc}
\usepackage{graphicx}
\usepackage{dcolumn}
\usepackage{bbold}
\usepackage{bm}
\usepackage[usenames]{xcolor}
\usepackage[colorlinks=true,linktocpage,bookmarks=true,citecolor=blue,linkcolor=blue,urlcolor=blue,hyperfootnotes=true,pageanchor=true]{hyperref}
\usepackage[caption=false,labelformat=simple]{subfig}
\usepackage{accents}
\usepackage[export]{adjustbox}
\usepackage{mathtools}
\usepackage{float}
\usepackage{oplotsymbl}
\usepackage{mathrsfs}

\newcommand{\mc}{\mathcal}
\newcommand{\mb}{\mathbf}

\newcommand{\be}{\begin{equation}}
\newcommand{\ee}{\end{equation}}
\newcommand{\bea}{\begin{eqnarray}}
\newcommand{\eea}{\end{eqnarray}}

\begin{document}

\title{Multipolar spin liquid in an exactly solvable model for $j_\mathrm{eff} = \frac{3}{2}$ moments}

\author{Vanuildo S. de Carvalho}
\affiliation{Instituto de F\'{i}sica, Universidade Federal de Goi\'as, 74.001-970, Goi\^ania-GO, Brazil}
\author{Hermann Freire}
\affiliation{Instituto de F\'{i}sica, Universidade Federal de Goi\'as, 74.001-970, Goi\^ania-GO, Brazil}
\author{Rodrigo G. Pereira}
\affiliation{International Institute of Physics and Departamento de F\'isica Te\'orica e Experimental, Universidade Federal do Rio Grande do Norte, Campus Universit\'ario, Lagoa Nova,  Natal, RN, 59078-970, Brazil}

\date{\today}

\begin{abstract}
We study  an exactly solvable  model with bond-directional quadrupolar and octupolar  interactions between  spin-orbital entangled  $j_{\mathrm{eff}} = \frac{3}{2}$ moments on the honeycomb lattice. We show that this model features a  multipolar spin liquid phase with gapless fermionic excitations. In the presence of perturbations that break time-reversal and   rotation symmetries, we find  Abelian and non-Abelian topological phases in which the Chern number evaluates to $0$, $\pm 1$, and $\pm 2$. We also  investigate  quantum phase transitions out of the multipolar spin liquid   using a parton mean-field approach and   orbital wave theory. In the regime of strong integrability-breaking interactions, the multipolar spin liquid gives way to  ferroquadrupolar-vortex and antiferro-octupolar ordered phases that harbor  a hidden spin-$\frac{1}{2}$ Kitaev spin liquid. Our work unveils mechanisms for unusual multipolar orders and quantum spin liquids in Mott insulators with strong spin-orbit coupling. 

\end{abstract}

\maketitle

\section{Introduction}\label{Sec_I}
\noindent

Quantum spin liquids (QSLs) refer to disordered ground states of interacting spin systems which are characterized by the emergence of fractionalized excitations, topological order, and long-range entanglement \cite{Balents-N(2010),Senthil-S(2020)}. The quest for QSLs stands as one of the most prolific topics of  investigation in condensed matter physics, as demonstrated  by  the growing list of    QSL candidates and by  notable advances in our theoretical understanding of  phases beyond the Landau paradigm   \cite{Savary-RPP(2017),Takagi-NRP(2019),Trebst-PR(2022)}.

A guiding principle in the search for new  materials  is rooted in the engineering of bond-dependent interactions between $j_\mathrm{eff} = \frac{1}{2}$ moments, as found in the exactly solvable model proposed by Kitaev for the honeycomb lattice \cite{Kitaev-AP(2006)}. This model exhibits a genuine QSL ground state resulting from the fractionalization of   $j_\mathrm{eff} = \frac{1}{2}$ moments into Majorana fermions and a $\mathbb Z_2$ gauge field. Kitaev interactions appear as the leading term in effective models for spin-orbit-assisted Mott insulators with $4d^5$ and $5d^5$ electronic configuration \cite{Jackeli-PRL(2009)}. Over the last decade, several experiments have indicated  the proximity of   Kitaev material candidates, such as the  iridates (Na$_{1 - x}$Li$_x$)$_2$IrO$_3$ \cite{Singh-PRB(2010),Singh-PRL(2012),Cao-PRB(2013)} and H$_3$LiIr$_2$O$_6$ \cite{Kitagawa-Nature(2018)}, to a  QSL phase. Special attention has been devoted to  the compound  $\alpha$-RuCl$_3$ \cite{Plumb-PRB(2014),Kasahara-Nature(2018),Tanaka-NP(2022)}, for which a half-quantized thermal Hall conductance has been reported \cite{Kasahara-Nature(2018)}.

In recent years, the search for new systems with QSL ground states has expanded to Mott insulators with effective magnetic moments beyond the $j_\mathrm{eff} = \frac{1}{2}$ picture \cite{Balents-ARCMP(2014),Takagi-JPSJ(2021)}. In that regard, higher-spin Kitaev models were microscopically derived for the honeycomb Mott insulators NiI$_2$ ($d^8$ configuration and spin $S = 1$) \cite{Stavropoulos-PRL(2019)} and CrI$_3$ ($d^3$ configuration and $S = \frac{3}{2}$) \cite{Xu-PRL(2020),Xu-NPJ(2018),Stavropoulos-PRB(2021)}.

In contrast to the original spin-$\frac{1}{2}$ Kitaev model, these generalizations are not integrable.  Their properties have been studied by  semiclassical analysis \cite{Baskaran-PRB(2008)}, exact diagonalization \cite{Stavropoulos-PRL(2019),Koga-JPSJ(2018),Hickey-PRR(2020)}, the density matrix renormalization group method \cite{Dong-PRB(2020),Zhu-PRR(2020),Khait-PRR(2021)}, and parton mean-field  theory \cite{Jin-NC(2022)}. Remarkably,   all these methods point to the existence of a QSL ground state for both   $S = 1$ and $S = \frac{3}{2}$. In addition, Mott insulators with $4d^1$ and $5d^1$   configurations have also attracted   attention  \cite{Balents-ARCMP(2014),Takagi-JPSJ(2021)}. The effective spin Hamiltonians for these systems   are written in terms of  $j_\mathrm{eff} = \frac{3}{2}$ moments \cite{Balents-PRB(2011)} and contain multipolar interactions which can give rise to exotic hidden multipolar orders and even spin-orbital liquid phases \cite{Pereira-PRL(2016),Natori-PRB(2017),Jackeli-PRL(2017),Natori-PRB(2018),Yamada-PRL(2018),Ishikawa-PRB(2019),Pereira-PRB(2020),Chen-arXiv(2021),Natori-arXiv(2023)}.

In this work, we propose an exactly solvable model for $j_\mathrm{eff} = \frac{3}{2}$ moments located on the sites of a honeycomb lattice which could potentially describe some properties of $4d^1$ and $5d^1$ materials. This model involves a combination of nearest-neighbor bond-directional quadrupole-quadrupole and octupole-octupole  interactions. The representation of the quadrupolar and octupolar operators in terms of Majorana fermions maps the model to noninteracting fermions hopping on the background of a static $\mathbb{Z}_2$ gauge field. Since this model exhibits a  gapless  quantum disordered ground state and   spin-orbital correlations associated with the fractionalization of higher magnetic multipoles, it epitomizes all the features of a \emph{multipolar spin liquid} (MSL) \cite{Takagi-JPSJ(2021),Chen-PRB(2017),Sibille-NP(2020),Rayyan-PRB(2023)}.

Our model should be viewed as an integrable point in the parameter space for the most general model describing $j_\mathrm{eff} = \frac{3}{2}$ spin-orbital systems. In this sense, the existence of the MSL offers a different avenue of research in the quest for materials with unusual QSL properties. Indeed, we show that, under the action of rotational and time-reversal-symmetry-breaking fields, our exactly solvable model realizes non-trivial Abelian and non-Abelian topological phases with gapless chiral Majorana modes on the edge. Furthermore, we study the robustness of the MSL state against the formation of long-range multipolar  orders driven by integrability-breaking  interactions. Interestingly, even in the strong-coupling regime the system exhibits a hidden QSL state described by a spin-$\frac{1}{2}$ Kitaev model  on top of either a ferroquadrupolar-vortex (FQV) or an antiferro-octupolar (AO) ordered state.

The remainder of our paper is structured as follows. In Sec. \ref{Sec_II}, we introduce and provide the motivation for the $j_\mathrm{eff} = \frac{3}{2}$ model with bond-directional quadrupolar and octupolar interactions. In Sec. \ref{sec:solvable}, we study the properties of the MSL in the exactly solvable limit of the model. While the fully symmetric MSL is gapless, we show that magnetic and strain fields can give rise  to  Abelian and non-Abelian topological phases. Section \ref{Sec_IV} focuses on the stability of the MSL and the quantum phase transitions out of this state,  analyzed using  parton mean-field theory and orbital wave theory. Finally, we present our concluding remarks in Sec. \ref{Sec_V}. We leave to Appendixes \ref{App:GS_Energy} and \ref{App:MF} some technical details on the identification of the lower-energy gauge-flux sector of the MSL and also on the parton mean-field approach employed here.

\section{Model}\label{Sec_II}
\noindent

We consider interacting local moments with effective total angular momentum $j_\mathrm{eff} = \frac{3}{2}$, which appear in Mott insulators with $4d^1$ or $5d^1$ transition-metal ions \cite{Balents-PRB(2011),Takagi-JPSJ(2021),Khomskii-CR(2021)}. The single electron in the open shell   occupies the triply degenerate $t_{2g}$ orbitals with effective orbital angular momentum $l_{\rm eff} = 1$, which in turn are split by the spin-orbit coupling into a higher-energy $j_{\rm eff} = 1/2$ doublet and a lower-energy $j_{\rm eff} = 3/2$ quartet. The latter  is formed by two Kramers doublets, which may remain degenerate at low temperatures because  the strong spin-orbit coupling suppresses the Jahn-Teller effect \cite{Streltsov-PRX(2020)}.

An effective spin-orbital Hamiltonian for $j_\mathrm{eff} = \frac{3}{2}$ moments  was derived in Refs. \cite{Yamada-PRL(2018),Natori-PRB(2018)} considering the dominant exchange process on the honeycomb lattice. The  starting point is  a multiorbital Hubbard-Kanamori model  that combines the  effects of the  crystal electric field, spin-orbit coupling,  and electronic correlations, assuming  a single hopping  path for 90$^\circ$ bonds formed by edge-sharing octahedra. The resulting Hamiltonian   contains  quadrupole-quadrupole and octupole-octupole interactions with coupling constants of the same order as the coupling between dipole moments.  More generally, one should    take into account subleading hopping paths known to be important in the generic  model for Kitaev materials \cite{Rau-PRL(2014)}. Furthermore, the electrostatic interaction between orbitals  with different charge distributions and  the  virtual  exchange of optical phonons in the cooperative Jahn-Teller effect can also renormalize the quadrupole-quadrupole interactions   \cite{Baker-RPP(1971),Gehring-RPP(1975)}.

Out of the most general Hamiltonian for $j_\mathrm{eff} = \frac{3}{2}$ moments, here we shall focus on the regime of parameter space where the leading  interactions involve  quadrupole   and   octupole moments. Such a hierarchy of interactions could in principle  be realized in  layered  materials where the relative strength of the couplings can be tuned by  strain \cite{Voleti-PRB(2021),Yadav-PRB(2018)}.

\begin{table}
\caption{\label{Tab_Multipoles} Quadrupolar and octupolar operators acting on $j_{\textrm{eff}} = \frac{3}{2}$ states. The overlines in $\overline{J^{\alpha} J^{\beta}}$ and $\overline{J^{\alpha} J^{\beta} J^{\gamma}}$ stand for  a symmetrization with respect to the indices of the dipolar operators $J^{\alpha}$, e.g., $\overline{J^{x}(J^{y})^{2}} \equiv J^{x}(J^{y})^{2}+J^{y}J^{x}J^{y}+(J^{y})^{2}J^{x}$. Here we omit the octupolar operators that transform in the $\Gamma_4$  representation (see, e.g.,  Ref. \cite{Santini-RMP(2009)}). 
}
\begin{tabular}{ccc}
\hline 
\hline 
Moment  & Symmetry  & Operators \tabularnewline 
\hline 
\tabularnewline 
Quadrupole  & $\Gamma_{3}$  & $O_3\equiv O^{3z^2-r^2}=\frac1{3}[3(J^{z})^{2}-\textbf{J}^{2}]$ \tabularnewline \tabularnewline 
 &  & $O_2\equiv O^{x^2-y^2}=\frac1{\sqrt3}[(J^{x})^{2}-(J^{y})^{2}]$  \tabularnewline \tabularnewline 
 & $\Gamma_{5}$  & $O^x \equiv O^{yz}=\frac1{\sqrt3}\overline{J^{y}J^{z}}$ \tabularnewline \tabularnewline
 &  & $O^y \equiv O^{zx}=\frac1{\sqrt3}\overline{J^{x}J^{z}}$ \tabularnewline \tabularnewline 
 &  & $O^z \equiv O^{xy}=\frac1{\sqrt3}\overline{J^{x}J^{y}}$ \tabularnewline \tabularnewline 

Octupole  & $\Gamma_{2}$  & $T^{xyz}= \frac{2}{3\sqrt{3}}\overline{J^{x}J^{y}J^{z}}$  \tabularnewline \tabularnewline 
  
 & $\Gamma_{5}$  & $T^x \equiv T^x_\beta = \frac{2}{3\sqrt3}[\overline{J^{x}(J^{y})^{2}}-\overline{(J^{z})^{2}J^{x}}]$ \tabularnewline \tabularnewline 
 &  & $T^y \equiv T^y_\beta = \frac{2}{3\sqrt3}[\overline{J^{y}(J^{z})^{2}}-\overline{(J^{x})^{2}J^{y}}]$ \tabularnewline \tabularnewline 
 &  & $T^z \equiv T^z_\beta = \frac{2}{3\sqrt3}[\overline{J^{z}(J^{x})^{2}}-\overline{(J^{y})^{2}J^{z}}]$ \tabularnewline \tabularnewline 
\hline
\hline
\end{tabular}
\end{table}

We can label the multipolar operators   acting on $j_\mathrm{eff} = \frac{3}{2}$ states  by irreducible representations of the octahedral group \cite{Santini-RMP(2009)}. As shown in Table \ref{Tab_Multipoles}, the quadrupole moment has three components, $O^\gamma$, with $\gamma\in\{1,2,3\}\equiv\{x,y,z\}$, which forms a $\Gamma_5$ representation, and  two components, $O_2$ and $O_3$,  which transform in the $\Gamma_3$ representation. From the latter, we define the linear combinations \be
\tilde O^\gamma =\cos\left(\frac{2\pi\gamma}3\right)O_3+\sin\left(\frac{2\pi\gamma}3\right)O_2,
\ee
so that a $C_3$ rotation about the $(111)$ axis perpendicular to the lattice  plane acts as $\gamma\mapsto \gamma+1$ for both $O^\gamma$ and $\tilde O^\gamma$.
We then consider bond-dependent quadrupole-quadrupole interactions of the form\be
H_q=\sum_{\gamma=x,y,z}\sum_{\langle jl\rangle_\gamma}(K_qO_j^\gamma O_l^\gamma+K_q'\tilde O_j^\gamma \tilde O_l^\gamma),\label{quadrupole}
\ee
where $\langle jl \rangle_\gamma$ refers to   nearest-neighbor sites along a $\gamma $ bond of the honeycomb lattice, as in the standard notation for the Kitaev model \cite{Trebst-PR(2022)}. The first term in $H_q$ can be generated by integrating out vibronic couplings to  trigonal distortions, which are significant in some cubic $d^1$ systems \cite{Iwahara-PRB(2018),Khomskii-CR(2021)}. This term was considered in the model for a quadrupolar spin liquid defined in Ref. \cite{Pereira-PRB(2020)}. The   model with the pure $K_q$  interaction has a macroscopically  degenerate ground-state manifold \cite{Pereira-PRB(2020)}. Adding perturbations to this special point in the phase diagram, we can access different phases with exotic multipolar orders or spin-orbital liquid behavior.  On the other hand, the second term in Eq. (\ref{quadrupole}) can be generated  by coupling to tetragonal distortions or by electrostatic interactions   \cite{Balents-PRB(2011)}. This term is known to favor quadrupolar  order with a vortex pattern \cite{Wu-PRL(2008),Khaliulin-PRR(2021)}.

We now add octupole-octupole interactions to our model. To reduce the number of parameters, we focus on interactions that involve the  components   transforming  in the $\Gamma_2$ and $\Gamma_5$ representations shown in Table  \ref{Tab_Multipoles}. This means that we neglect the octupolar operators, which, like the dipole moment $\mathbf{J}$, transform in the $\Gamma_4$ representation \cite{Santini-RMP(2009)}. As we shall see in Sec. \ref{sec:solvable}, this choice is convenient for the purpose of identifying exactly solvable limits of the general model. We then consider \be
H_{o}=\sum_{\gamma=x,y,z}\sum_{\langle jl\rangle_\gamma}(K_oT_j^\gamma T_l^\gamma+K_o'T^{xyz}_j T^{xyz}_l).\label{octupolar}
\ee
Note that the octupolar operators are odd under time reversal, but the interactions are time-reversal invariant. A bond-dependent octupolar interaction similar to the first term in Eq. (\ref{octupolar})    appears in phenomenological  models for the hidden multipolar order in the heavy-fermion compound Ce$_x$La$_{1 - x}$B$_6$ \cite{Kubo-JPSJ(2003),Kubo-JPSJ(2004)}. The second term in   Eq. (\ref{octupolar})    favors an octupolar order without  lattice distortions analogous to the one proposed  for  double perovskites with $j_{\rm eff}=2$ local moments \cite{Paramekanti-PRB(2020),Churchill-PRB(2022)}. Altogether, the Hamiltonian $H=H_q+H_o$  captures the competition between different types of multipolar order.

To analyze the multipolar interactions, it is useful to express the  $j_\mathrm{eff} = \frac{3}{2}$ states in terms of pseudospin and a pseudo-orbital degrees of freedom  \cite{Shiina-JPSJ(1997),Pereira-PRL(2016),Natori-PRB(2017),Pereira-PRB(2020)}. The local Hilbert space  is spanned by the eigenstates $\vert m_J \rangle$ of  $J^z$, which obey $J^z \vert m_J \rangle = m_J \vert m_J \rangle$, with $m_J \in \{ \pm \frac{1}{2}, \pm \frac{3}{2}\}$. We define the mapping 
\begin{align}
\Big \vert m_J = \frac{3}{2} \Big\rangle & = \Big \vert s^z = - \frac{1}{2}, \tau^z = \frac{1}{2} \Big\rangle, \\
\Big \vert m_J = \frac{1}{2} \Big\rangle & = - \Big \vert s^z = \frac{1}{2}, \tau^z = - \frac{1}{2} \Big\rangle, \\
\Big \vert m_J = - \frac{1}{2} \Big\rangle & = \Big \vert s^z = - \frac{1}{2}, \tau^z = - \frac{1}{2} \Big\rangle, \\
\Big \vert m_J = - \frac{3}{2} \Big\rangle & = - \Big \vert s^z = \frac{1}{2}, \tau^z = \frac{1}{2} \Big\rangle,
\end{align}
where $\mathbf{s}_j$ and $\boldsymbol{\tau}_j$ are, respectively, pseudospin and pseudo-orbital operators, which obey the SU(2) algebra $[ s^\alpha_j, s^\beta_k ] = i \epsilon^{\alpha \beta \gamma} \delta_{j k} s^\gamma_j$ and $[ \tau^\alpha_j, \tau^\beta_k ] = i \epsilon^{\alpha \beta \gamma} \delta_{j k} \tau^\gamma_j$ and the relation $[ s^\alpha_j, \tau^\beta_k ] = 0$. In the $\{| s^z, \tau^z \rangle \}$ basis, the quadrupolar and octupolar operators given in Table \ref{Tab_Multipoles}  become \bea
O^\gamma&=&-4s^\gamma\tau^y,\qquad O_3=2\tau^z,\qquad O_2 =2\tau^x,\nonumber\\
T^{\gamma}&=&-4s^\gamma\mathbf v^\gamma\cdot \boldsymbol\tau,\qquad T^{xyz}=2\tau^y,\label{opsST}
\eea
where $\mathbf{v}^\gamma =\cos(2\pi\gamma/3)\hat{\mathbf x} -\sin(2\pi\gamma/3)\hat{\mathbf z}$ are unit vectors in the $xz$ plane of pseudo-orbital space.

\section{Exactly solvable model  \label{sec:solvable}}

In this section, we analyze the Hamiltonian $H=H_q+H_o$ in the special limit  $K_q'=K_o'=0$, in which we obtain\be
H_s=\sum_{\gamma=x,y,z}\sum_{\langle jl\rangle_\gamma}(K_qO_j^\gamma O_l^\gamma+K_oT_j^\gamma T_l^\gamma).\label{Eq_MSL_Ham}
\ee
We show that this model can be solved exactly  using a Majorana fermion representation for the multipolar operators and exhibits a MSL  ground state.

\begin{figure}[t]
\centering
\includegraphics[width=0.75\linewidth]{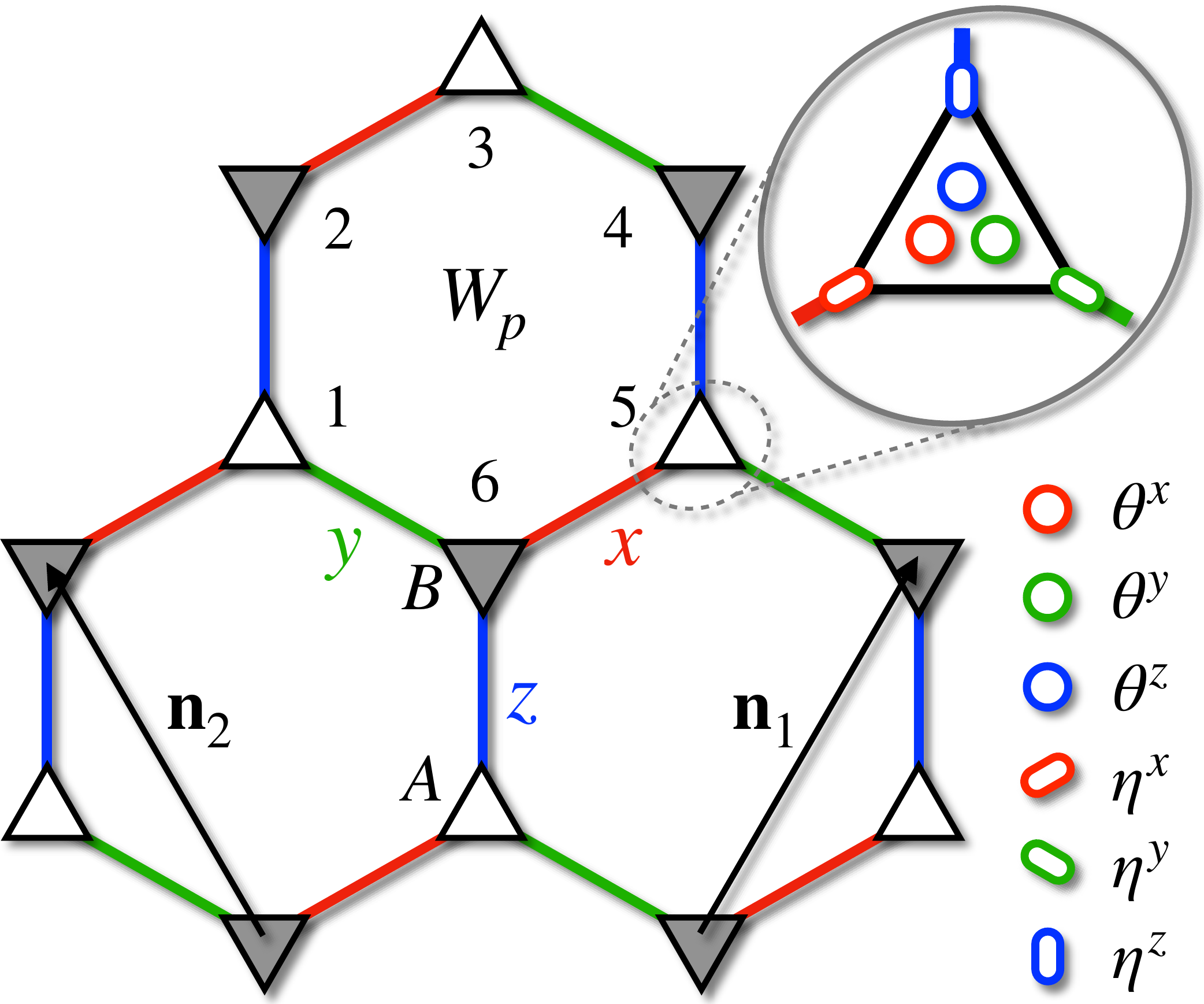}
\caption{Schematic representation of fractionalized  $j_{\mathrm{eff}} = \frac{3}{2}$ moments on the honeycomb lattice. Here  $x$, $y$, and $z$ bonds are represented by red, green, and blue solid lines, respectively. The two sublattices, A and B, are represented by up-pointing and down-pointing triangles.  The inset in the top right corner shows  the matter ($\theta^{\gamma}$) and gauge ($\eta^{\gamma}$) Majorana fermions. Each plaquette is associated with the conserved quantity $\hat{W}_{\small p} = - e^{i \pi (J^{x}_1 + J^{y}_2 + J^{z}_3 + J^{x}_4 + J^{y}_5 + J^{z}_6)}$.  }\label{Honeycomb_Lattice}
\end{figure}

We can write the pseudospin and pseudo-orbital operators in terms of six Majorana fermions as \cite{Pereira-PRL(2016),Pereira-PRB(2020)}
\begin{align}
s_j^\gamma&=-\frac{i}4\epsilon^{\alpha\beta\gamma}\eta_j^\alpha\eta_j^\beta, \label{Eq_Majorana_01}\\
\tau_j^\gamma&=-\frac{i}4\epsilon^{\alpha\beta\gamma}\theta_j^\alpha\theta_j^\beta, \label{Eq_Majorana_02} 
\end{align}
where $\epsilon^{\alpha\beta\gamma}$ is the Levi-Civita tensor and $\eta^{\alpha}$ and $\theta^{\alpha}$ are Majorana fermion operators which obey the algebra $\{\eta^{\alpha}_j, \eta^{\beta}_k\} = 2 \delta_{j k}\delta^{\alpha \beta}$, $\{\theta^{\alpha}_j, \theta^{\beta}_k\} = 2 \delta_{j k}\delta^{\alpha \beta}$, and $\{\eta^{\alpha}_j, \theta^{\beta}_k\} = 0$. This Majorana representation enlarges the   Hilbert space, introducing  unphysical states. To avoid that, we impose the local constraint \be
D_j \equiv i\eta_j^x\eta_j^y\eta_j^z\theta_j^x\theta_j^y\theta_j^z=1\ee
for every site $j$. This is equivalent to obtaining a physical state $\vert \Psi \rangle_\mathrm{phys}$ from the Majorana wave function $\vert \Psi_0 \rangle$ by implementing the projection $\vert \Psi \rangle_\mathrm{phys} = {\Pi_j [\frac12(1+D_j )] \vert \Psi_0 \rangle}$. Using the local constraint and the expressions in Eq. (\ref{opsST}), we can write the multipolar operators as Majorana bilinears:  \bea
O^\gamma&=&i\eta^\gamma\theta^y,\qquad O_3=-i\theta^x\theta^y,\qquad O_2 =-i\theta^y\theta^z,\nonumber\\
T^{\gamma}&=&-i\eta^\gamma(\mathbf v^\gamma\cdot\boldsymbol\theta),\qquad T^{xyz}=-i\theta^z\theta^x.\label{multipolesMajorana}
\eea
Note that $\mathbf v^\gamma\cdot\boldsymbol\theta=\cos(2\pi\gamma/3)\theta^x-\sin(2\pi\gamma/3)\theta^z$ only involves only  $\theta^x$ and $\theta^z$.

\begin{figure*}[t]
\centering
\centering \includegraphics[width=0.45\linewidth,valign=t]{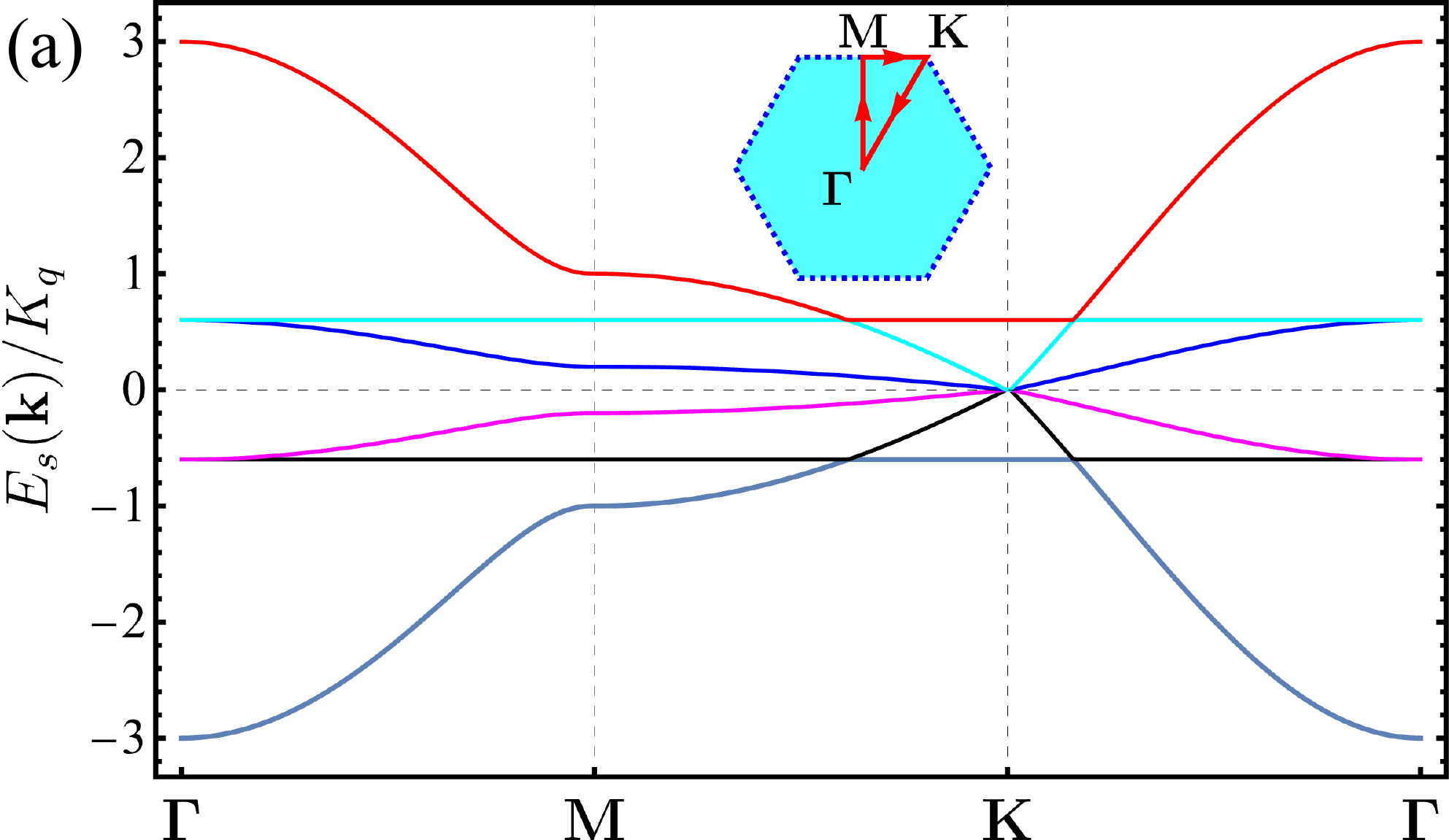} \hfil{} \includegraphics[width=0.45\linewidth,valign=t]{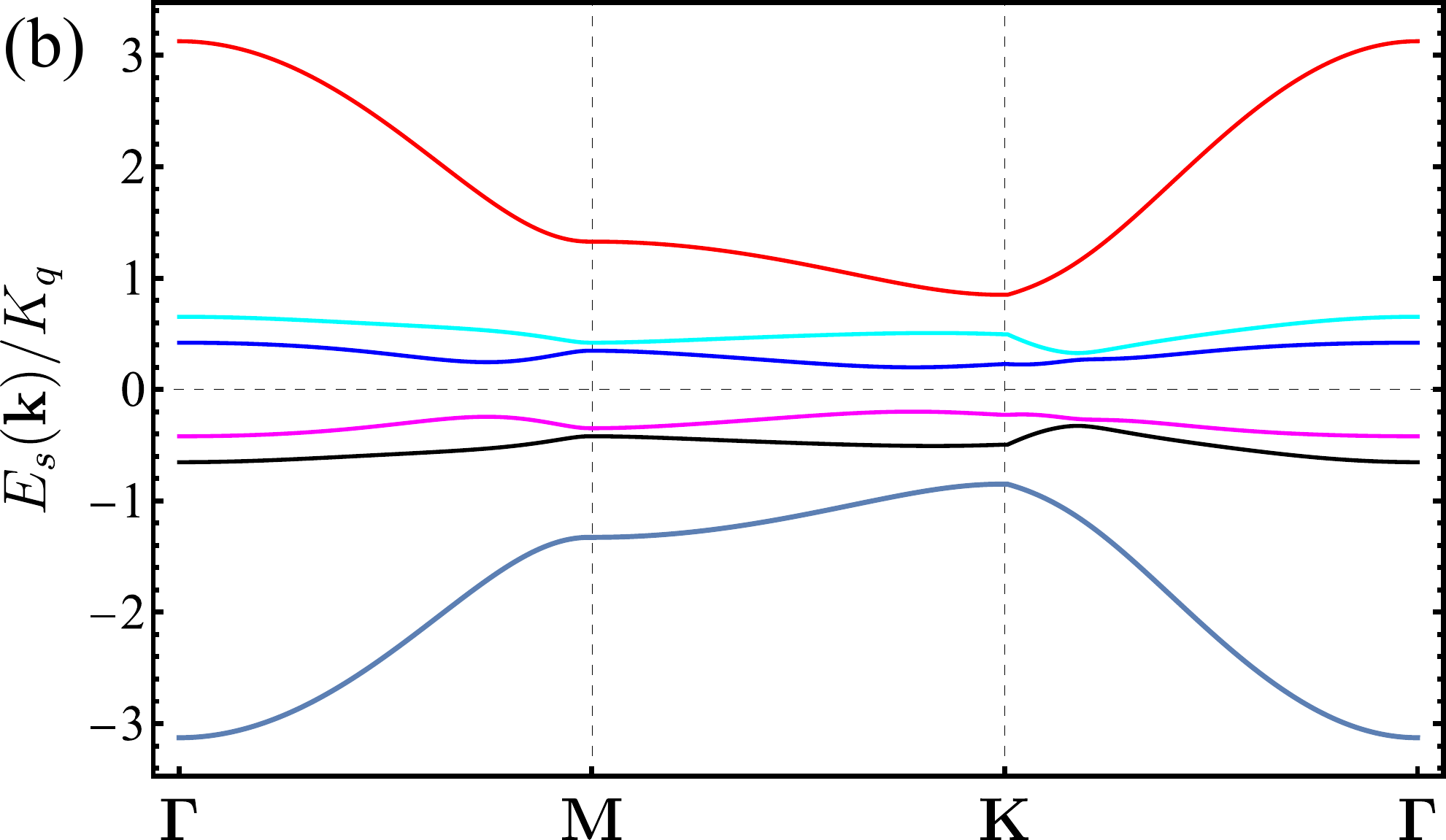}
\caption{Dispersion relation for  Majorana fermion excitations in  the exactly solvable  MSL model.  Here we set $K_o=0.4K_q$.   (a)  For the unperturbed model in   Eq. (\ref{Eq_MSL_Ham}), the system exhibits a gapless spectrum with  Dirac nodes at the $\mathbf{K}$ point. (b) In the presence of strain fields  $\varepsilon_2$ and $\varepsilon_{3}$ and time-reversal-symmetry-breaking field $h$ [see Eq. (\ref{Eq_BS_MSL_Ham_01})], the excitation spectrum becomes gapped. Here we set $\varepsilon_{2} =\varepsilon_{3}= 0.4K_q$ and $h = 0.1 K_q$.}\label{MSOL_Dispersions}
\end{figure*}

Using the Majorana representation, we cast the  Hamiltonian in the form
\begin{equation}\label{Eq_MSL_Maj_Ham}
H_s = i\sum_{\gamma }\sum_{\langle j l \rangle_{\gamma}} \hat{u}_{\langle jl\rangle_\gamma} [ K_q \theta^y_j\theta^y_l + K_o (\mathbf{v}^\gamma \cdot \boldsymbol{\theta}_j)(\mathbf{v}^\gamma \cdot \boldsymbol{\theta}_l) ],
\end{equation}
where $\hat{u}_{\langle j l \rangle_{\gamma}} \equiv - i\eta^\gamma_j\eta^\gamma_l$ are antisymmetric bond operators obeying $\hat{u}_{\langle j l \rangle_{\gamma}} = - \hat{u}_{\langle l j \rangle_{\gamma}}$.  These   operators commute with each other and define $\mathbb{Z}_2$ gauge fields since $H_s$ is invariant with respect to the local transformations $\theta^{\gamma}_j \mapsto \Lambda_j \theta^{\gamma}_j$ and $\hat{u}_{\langle j l \rangle_{\gamma}} \mapsto \Lambda_j \hat{u}_{\langle j l \rangle_{\gamma}} \Lambda_l$, with $\Lambda_j = \pm 1$. Like what happens for the dipolar Kitaev models \cite{Kitaev-AP(2006),Baskaran-PRB(2008)}, the Hamiltonian in Eq. \eqref{Eq_MSL_Ham}   possesses a conserved   operator $\hat{W}_{\small p} \equiv - e^{i \pi (J^{x}_1 + J^{y}_2 + J^{z}_3 + J^{x}_4 + J^{y}_5 + J^{z}_6)}$ for  each hexagonal  plaquette $p$. In the Majorana representation, the plaquette operators  read  $\hat{W}_{\small p} = \prod_{\small \langle j, l \rangle_{\gamma} \in p} \hat{u}_{\langle j, l \rangle_{\gamma}}$ \cite{Pereira-PRB(2020),Jin-NC(2022)}. Due to the form of the Hamiltonian in Eq. (\ref{Eq_MSL_Maj_Ham}), we refer to    $\eta^\gamma$ as the gauge  fermions and to $\theta^\gamma$ as the matter fermions.    The fractionalization of the multipolar operators into  Majorana fermions is illustrated in Fig. \ref{Honeycomb_Lattice}.

We verified numerically that the ground state of $H_s$ lies in the sector where  the eigenvalues of all conserved plaquette operators are set to $W_p = + 1$ (see Appendix \ref{App:GS_Energy}). Fixing a  gauge configuration with ${u}_{\langle j l \rangle_{\gamma}}=1$ for sites $j$ in sublattice A and $l$ in sublattice B,  we obtain from Eq. (\ref{Eq_MSL_Maj_Ham}) a translation-invariant quadratic Hamiltonian for the matter   fermions moving in the background of the static $\mathbb Z_2$ gauge field. Note that the $\theta^y$ fermion decouples from $\theta^x$ and $\theta^z$. The dynamics of $\theta^y$ is governed by the quadrupolar coupling $K_q$, whereas the dynamics of $\theta^x$ and $\theta^z$ depends on the octupolar coupling $K_o$. We diagonalize the Hamiltonian using  the Fourier transform \begin{equation}
\theta^\gamma_b (\mathbf{R}) = \sqrt{\frac{2}{N}} \sum_{\mathbf{k} \in \frac{1}{2}(\mathrm{BZ})} [ e^{- i\mathbf{k}\cdot\mathbf{R}} \theta^\gamma_b (\mathbf{k}) + e^{ i\mathbf{k}\cdot\mathbf{R}} \theta^{\gamma \dagger}_b (\mathbf{k}) ], 
\end{equation}
where $\mb R$ is the position of the unit cell, $b \in \{A, B\}$ refers to the sublattice index, $N$ is the number of sites,  and the momentum sum runs  over half of the Brillouin zone (BZ). The bands associated with the $\theta^y$ fermions have the dispersion relation $E_s(\mb k)=\pm |K_qg_y(\mb k)|$, where $g_y(\mb k)= 1+e^{i\mb k\cdot \mb n_1}+e^{i\mb k\cdot \mb n_2}$, with lattice vectors $\mathbf{n}_{1}=\frac{\sqrt{3}}{2}(1,\sqrt{3})$ and  $\mathbf{n}_{2}=\frac{\sqrt{3}}{2}(-1,\sqrt{3})$. For the bands associated with $\theta^x$ and $\theta^z$, we obtain   \begin{widetext}
\be
E_s(\mb k)=\pm \frac{|K_o|}{\sqrt{2}} \sqrt{|g_z(\mathbf{k})|^2 + |g_x(\mathbf{k})|^2 + 2 |g_{xz}(\mathbf{k})|^2 \pm \sqrt{[|g_z(\mathbf{k})|^2 - |g_x(\mathbf{k})|^2]^2 + 4 |g_x(\mathbf{k}) g^*_{xz}(\mathbf{k}) + g^*_z(\mathbf{k}) g_{xz}(\mathbf{k})|^2}}, \label{Esk}
\ee
\end{widetext}
where we define  the functions $g_z(\mathbf{k}) ={ \frac{3}{4}(e^{i \mathbf{k} \cdot \mathbf{n}_1} +  e^{i \mathbf{k} \cdot \mathbf{n}_2})}$, $g_x(\mathbf{k}) =1+ {\frac{1}{4}( e^{i \mathbf{k} \cdot \mathbf{n}_1} + e^{i \mathbf{k} \cdot \mathbf{n}_2})} $,   and $g_{xz}(\mathbf{k}) ={ \frac{\sqrt{3}}{4}(  e^{i \mathbf{k} \cdot \mathbf{n}_1} - e^{i \mathbf{k} \cdot \mathbf{n}_2})}$.

The band structure of the Majorana fermion excitations  is shown in Fig. \ref{MSOL_Dispersions}(a). The spectrum exhibits gapless modes with Dirac cones at the $\mathbf{K}$ point  in    both  sectors associated with  $\theta^y$ fermions and  mixed $\theta^x$ and $\theta^z$ fermions. The latter sector also displays a completely flat gapped  band. This  peculiar feature holds for only the model with spatially isotropic  couplings; the gapped band becomes dispersive as soon as  we introduce different octupolar couplings $K_o^{(\gamma)}$ for each bond direction $\gamma$. Importantly, by generating a dispersion for $\theta^x$ and $\theta^z$ fermions, the octupolar interaction lifts the macroscopic degeneracy of the pure quadrupolar model \cite{Pereira-PRB(2020)}. Consequently, our exactly solvable model stabilizes  a unique MSL ground state.

In addition to   gapless Majorana fermions, the spectrum of $H_s$   contains vortex excitations (visons)  associated with  changing the eigenvalues of the plaquette operators to $W_p=-1$. These gapped excitations are created in pairs by the action of local operators which anticommute with the bond operators $\hat{u}_{\langle j l \rangle_{\gamma}}$. Remarkably, the quadrupole operators $O_2$ and $O_3$ and the octupolar operator $T^{xyz}$ do not excite visons since they commute with the gauge fermions [see Eq. \eqref{multipolesMajorana}]. As a result,  $O_2$,   $O_3$, and $T^{xyz}$ are the only on-site operators featuring  a  gapless spectrum in the corresponding dynamical structure factors probed by resonant  inelastic x-ray scattering \cite{Natori-PRB(2017)}. By contrast, the dynamical spin structure factor  for the pure  Kitaev model exhibits a flux gap \cite{Chalker-PRL(2014)}.

In analogy with  the Kitaev model, we can drive the MSL to topological phases   by adding perturbations that   break symmetries and gap out the fermionic spectrum. We can break the $\mathbb Z_3$ symmetry (rotation in real and spin-orbital space) and time-reversal symmetry while still preserving the integrability of the model if we consider the perturbations 
\begin{align}
\delta H_s & =  - \sum_j (\varepsilon_2 O_{2,j} +\varepsilon_3 O_{3,j} + hT^{x y z}_j)\label{Eq_BS_MSL_Ham_01}.
\end{align}
 Here $\varepsilon_2$ and $\varepsilon_3$ describe tetragonal strain fields and  $h$ induces an octupole moment, allowed by symmetry in the presence of an external magnetic field. According to Eq. (\ref{multipolesMajorana}), the total Hamiltonian remains quadratic in the matter fermions, but the  strain fields couple  $\theta^y$ to $\theta^x$ and $\theta^z$.  As shown in Fig. \ref{MSOL_Dispersions}(b), we find a fully gapped spectrum for generic values of $\varepsilon_{2}$, $\varepsilon_{3}$, and $h$.  However, there exist critical values of these parameters for which  the Majorana gap closes and then reopens, which is suggestive of  topological phase transitions.

\begin{figure}[t]
\centering
 \includegraphics[width=0.90\columnwidth]{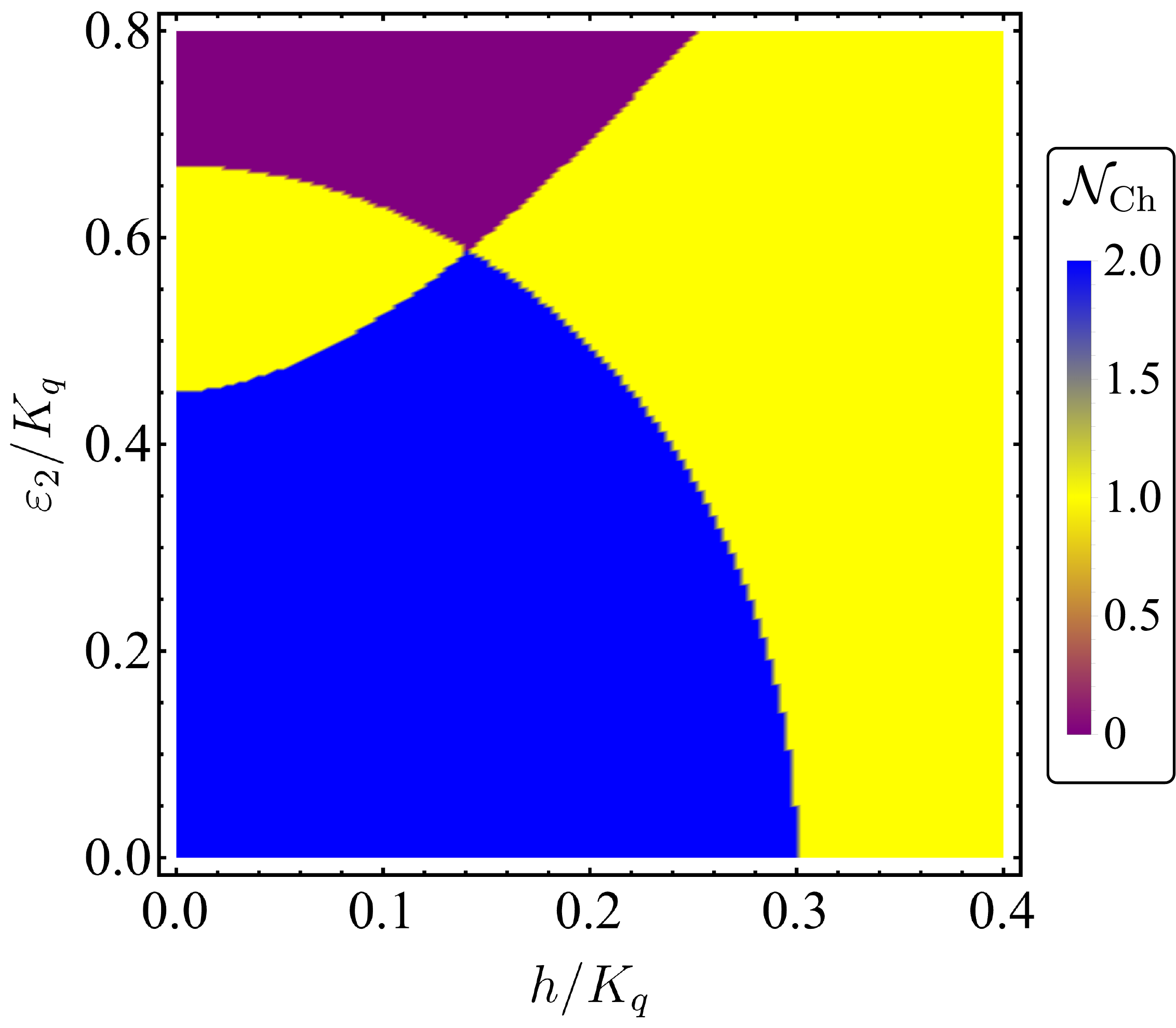} 
\caption{Topological phase diagram of the MSL as a function of the symmetry-breaking fields $\varepsilon_2$ and $h$. Here we set $K_o = 0.2 K_q$ and $\varepsilon_2=\varepsilon_3$. The purple, yellow, and blue regions correspond to Chern numbers 0, 1, and 2, respectively.}\label{Chern_Number}
\end{figure}

To investigate the topological nature of the gapped phases, we  evaluate the Chern number
\begin{equation}
\mathcal{N}_\text{Ch} \equiv \frac{1}{2 \pi} \sum\limits_{n} \int_\text{BZ} d^2 \mathbf{k} \,\mathcal{B}_{n}(\mathbf{k}),
\end{equation}
where ${\mathcal{B}}_{n}(\mathbf{k}) = [\boldsymbol{\nabla \times} \boldsymbol{\mathcal{A}}_{n}(\mathbf{k})]_z$ is the Berry curvature  associated with the $n$th energy band; $\boldsymbol{\mathcal{A}}_{n}(\mathbf{k}) = i \langle u_n(\mathbf{k})| \boldsymbol{\nabla}_{\mathbf{k}} \vert u_n(\mathbf{k}) \rangle$ is the Berry connection, with  $|u_n(\mathbf{k}) \rangle$ being the eigenvectors of the Hamiltonian; and the sum on the right-hand-side runs over the occupied   bands. We determine $\mathcal{N}_\text{Ch}$ numerically following the approach devised in Ref. \cite {Fukui-JPSJ(2005)}. The dependence of $\mathcal{N}_\text{Ch}$ on the symmetry-breaking fields   is shown in Fig. \ref{Chern_Number}. Remarkably,  the interplay between strain and magnetic fields   realizes five of  Kitaev's 16-fold-way   phases \cite{Kitaev-AP(2006),Batista-PRR(2020),Chulliparambil-PRB(2020)}: a trivial phase with   $\mathcal{N}_\text{Ch} = 0$, two non-Abelian topological phases with $\mathcal{N}_\text{Ch} = \pm 1$, and two  Abelian topological phases with  $\mathcal{N}_\text{Ch} = \pm 2$. For  $0<|\varepsilon_2|,|\varepsilon_3|$ and $|h|\ll |K_q|$, the system is in  an Abelian phase with $\mathcal{N}_\text{Ch} = \pm 2$, depending on the sign of $h$. We  can reach the non-Abelian phase by increasing the strength of the symmetry-breaking fields, whereas the trivial phase appears only for strong strain fields. Varying the ratio $K_o/K_q$, we observe that enhancing  the octupolar interaction favors the Abelian phase.    We have   verified that the Majorana  gap closes  at   topological  phase transitions where the Chern number changes. The momentum at which the gap closes depends on the interactions and is not attached to the $\mathbf{K}$ point. A direct transition from $\mathcal{N}_\text{Ch} = \pm 2$ to $\mathcal{N}_\text{Ch} = 0$ is possible by fine tuning so that the gap closes simultaneously at two different points in the Brillouin zone.

\begin{figure}[t]
\centering
\centering \includegraphics[width=0.95\columnwidth,valign=t]{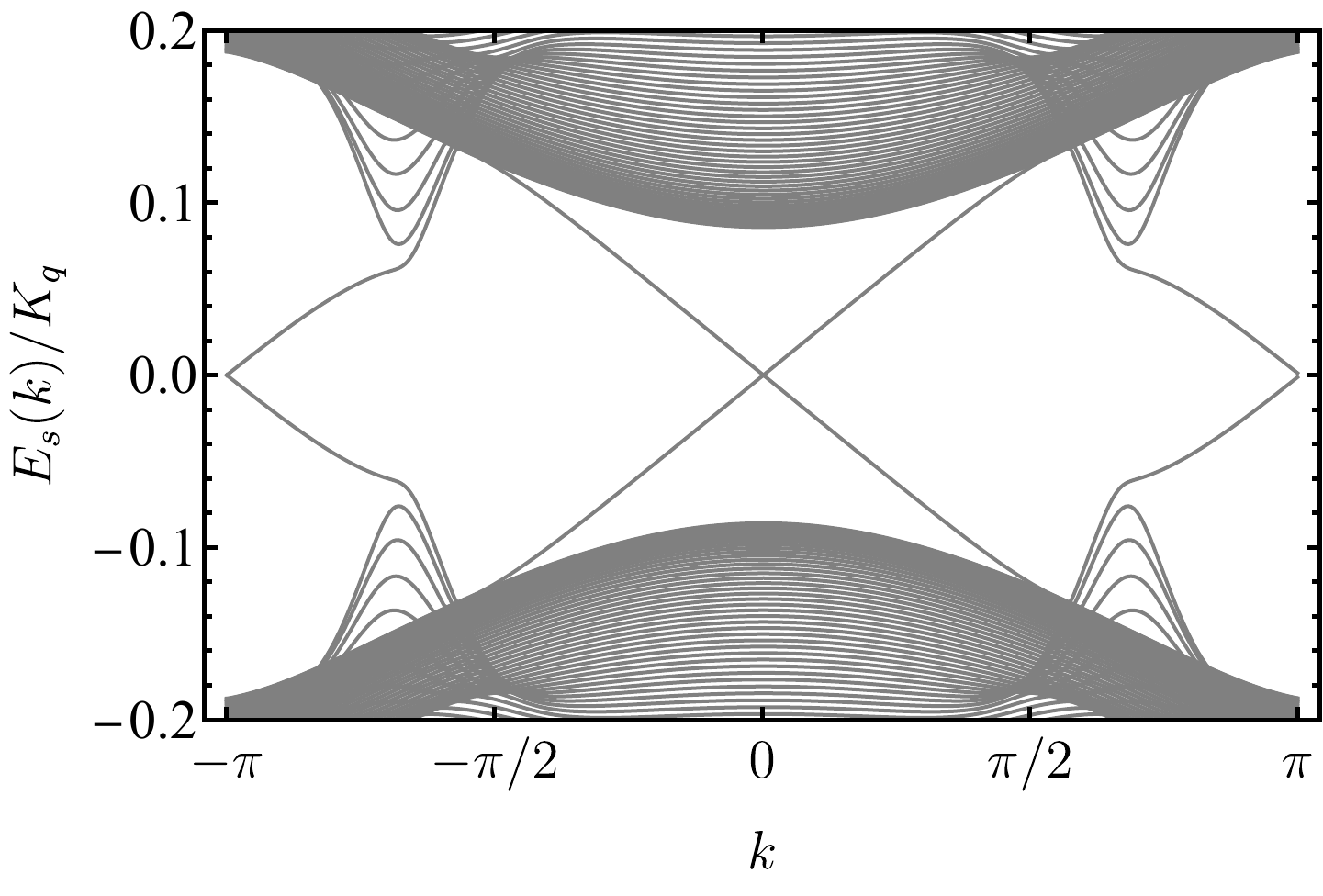}
\caption{Edge states of the MSL  for a honeycomb lattice with zigzag edges. Here we use a cylinder   with $L = 50$ unit cells along the open direction. We fix the parameters as  $K_o =h =\varepsilon_{2} = \varepsilon_{3} = 0.2 K_q$ in the regime of   $\mathcal{N}_\text{Ch} = \pm 2$ (see  Fig. \ref{Chern_Number}).}\label{Zigzag_Edge_States}
\end{figure}

As a  signature of the topological phases, we   look for  chiral  edge states  in a cylinder geometry with zigzag edges. The result  matches the Chern number, in accordance with the bulk-boundary correspondence. Figure \ref{Zigzag_Edge_States} shows the spectrum with two chiral edge states in the regime where $\mc N_{\rm Ch}=\pm 2$.   The chiral edge states can be probed by  the thermal Hall conductance $\kappa_{xy}$, which for low enough temperatures behaves as $\kappa_{xy} = \frac{\mathcal{N}_{\text{Ch}}}{2} \frac{\pi k^2_B}{6 \hbar}T$.

\section{Multipolar orders from integrability-breaking interactions}\label{Sec_IV}

In this section, we address the effect of  the interactions $K_q'$ and $K_o'$  in  Eqs. (\ref{quadrupole}) and (\ref{octupolar}) on the ground state and elementary excitations of the spin-orbital system.  We define\be
H_I=\sum_{\gamma}\sum_{\langle jl\rangle_\gamma}(K'_q\tilde{O}_j^\gamma \tilde{O}_l^\gamma+K_o'T^{xyz}_j T^{xyz}_l),
\ee
so that $H=H_q+H_o=H_s+H_I$. Hereafter we assume that all coupling constants are positive. In the Majorana representation in Eq. (\ref{multipolesMajorana}), $H_I$ gives rise to   interaction terms which  spoil the integrability of the model. For this reason, we   resort to two complementary analytical  approximations which    provide a qualitative picture of the excitation spectrum, namely  parton mean-field theory and orbital wave theory.

\subsection{Parton mean-field theory}\label{Sec_IV.1}

For small $K_q'$ and $K_o'$, we can start from  the solution  in Sec. \ref{sec:solvable}   to investigate  quantum phase transitions out of the MSL state. We first notice that the quartic terms contained in $H_I$ involve only the matter fermions $\theta^{\gamma}$. Thus, at least at weak coupling they do not alter the gauge-flux configuration in the ground state, and  we  can  work in the sector with fixed $W_p=1$ for all plaquettes.

\begin{figure}[t]
\centering
\includegraphics[width=0.60\linewidth]{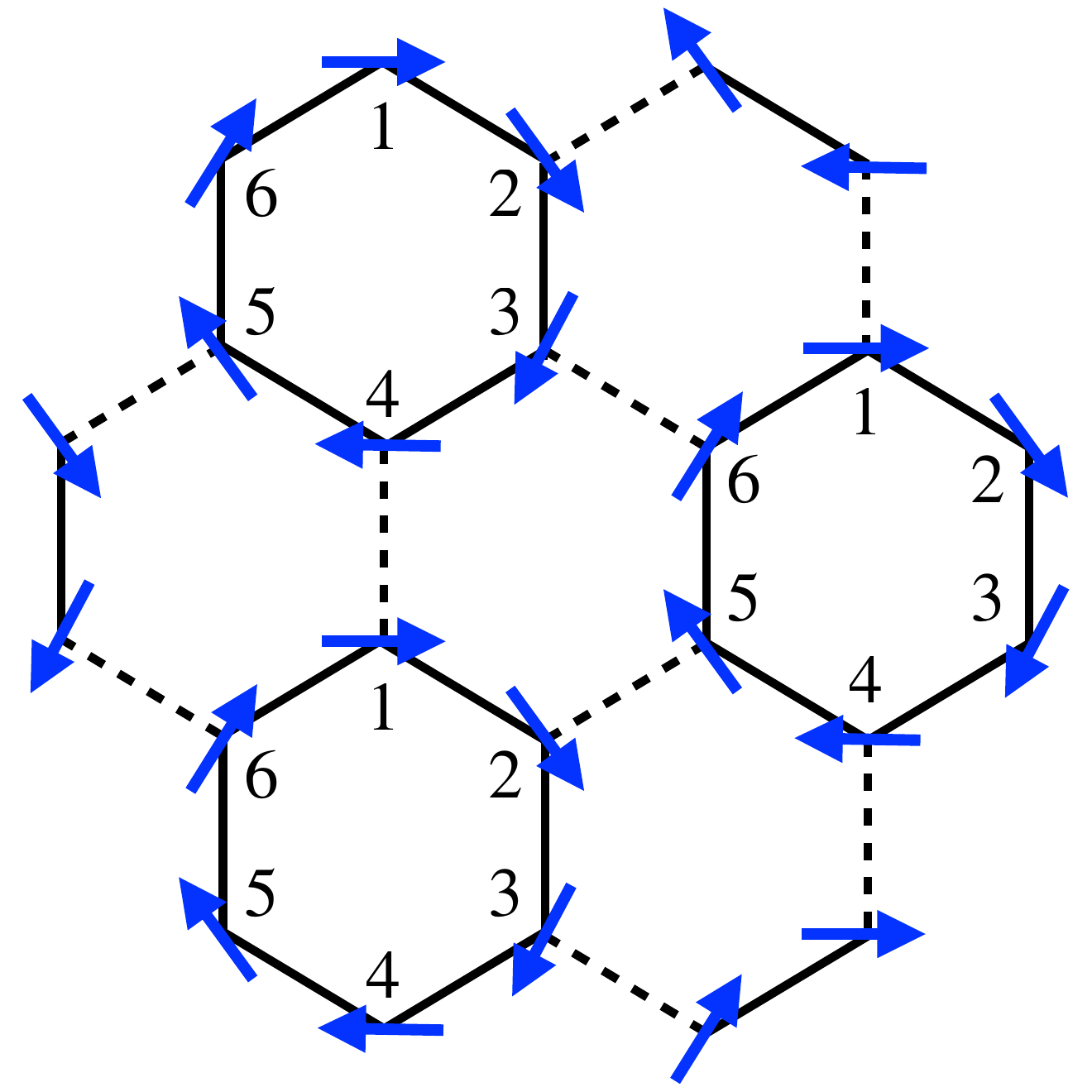}
\caption{Schematic representation of the ferroquadrupolar-vortex state on the honeycomb lattice. The arrows represent the direction of the  vector defined by $(\langle O_2\rangle,\langle O_3\rangle)=2(\langle \tau^x\rangle,\langle \tau^z\rangle)\propto (\cos\phi_n,-\sin\phi_n)$, where $\phi_n=(n-1)\pi/3$, with $n=1,\dots, 6$ being the sublattice index.}\label{Orbital_Configuration}
\end{figure}

Next, we treat the interactions  within a parton mean-field  approach \cite{Savary-RPP(2017),Jin-NC(2022)}. To search for multipolar orders, we use the mean-field decoupling
\bea
\tilde{O}^\gamma_j\tilde{O}^\gamma_l & \mapsto&  \langle \tilde{O}^\gamma_j\rangle \tilde{O}^\gamma_l  +\langle \tilde{O}^\gamma_l\rangle \tilde{O}^\gamma_j  -\langle \tilde{O}^\gamma_j\rangle \langle \tilde{O}^\gamma_l \rangle, \\
T^{xyz}_j T^{xyz}_l & \mapsto &\langle T^{xyz}_j \rangle T^{xyz}_l + T^{xyz}_j \langle T^{xyz}_l \rangle \nonumber \\
&& - \langle T^{xyz}_j \rangle \langle T^{xyz}_l \rangle.\label{decouplT}
\eea
The quadrupolar interaction $K_q'$ is equivalent to an orbital compass model on the honeycomb lattice \cite{Wu-PRL(2008),Khaliulin-PRR(2021)} and favors a six-sublattice  FQV state  that  breaks   rotational and translational symmetries (see Fig. \ref{Orbital_Configuration}).  We parametrize the quadrupolar  order parameter  as
\begin{equation}\label{Eq_Ferro_Orb_Vort}
\langle  \tilde {O}^\gamma_j \rangle = \rho \sum^{2}_{\mu = 1}  f^\gamma_\mu(\varphi) \cos(\mathbf{Q}_\mu \cdot \mathbf{R}_j).
\end{equation}
Here $\rho$ and $\varphi$ are   variational  parameters, the functions $ f^\gamma_\mu(\varphi)$ are given by $ f^\gamma_1(\varphi)=\sin(2\pi\gamma/3-\varphi)$ and $ f^\gamma_2(\varphi)=\cos(2\pi\gamma/3-\varphi)$, and  the wave vectors $\mathbf{Q}_1 = \frac{\pi}{\sqrt{3}}(\sqrt{3}, 1)$ and $\mathbf{Q}_2 = \frac{\pi}{\sqrt{3}}(- \sqrt{3}, 1)$ connect high-symmetry points of the reciprocal   lattice \cite{Fernandes-ARCMP(2019)}. On the other hand, for $K_o' > 0$ the octupolar order parameter $\langle T^{xyz}_j \rangle$ is constrained to having the familiar form for   antiferromagnetic order on a bipartite lattice, i.e.,
\begin{equation}\label{Eq_Anti_Oct}
\langle T^{xyz}_j \rangle =  
\begin{dcases} 
+ \chi \; \text{if $j \in \text{A}$},\\
- \chi \; \text{if $j \in \text{B}$}.
\end{dcases}
\end{equation}
A nonzero value of $\chi$ implies a phase with  AO order.

\begin{figure}[t]
\centering
\centering \includegraphics[width=0.95\linewidth,valign=t]{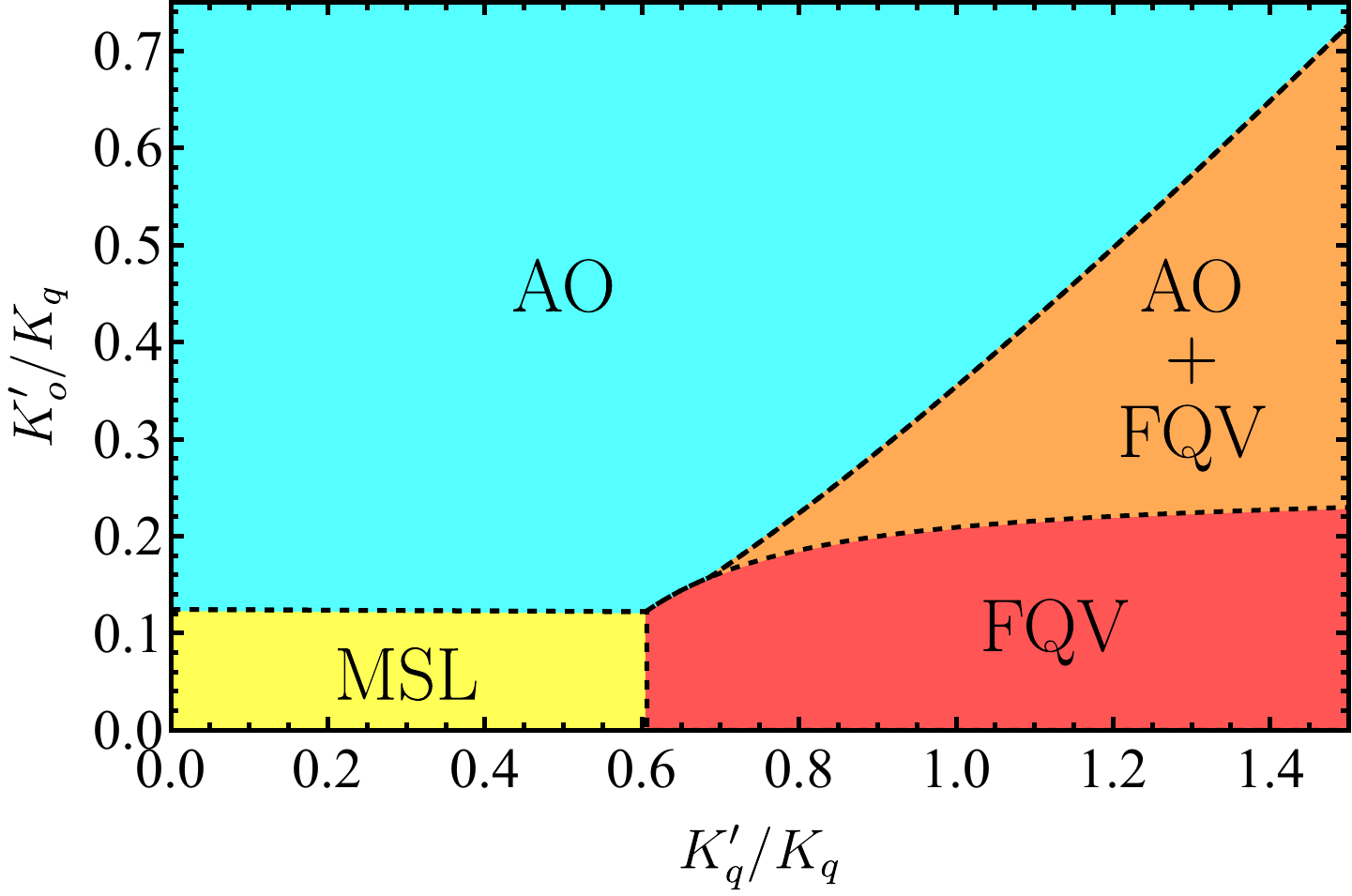} 
\caption{Ground-state phase diagram obtained from the parton mean-field theory  setting $K_o=0.2K_q$.  Here AO and FQV refer to the antiferro-octupolar and ferroquadrupolar-vortex phases, respectively. In addition, the system exhibits a phase with coexisting AO and FQV orders. }\label{phasediagram}
\end{figure}

The precise form of the mean-field Hamiltonian can be found in Appendix \ref{App:MF}. From this Hamiltonian, we obtain the free energy  
\bea\label{Eq_Free_Energy}
\mathcal{F}_\text{MF} & =& - T \sum^{18}_{\ell = 1} \int_{\frac{1}{6}(\text{BZ})} \frac{d^2 \mathbf{k}}{\mathcal{A}_\text{BZ}} \ln \big[1 + e^{-\beta E_{\ell}(\mathbf{k})} \big] \nonumber \\
&& + K_q' \big [ \mathcal{A}_0 - \mathcal{B}_0 \sin(2 \varphi) \big] \rho^2 + 3 K_o' \chi^2,
\eea
where $\beta=1/T$ is the inverse temperature and
\begin{align}
\mathcal{A}_0 & \equiv \bigg | 2 \cos\bigg(\frac{\pi}{\sqrt{3}} \bigg) + \cos\bigg(\frac{2\pi}{\sqrt{3}} \bigg) \bigg | \approx 1.365, \\
\mathcal{B}_0 & \equiv \sqrt{3} \sin \bigg (\frac{\pi}{2 \sqrt{3}} \bigg) \sin \bigg (\frac{\sqrt{3} \pi}{2} \bigg) \approx 0.557.
\end{align}
In addition, $\mathcal{A}_\text{BZ} = 8 \pi^2/(3 \sqrt{3} )$ refers to the area of the BZ and $E_{\ell}(\mathbf{k})$ are the   dispersion relations obtained by numerical diagonalization of the mean-field Hamiltonian.

\begin{figure*}[t]
\centering
\centering \includegraphics[width=0.32\linewidth,valign=t]{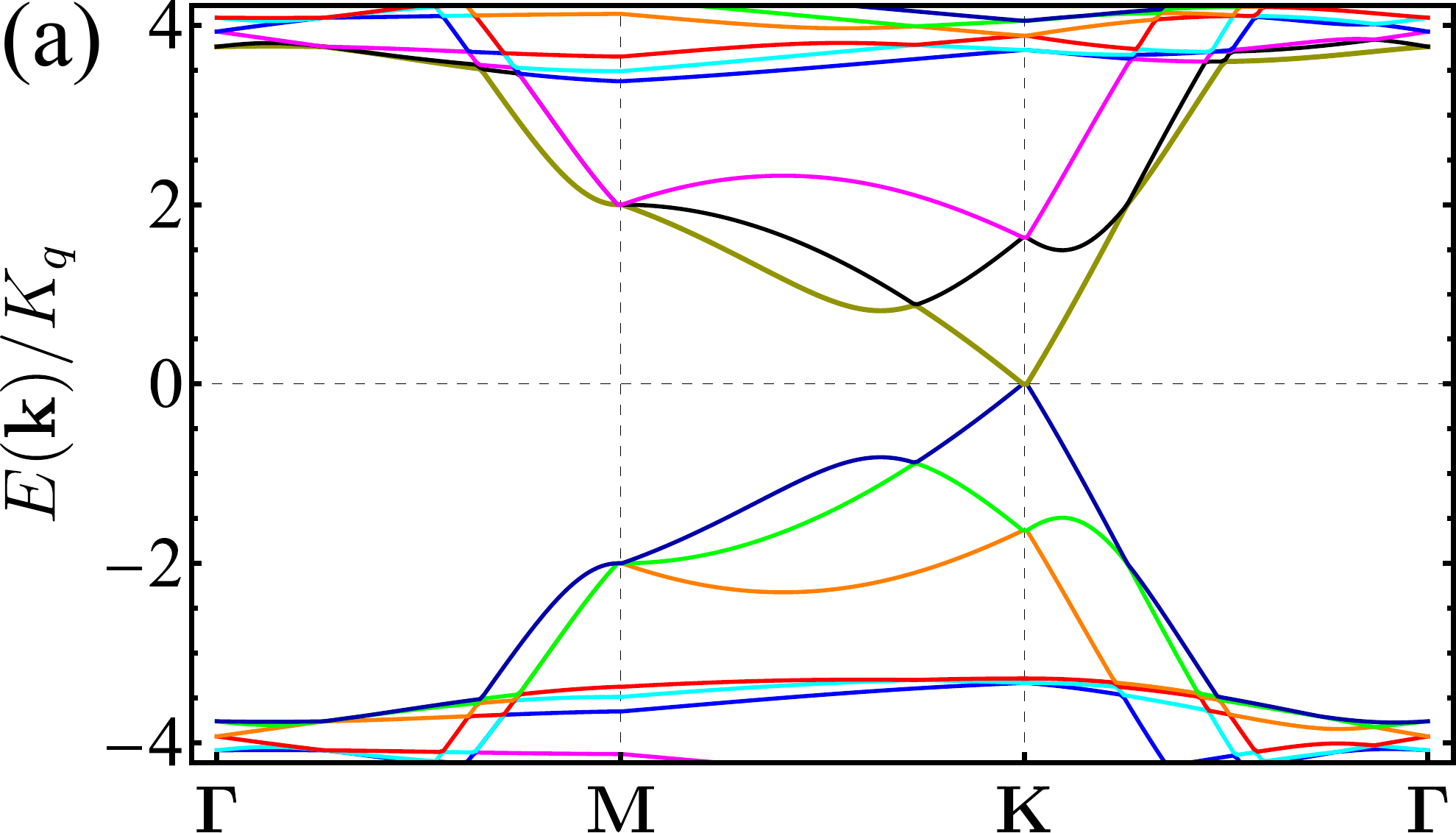}\hfil{} \includegraphics[width=0.32\linewidth,valign=t]{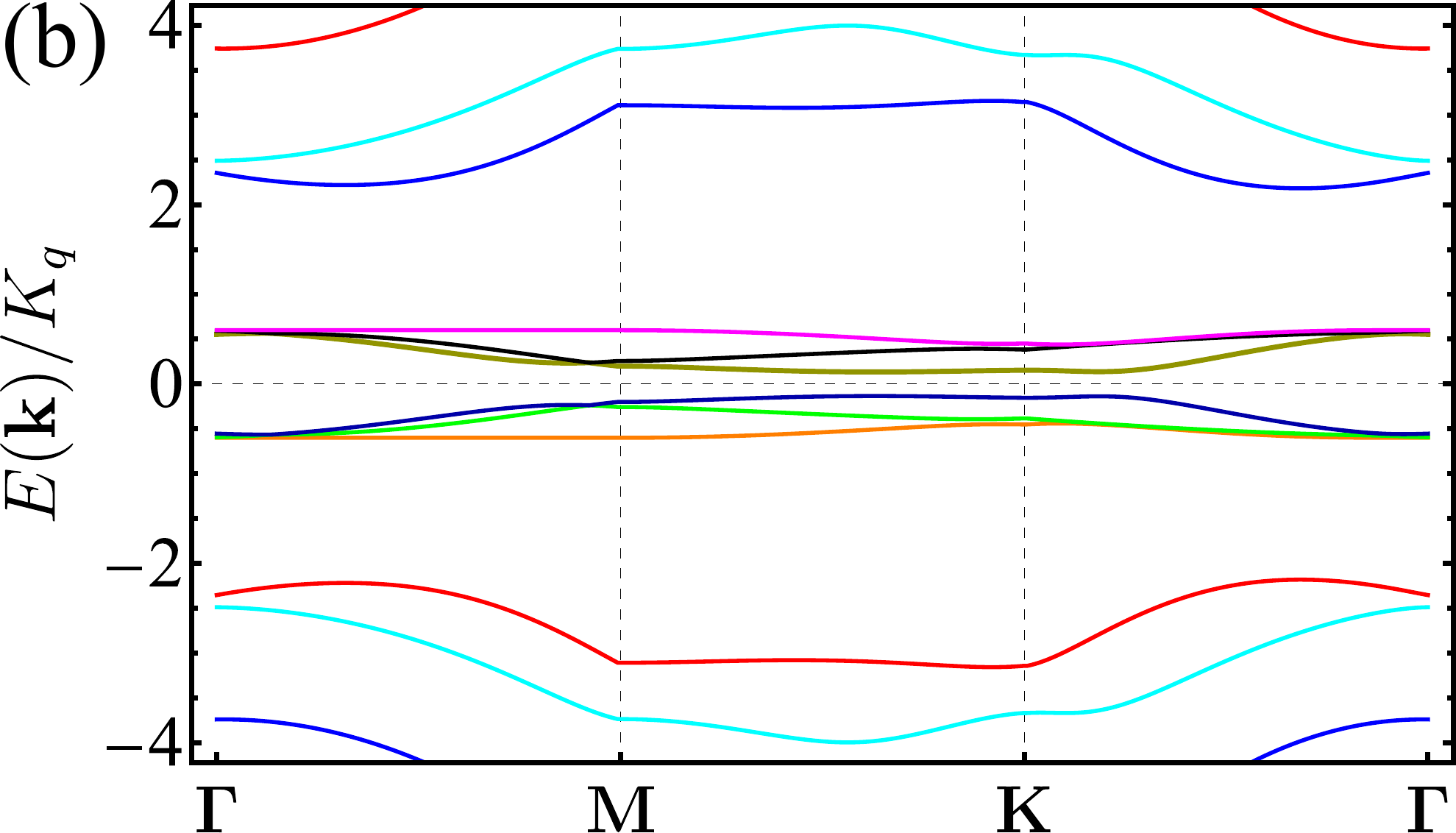}\hfil{}\includegraphics[width=0.32\linewidth,valign=t]{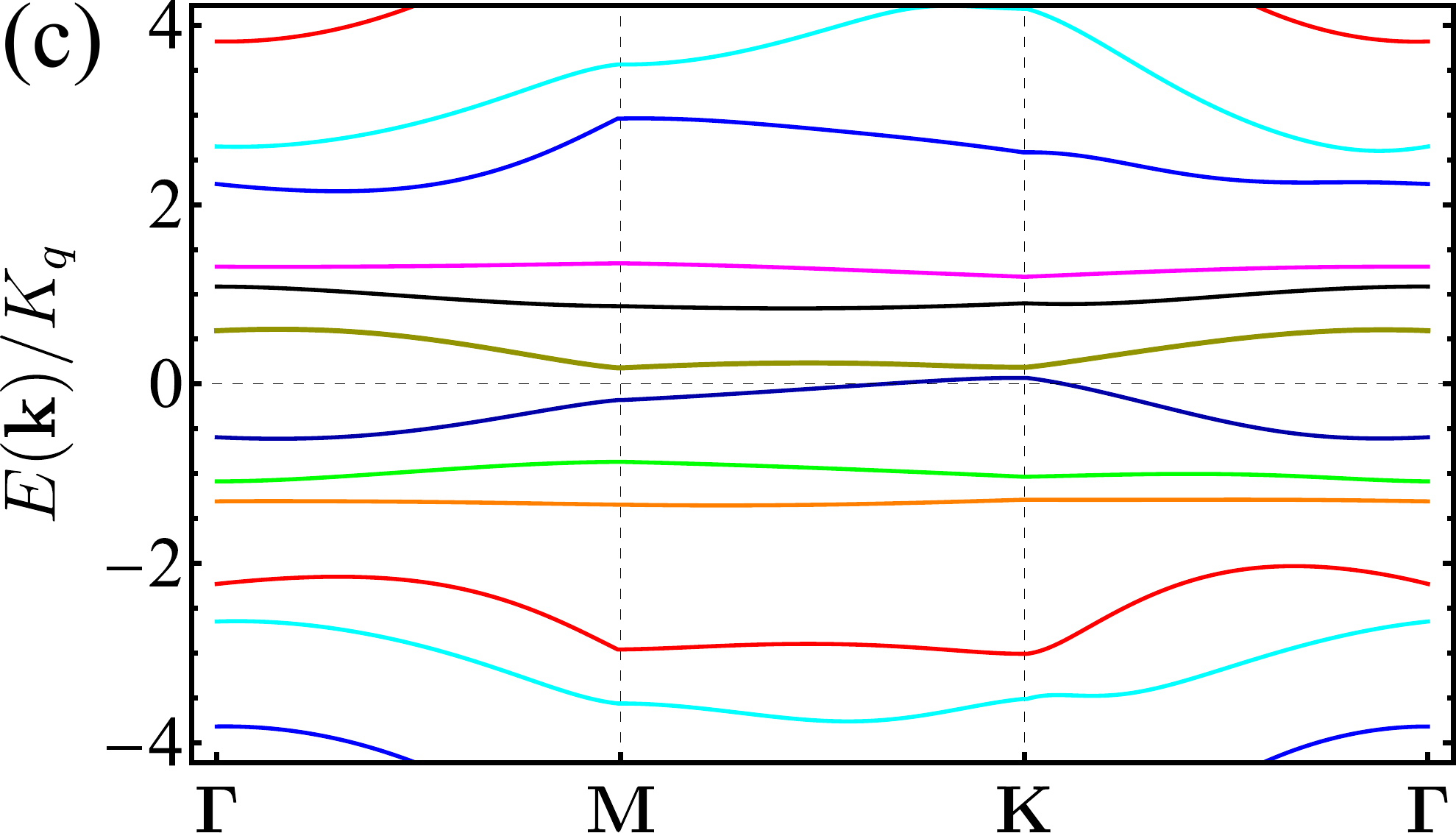}
\caption{Energy dispersion of the mean-field Hamiltonian  for $K_o = 0.2 K_q$, $K_q'=1.3 K_q$, and three different values of $K_o'$ in the broken-symmetry phases:   (a)  $K_o'=0.65 K_q$ in the AO phase, (b) $K_o'=0.10K_q$ in  the FQV phase, and (c) $K_o'=0.40K_q$ in the coexistence region. Here the Brillouin zone is defined with respect to the  unit cell shown in Fig. \ref{Honeycomb_Lattice}. \label{disps}}
\end{figure*}

To determine the mean-field parameters $\rho$, $\varphi$, and $\chi$, we numerically minimize the free energy in the zero-temperature limit by means of a random-search algorithm which finds the lowest value of $\mathcal{F}_\text{MF}$ from a set of   randomly chosen initial conditions.   The   ground-state phase diagram as a function of the integrability-breaking perturbations for fixed $K_o=0.2K_q$ is shown in Fig. \ref{phasediagram}. First, we note that the MSL phase is stable   as long as   $K_q$ and $K_o$ are both nonzero and there are no zero-energy flat bands in the spectrum of the unperturbed model. As we increase $K'_o$, the MSL undergoes a quantum phase transition to a state with   AO order, $\chi\neq 0$, which spontaneously breaks time-reversal symmetry. On the other hand, if we start from the solvable model and increase $K'_q$, the system develops  FQV order with $\rho\neq0$, breaking lattice rotation and translation symmetries. In this regime, we find that the      parameter $\varphi$ evaluates to either $\varphi= \pi/4$ or $\varphi = 5 \pi/4$. This degeneracy  is associated with the invariance of the Hamiltonian under inversion of all quadrupole moments, which follows from $f^\gamma_\mu(\varphi+\pi)=-f^\gamma_\mu(\varphi)$. Finally, when $K_q'$ and $K_o'$ are comparable, the phase diagram exhibits a coexistence region between the FQV and AO phases. Within our mean-field approach,  all phase transitions   are  of   first order.

We now analyze the dispersion relation of the Majorana fermions in the ordered phases. Representative results for the AO, FQV, and coexistence phases are shown in Fig. \ref{disps}.  In the AO phase, the mean-field decoupling with $\chi\neq0$  introduces a staggered on-site term that couples $\theta^x$ and $\theta^z$ without affecting $\theta^y$. As a result, the spectrum remains gapless with a Dirac node at the $\mathbf{K}$ point [see Fig. \ref{disps}(a)]. By contrast, in the  FQV phase the  mean-field Hamiltonian with $\rho\neq0$ couples all three flavors of Majorana fermions. In this case, we obtain a fully gapped spectrum [see Fig. \ref{disps}(b)]. In the coexistence region, we find that the gap closes as one of the bands crosses zero energy, forming a Majorana Fermi surface [see Fig. \ref{disps}(c)].   Note that the spectrum is not particle-hole symmetric because the on-site terms break the chiral symmetry in the mean-field Hamiltonian.

\subsection{Orbital wave theory}\label{Sec_IV.2}
\noindent

In the regime $K_q',K_o' \gg K_q,K_o$, the mean-field approximation used in the previous section is no longer reliable  because the strong integrability-breaking interactions  are able to create vison excitations.  However, we can go the other way and consider the MSL Hamiltonian $H_s$ to be a small perturbation to $H_I$. 
It is instructive to  rewrite the Hamiltonian in terms of the pseudospin and pseudo-orbital operators using Eq. (\ref{opsST}):
\bea
H_I  &=& 4K_q' \sum_{\gamma } \sum_{\langle j l \rangle_\gamma}(\mathbf{u}^\gamma\cdot \boldsymbol{\tau}_j)(\mathbf{u}^\gamma\cdot \boldsymbol{\tau}_l) + 4 K_o' \sum_{\langle j l \rangle} \tau^{y}_j \tau^{y}_l,\nonumber\\
H_s & =& 16K_q \sum_{\gamma } \sum_{\langle j l \rangle_\gamma} (s^\gamma_j \tau^y_j)(s^\gamma_l \tau^y_l) \nonumber \\ 
&& + 16K_o \sum_{\gamma } \sum_{\langle j l \rangle_\gamma} (s^\gamma_j \mathbf{v}^\gamma \cdot \boldsymbol{\tau}_j)(s^\gamma_l \mathbf{v}^\gamma \cdot \boldsymbol{\tau}_l), \label{Eq_H_s_tau}
 \label{Eq_H_I_tau}
\eea
where $\mb u^\gamma=  {\mb v}^\gamma\times \hat{\mb y}=\sin(2\pi\gamma/3)\hat{\mathbf x} +\cos(2\pi\gamma/3)\hat{\mathbf z}$. 
Clearly, $H_I$  describes   interactions within the pseudo-orbital sector. For  $K_q',K_o' \gg K_q,K_o$, we expect  $\boldsymbol \tau_j$ to develop long-range order first,  and the resulting ordered state determines the effective interactions in the pseudospin sector at a lower energy scale.

Let us consider the classical ground states of $H_I$,  with pseudo-orbitals treated as vectors with length $\tau$. The FQV phase corresponds to a vortex state with  $\boldsymbol \tau$ vectors  in the $xz$ plane, as illustrated in Fig. \ref{Orbital_Configuration}. In the  AO phase, the $\boldsymbol \tau$ vectors  point along the $y$ axis. If we consider a state that interpolates between FQV and AO configurations, with   orbital vectors forming an angle $\theta$ with the $\pm \hat{\mb y}$ direction (alternating between A and B sublattices), the corresponding energy per site is\be
E_{\rm cl}(\theta)=-3\tau^2 (K_q'\sin^2\theta+2K_o'\cos^2\theta). \label{classical}
\ee
Minimizing this energy, we obtain a direct transition between AO order ($\theta=0$) and FQV order ($\theta=\pi/2$) at $K_o'=K'_q/2$. In contrast to the mean-field theory in Sec. \ref{Sec_IV.1}, this classical analysis in the strong-coupling limit does not support the presence of   a coexistence region in the phase diagram.

We can go beyond the classical analysis and compute quantum corrections using  linear orbital wave theory \cite{Murthy-PRB(1997),Wu-PRL(2008),Liu-PRL(2008)}. The first step is to rotate the pseudo-orbital operators to the local polarization axis in either FQV or AO states.  We write $\boldsymbol \tau_j =R^T_j \tilde {\boldsymbol\tau}_j$, where $R_j$ is a  $3\times3$ orthogonal matrix that implements the appropriate position-dependent rotation.  In the rotated frame, we employ the Holstein-Primakoff transformation\bea
\tilde \tau^z_j&=&\tau-a^\dagger_ja^{\phantom\dagger}_j,\\
\tilde \tau^-_j&=&  a^\dagger_j (2\tau-a^\dagger_ja^{\phantom\dagger}_j)^{1/2},
\eea
where $a^\dagger_j$ and $a^{\phantom\dagger}_j$ are bosonic creation and annihilation operators. The effective Hamiltonian to order $\tau$ is quadratic in the bosonic operators and can be diagonalized by a Bogoliubov transformation. We then obtain 
\be
H_I\approx \sum_{\nu, \mb k}\omega_\nu(\mb k) \alpha_\nu^\dagger(\mb k)\alpha_\nu^{\phantom\dagger}(\mb k)+E_0,
\ee
where  $\omega_\nu(\mb k)$ is the dispersion relation of the quantized pseudo-orbital waves in the FQV or AO states, with $\nu$ being the band index and $\alpha_\nu(\mb k)$ being the corresponding boson annihilation operator, and \be
E_0=-3N\tau(\tau+1)\mc E+\frac12\sum_{\nu,\mb k}\omega_\nu(\mb k)\label{E0orbitalwave}
\ee
is the ground state energy including the quantum correction, with  $\mc E=K_q'$   in the FQV state and $\mc E=2K_o'$ in the AO state.

\begin{figure}[t]
\centering
\centering \includegraphics[width=0.95\linewidth,valign=t]{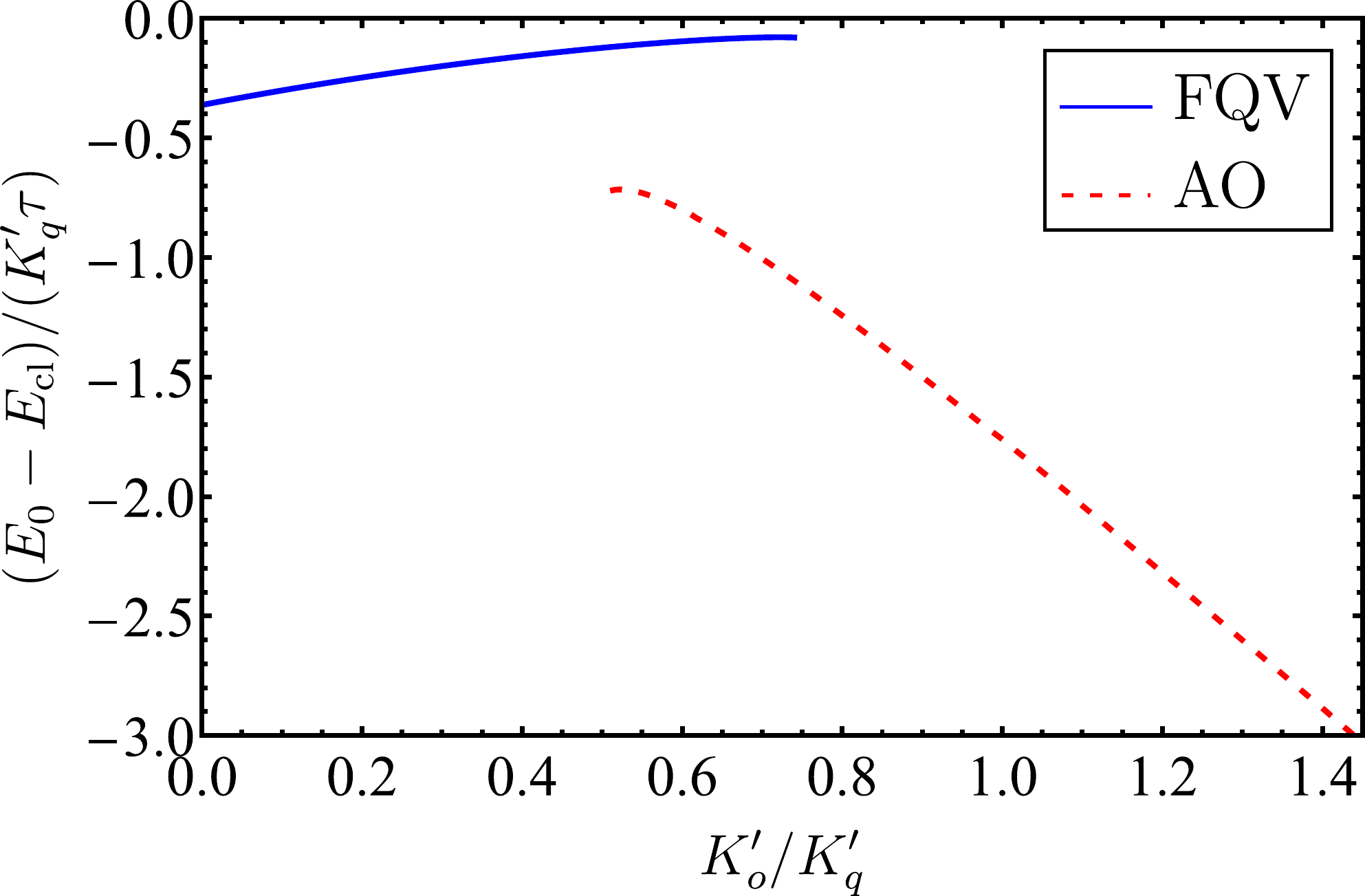}
\caption{Quantum correction to the ground-state energy of FQV and AO states [see Eqs. \eqref{classical} and \eqref{E0orbitalwave}]. The dispersion relation of pseudo-orbital waves in the FQV state becomes complex for $K_o'>0.746K_q'$, signaling an instability. Similarly, the AO state becomes unstable for $K_o'<0.5K_q'$. \label{energies}}
\end{figure}

Perturbative stability of the ordered states against quantum fluctuations requires $\omega_\nu(\mb k)$ to be real and positive  for all bands. Using this criterion, we find that the FQV state remains locally stable for $K_o'<0.746K_q'$, beyond the point $K_o'=K_q'/2$ where its classical energy becomes higher than that of  the AO state. Consistently, the quantum correction to the ground-state energy of the AO state is real for $K_o'>K_q'/2$ (see Fig. \ref{energies}). In the AO phase, the pseudo-orbital wave spectrum is gapped, as expected since the ground state breaks only discrete symmetries. In the FQV phase, the spectrum contains a zero-energy flat band. As emphasized in Ref. \cite{Wu-PRL(2008)}, the classical FQV state has an SO(2) degeneracy which is lifted by quantum fluctuations, and the quantum corrections select   states that maximize the number of zero-energy modes. The gap in the  low-energy  band in the FQV phase appears only beyond the linear orbital wave theory \cite{Khaliulin-PRR(2021)}. Since the excitation spectrum indicates a perturbative stability of the ordered states in the corresponding regimes, with an energy level crossing around $K_o'=K_q'/2$, we conclude that the orbital wave theory points to  a first-order transition between FQV and AO phases.

We can now derive an effective Hamiltonian for the pseudospins using a mean-field decoupling of $H_s$ in which  the pseudo-orbital degrees of freedom are frozen  in the ground state of $H_I$. This is a standard approach to analyze the low-energy excitations of spin-orbital models   \cite{Brink-PRB(1998)}. We obtain from Eq. (\ref{Eq_H_I_tau}) 
\begin{equation}\label{Eq_Strong_Kitaev_Ham}
 H^{\text{eff}}_{s} = \sum_{\gamma = x, y, z} \sum_{\langle j l \rangle_\gamma} \mathcal{K}^{\gamma}_{j l} s^\gamma_j s^\gamma_l,
\end{equation}
where
\begin{equation}\label{Eq_Strong_Kitaev_Int}
\mathcal{K}^{\gamma}_{j l} = - 16 K_o \langle \tau^y_j \tau^y_l \rangle - 16 K_q \langle (\mathbf{v}^\gamma \cdot \boldsymbol{\tau}_j)(\mathbf{v}^\gamma \cdot \boldsymbol{\tau}_l) \rangle.
\end{equation}
The effective Hamiltonian in Eq. (\ref{Eq_Strong_Kitaev_Ham}) has the precise form of a Kitaev model for pseudospin-$\frac{1}{2}$ moments. As a consequence, the fate of the   system in the regime $K_q',K_o'\gg K_q,K_o$ is to exhibit a \emph{hidden} Kitaev spin liquid on top of   classical AO or FQV   phases. This feature bears some resemblance to the behavior of the spin-$\frac{3}{2}$ Kitaev model  studied in Ref. \cite{Jin-NC(2022)}, in which a spin-$\frac{1}{2}$ Kitaev model emerges in the presence of a single-ion anisotropy that induces a  uniform spin quadrupole moment. In our case, the Kitaev spin liquid arises either  from octupolar order for $K'_o\gg K_q'$ or from translation-symmetry-breaking quadrupolar order  for $K'_q\gg K_o'$.

The  pseudospin excitation spectrum depends on the effective Kitaev couplings   in Eq. (\ref{Eq_Strong_Kitaev_Int}). We determine these couplings by computing the pseudo-orbital correlations  within the linear orbital wave theory. In the AO phase, which preserves the $\mathbb Z_3$  symmetry,  we obtain homogeneous antiferromagnetic coupling   $\mathcal{K}^{\gamma}_{j l} =\mc K>0$. As a consequence, in the AO phase the Kitaev spin liquid harbors gapless Majorana fermion excitations  with linear dispersion about the $\mathbf{K}$ point. This conclusion agrees with the result from the parton mean-field approach in Sec. \ref{Sec_IV.1}. Remarkably, the Kitaev spin liquid remains gapless despite the spontaneous breakdown of time-reversal symmetry. In contrast to the effect of a magnetic field in the original Kitaev model \cite{Kitaev-AP(2006)},  here the AO order parameter preserves the symmetry   combining time reversal with a $C_2$ rotation that exchanges A and B sublattices. This symmetry rules out the three-spin interaction that would drive a nonzero spin chirality in the Kitaev spin liquid and generate a mass term for the Majorana fermions \cite{Kitaev-AP(2006)}.

In the FQV phase, the effective Kitaev couplings $\mc K_{jl}^\gamma$ become bond dependent.  While the FQV order breaks the $\mathbb Z_3$ symmetry, it preserves a $\mathbb Z_6$ symmetry corresponding to a rotation in real and internal space around the center of the hexagon that defines the enlarged unit cell in Fig. \ref{Orbital_Configuration}. This symmetry implies that there are only two types of Kitaev couplings: $\mathcal{K}^{\gamma}_{j l} =\mc K_{\rm in}$ for bonds inside  the unit cell, represented by solid lines in Fig. \ref{Orbital_Configuration}, and $\mathcal{K}^{\gamma}_{j l} =\mc K_{\rm out}$ for bonds connecting neighboring unit cells, represented by dashed lines. Using the linear orbital wave theory, we find $\mc K_{\rm in}>0$ and $\mc K_{\rm out}>0$ with $\mc K_{\rm out}\ll \mc K_{\rm in}$ regardless of the ratio between $K_o$ and $K_q$.   In this regime, the Majorana spectrum is gapped with weakly dispersive bands. This gapped FQV phase is consistent with the result from the parton mean-field approach. We note that the  effect of FQV order on the Majorana fermions in the hidden Kitaev spin liquid  is analogous to a Kekul\'e distortion in graphene, which also triples the size of the unit cell and opens a gap in the spectrum   \cite{Hou-PRL(2007),Qu-SciAdv(2022)}.

\section{Conclusions}\label{Sec_V}
\noindent

In summary, we investigated the properties of a  model for  $j_{\mathrm{eff}} = \frac{3}{2}$   local moments on a honeycomb lattice with  bond-directional quadrupolar and octupolar interactions.  For special values of the parameters, we obtained an exact solution  by representing the $j_{\mathrm{eff}} = \frac{3}{2}$ multipolar operators in terms of two  sets of Majorana fermions.  This representation  allowed us to map the  model onto noninteracting fermions hopping in the  background of a static  $\mathbb{Z}_2$ gauge field. This  model  provides a prime example of an interacting system with a nondegenerate   ground state, gapless fermionic excitations, and QSL correlations due to the fractionalization of higher-order multipoles.

We  demonstrated  the emergence of nontrivial topological phases by explicitly breaking   rotational and time-reversal symmetries in the   MSL. Applying both strain and magnetic fields, we found Abelian and non-Abelian  phases with Chern numbers of $0$, $\pm 1$, and $\pm 2$. The Abelian topological phase with $\mathcal{N}_{\mathrm{Ch}} = \pm 2$ appears  for arbitrarily weak perturbations. While Abelian and non-Abelian phases were reported in recent studies of the antiferromagnetic  Kitaev model  in a magnetic field \cite{Jiang-PRL(2020),Zhang-NC(2022)},  our model highlights the role of strain as an additional  control parameter for driving topological phase transitions between chiral spin-orbital liquid phases.

The stability of the MSL was  probed by considering integrability-breaking quadrupolar and octupolar interactions. In the weak-coupling regime, the application of parton mean-field theory showed that   the MSL remains stable over a wide range of interaction strengths. This stability can be   understood from a renormalization-group point of view by noting that in the fermionic representation the interactions are clearly irrelevant   due to the vanishing density of states of the Dirac nodes in the MSL. When a phase transition out of the MSL finally takes place, the system  develops either  ferroquadrupolar-vortex or antiferro-octupolar order. The parton mean-field theory leads to a coexistence region at intermediate couplings, but the orbital wave theory at strong coupling indicates a direct first-order transition between the two ordered phases. Despite the spontaneous symmetry breaking,  the low-energy spectrum  in the strong-coupling regime can be interpreted in terms of a hidden Kitaev spin liquid  with effective couplings between pseudospin degrees of freedom. Of course, an important open question is whether the regime of dominant quadrupolar and octupolar interactions considered  in this work can be reached in layered materials with $4d$ and $5d$ transition-metal ions \cite{Takagi-JPSJ(2021)}.

\begin{acknowledgments}

We would like to thank C. S. de Farias, E. Miranda, and W. Natori for interesting discussions with. H.F. acknowledges funding from the Brazilian agency CNPq under Grant No. 311428/2021-5. R.G.P. acknowledges funding  from CNPq under Grant No. 309569/2022-2.   This work was supported by a grant from the Simons Foundation (Grant No. 1023171, R.G.P.).  Research at IIP-UFRN is supported by the Brazilian ministries Minist\'erio da Educa\c{c}\~ao (MEC) and Minist\'erio da Ci\^encia, Tecnologia e Inova\c{c}\~ao (MCTI).

\end{acknowledgments}

\noindent 

\appendix

\section{\uppercase{Comparison between the energies of the zero- and $\pi$-gauge flux ground states}}\label{App:GS_Energy}

In the main text, we stated that the ground state (GS) of the MSL lies in the zero-gauge flux sector $W_p = + 1$, which is obtained by setting $u_{\langle jl\rangle_\gamma} = 1$ for sites $j$ in sublattice A and $l$ in sublattice B. To confirm that result, we will compare the GS energy of the MSL for both $W_p = + 1$ and $W_p = - 1$, in which the latter defines the so-called $\pi$-gauge flux sector for this model.

\begin{figure*}[t]
\centering
\includegraphics[width=0.85\linewidth,valign=t]{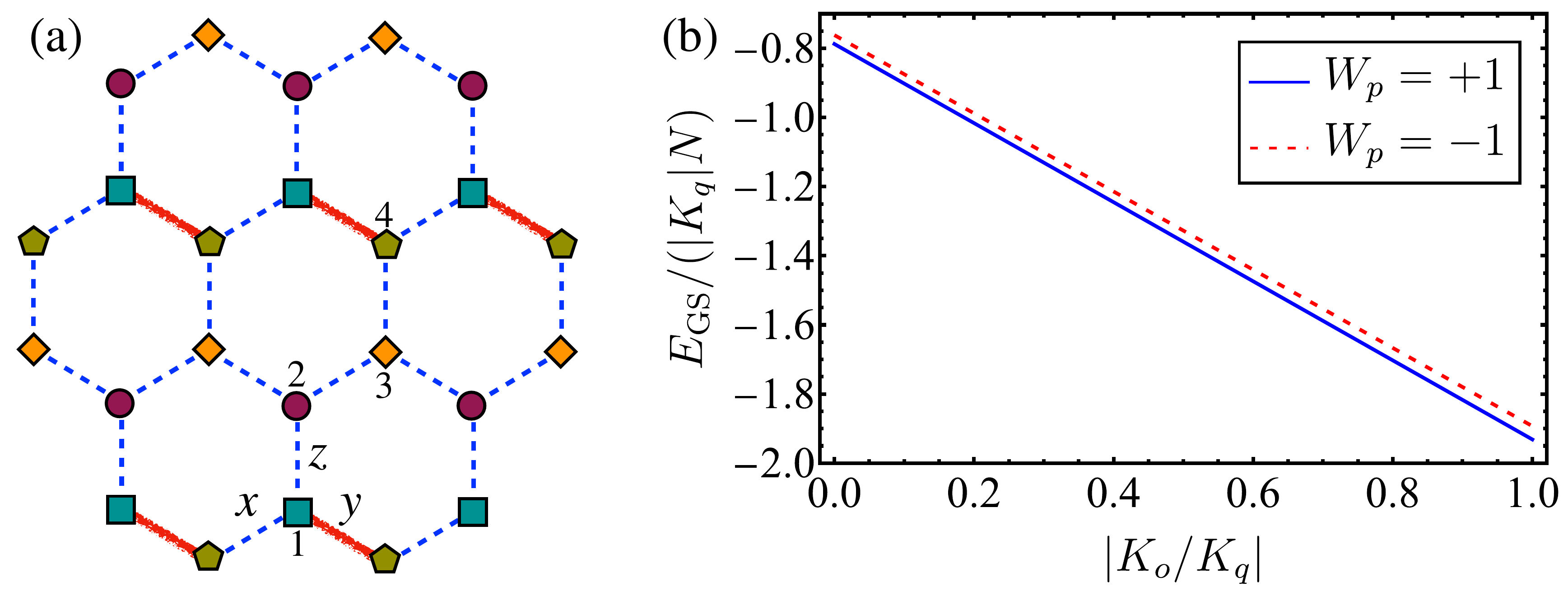}
\caption{(a) Honeycomb lattice for a translationally invariant $\pi$-gauge flux configuration $W_p = - 1$. The dashed blue and filled red bonds refer to the bond variables $u_{\langle jl\rangle_\gamma} = + 1$ and $u_{\langle jl\rangle_\gamma} = - 1$, respectively. In terms of this configuration, the physical unit cell contains four sublattices, which are indicated by numbers and also by different regular polygons. (b) Behavior of the MSL GS energies $E^{+}_{\mathrm{GS}} \equiv E_{\mathrm{GS}}(\{W_p = + 1\})$ and $E^{-}_{\mathrm{GS}} \equiv E_{\mathrm{GS}}(\{W_p = - 1\})$ for the zero- and $\pi$-gauge flux configurations, respectively. Note that $E^{+}_{\mathrm{GS}}$ is always lower than $E^{-}_{\mathrm{GS}}$, which is consistent with Lieb's theorem \cite{Lieb-PRL(1994)}.}\label{Pi_gauge_flux}
\end{figure*}

As exemplified in Fig. \ref{Pi_gauge_flux}(a), a translationally invariant honeycomb lattice with $W_p = - 1$ is obtained with a unit cell containing four sublattice sites. Since the gauge variables for this unit cell evaluate to
\begin{align}
& u_{\langle 3, 2\rangle_x} = + 1, \hspace{1.0cm} u_{\langle 3, 2\rangle_y} = + 1, \hspace{1.0cm} u_{\langle 1, 2\rangle_z} = + 1, \label{Eq_Bond_Var_01}\\
& u_{\langle 1, 4\rangle_x} = + 1, \hspace{1.0cm} u_{\langle 1, 4\rangle_y} = - 1, \hspace{1.0cm} u_{\langle 3, 4\rangle_z} = + 1, \label{Eq_Bond_Var_02}
\end{align} 
we find that the MSL Hamiltonian in Eq. \eqref{Eq_MSL_Maj_Ham} turns out to be
\begin{widetext} 
\begin{align}\label{Eq_Latt_Pi_Ham}
H_s = & \sum_{\mathbf{k} \in \overline{\mathrm{BZ}}}
\bar{\boldsymbol{\Theta}}^\dagger(\mathbf{k})
\begin{pmatrix} 
0 & i \Gamma^{z}(\mathbf{k}) & 0 & i[\Gamma^{x}(\mathbf{k}) - \Gamma^{y}(\mathbf{k})] \\
- i \Gamma^{z \dagger}(\mathbf{k}) & 0 & - i[\Gamma^{x \dagger}(\mathbf{k}) + \Gamma^{y \dagger}(\mathbf{k})] & 0  \\
0 & i[\Gamma^{x}(\mathbf{k}) + \Gamma^{y}(\mathbf{k})] & 0 & i \Gamma^{z}(\mathbf{k})  \\
- i [\Gamma^{x \dagger}(\mathbf{k}) - \Gamma^{y \dagger}(\mathbf{k})] & 0 & - i \Gamma^{z \dagger}(\mathbf{k}) & 0 
\end{pmatrix}
\bar{\boldsymbol{\Theta}}(\mathbf{k}),
\end{align}
where $\bar{\boldsymbol{\Theta}}(\mathbf{k}) \equiv \begin{pmatrix} \boldsymbol{\theta}_1(\mathbf{k}), & \boldsymbol{\theta}_2(\mathbf{k}), & \boldsymbol{\theta}_3(\mathbf{k}), & \boldsymbol{\theta}_4(\mathbf{k}) \end{pmatrix}^T$; $\overline{\mathrm{BZ}}$ refers to the Brillouin zone associated with the four-sublattice unit cell; and the functions $\Gamma^{\gamma}(\mathbf{k})$, with $\gamma \in \{x, y, z\}$, represent the following momentum-dependent matrices:
\begin{align}
&\Gamma^{x}(\mathbf{k}) = \Bigg[ K_q 
\begin{pmatrix} 
0 & 0 & 0 \\
0 & 1 & 0 \\
0 & 0 & 0
\end{pmatrix}
+ \frac{K_o}{4}
\begin{pmatrix} 
1 & 0 & \sqrt{3} \\
0 & 0 & 0 \\
\sqrt{3} & 0 & 3
\end{pmatrix}
\Bigg] e^{- i \mathbf{k} \cdot \boldsymbol{\delta}_1}, \\
&\Gamma^{y}(\mathbf{k}) = \Bigg[ K_q 
\begin{pmatrix} 
0 & 0 & 0 \\
0 & 1 & 0 \\
0 & 0 & 0
\end{pmatrix}
+ \frac{K_o}{4}
\begin{pmatrix} 
1 & 0 & - \sqrt{3} \\
0 & 0 & 0 \\
- \sqrt{3} & 0 & 3
\end{pmatrix}
\Bigg] e^{- i \mathbf{k} \cdot \boldsymbol{\delta}_2}, \\
&\Gamma^{z}(\mathbf{k}) = \Bigg[ K_q 
\begin{pmatrix} 
0 & 0 & 0 \\
0 & 1 & 0 \\
0 & 0 & 0
\end{pmatrix}
+ K_o
\begin{pmatrix} 
1 & 0 & 0 \\
0 & 0 & 0 \\
0 & 0 & 0
\end{pmatrix}
\Bigg] e^{- i \mathbf{k} \cdot \boldsymbol{\delta}_3}.
\end{align}
\end{widetext}
Here, $\boldsymbol{\delta}_{1}$, $\boldsymbol{\delta}_{2}$, and $\boldsymbol{\delta}_{3}$ are nearest-neighbor vectors pointing from an even- to an odd-sublattice site in the direction of the $x$, $y$, and $z$ bonds, respectively.

The diagonalization of $H_s$ in Eq. \eqref{Eq_Latt_Pi_Ham} yields $3 \bar{N}_s = 12$ energy dispersions $\bar{E}_{s, n}(\mathbf{k})$, but only four of them can be computed in closed form. Since the latter also have a complicated dependence, they will not be displayed here. Regardless, the GS energy $E^{-}_{\mathrm{GS}} \equiv E_{\mathrm{GS}}(\{W_p = - 1\})$ associated with this Hamiltonian can be computed numerically. In fact, its formula is given by
\begin{equation}\label{Eq_Ener_Wminus}
\frac{E^{-}_{\mathrm{GS}}}{N} = \frac{1}{\bar{N}_s} \sum\limits^{3 \bar{N}_s}_{n = 1} \int_{\overline{\mathrm{BZ}}} \frac{d^2 \mathbf{k}}{\bar{\mathcal{A}}_{\overline{\mathrm{BZ}}}} \bar{E}_n(\mathbf{k}) \varTheta[ - \bar{E}_{s, n}(\mathbf{k})],
\end{equation}
where $N$ refers to the number of sites of the honeycomb lattice, $\bar{\mathcal{A}}_{\overline{\mathrm{BZ}}} = 4 \pi^2/(3 \sqrt{3} a^2)$ is the area of the $\overline{\mathrm{BZ}}$, and $\varTheta(x)$ denotes the Heaviside step function. This formula should be contrasted with the equation
\begin{equation}\label{Eq_Ener_Wplus}
\frac{E^{+}_{\mathrm{GS}}}{N} = \frac{1}{N_s} \sum\limits^{3 N_s}_{n = 1} \int_{\mathrm{BZ}} \frac{d^2 \mathbf{k}}{\mathcal{A}_{\mathrm{BZ}}} E_n(\mathbf{k}) \varTheta[ - E_{s, n}(\mathbf{k})],
\end{equation}
which yields the GS energy $E^{+}_{\mathrm{GS}} \equiv E_{\mathrm{GS}}(\{W_p = + 1\})$ of the MSL in the zero-gauge flux configuration. In this case, the unit cell has $N_s = 2$ sublattices, and $\mathcal{A}_{\mathrm{BZ}} = 8 \pi^{2}/(3 \sqrt{3} a^2)$ gives the area of the corresponding BZ. In addition, $E_{s, n}(\mathbf{k})$ refers to $3 N_s = 6$ energy dispersions whose expressions can be found in the main text [see Eq. \eqref{Esk} and the paragraph above it].

In Fig. \ref{Pi_gauge_flux}(b), we show the behavior of both $E^{-}_{\mathrm{GS}}$ and $E^{+}_{\mathrm{GS}}$ as a function of the ratio $\vert K_o/K_q \vert$. Clearly, the zero-gauge flux configuration $W_p = + 1$ has the lowest GS energy. Moreover, we also investigate the behavior of the GS energy gap per lattice site $(E^{-}_{\mathrm{GS}} - E^{+}_{\mathrm{GS}})/N$. In fact, our numerical results reveal that $(E^{-}_{\mathrm{GS}} - E^{+}_{\mathrm{GS}})/N$ depends linearly on $\vert K_o/K_q \vert$. Consequently, we obtain
\begin{equation}
\frac{E^{-}_{\mathrm{GS}} - E^{+}_{\mathrm{GS}}}{N} = b_0 (\vert K_o \vert + 2 \vert K_q \vert),
\end{equation}
where $b_0 \approx 0.012867$. Remarkably, this means that the quadrupolar and octupolar degrees of freedom of the MSL contribute differently to $E^{-}_{\mathrm{GS}} - E^{+}_{\mathrm{GS}}$. In other words, the impact of the former is twice that of the latter.

Note that, according to Lieb's theorem \cite{Lieb-PRL(1994)}, the zero-gauge flux configuration yields the lowest-energy GS for mirror-symmetric models with nearest-neighbor interactions defined on a honeycomb lattice. In light of this result, consider the MSL Hamiltonian $H_s$ in Eq. \eqref{Eq_MSL_Maj_Ham}. The first term in $H_s$ involving the $\theta^y$ fermions is clearly invariant under a reflection exchanging the $x$ and $y$ bonds since $\theta^y$ transforms as the matter Majorana fermions in the Kitaev model for isotropic exchange interactions. However, the mirror symmetry of the term that involves $\theta^x$ and $\theta^z$ requires the additional transformation $\theta^z \to -\theta^z$, which can be traced back to the symmetry properties of the quadrupole components $(\tau^x,\; \tau^z)$. We can view $\theta^z \to -\theta^z$ as a gauge transformation which does not affect the gauge-invariant fluxes nor the spectrum of $H_s$. For this reason, we expected the result of Lieb's theorem to apply also to our model, which was confirmed by our numerical results.

\onecolumngrid  

\section{\uppercase{Mean-field Hamiltonian for Majorana fermions in the zero-gauge flux sector}}\label{App:MF}

After we fix the uniform gauge configuration for $u_{\langle jl\rangle}$, which is equivalent to setting $W_p = + 1$ for all plaquettes, the  exactly solvable Hamiltonian  can be written as\be
H_s+\delta H_s=\sum_{\mb k}  \boldsymbol{\Theta}^\dagger(\mathbf{k})\boldsymbol{\mc H}_s(\mb k) \boldsymbol{\Theta}(\mathbf{k}),
\ee
where $\boldsymbol{\Theta}(\mathbf{k}) \equiv ( \theta^x_A (\mathbf{k}), \theta^x_B (\mathbf{k}), \theta^y_A (\mathbf{k}), \theta^y_B (\mathbf{k}), \theta^z_A (\mathbf{k}), \theta^z_B (\mathbf{k}) )^T$ and we define the matrix
\bea
\boldsymbol{\mc H}_{s} (\mb k)= 
\begin{pmatrix} 
0 & i K_o g_x(\mathbf{k}) & i \varepsilon_{3} & 0 & - i h & i K_o  g_{xz}(\mathbf{k}) \\
- K_o g^*_x(\mathbf{k}) & 0 & 0 & i \varepsilon_{3} & - iK_o  g^*_{xz}(\mathbf{k}) & - i h \\
- i \varepsilon_{3} & 0 & 0 & i K_q g_y(\mathbf{k}) & i \varepsilon_{2} & 0 \\
0 & - i \varepsilon_{3} & - i K_q g^*_y(\mathbf{k}) & 0 & 0 & i \varepsilon_{2} \\
i h & i K_o g_{xz}(\mathbf{k}) & - i \varepsilon_{2} & 0 & 0 & iK_o  g_z(\mathbf{k}) \\
- i K_o g^*_{xz}(\mathbf{k}) & i h & 0 & - i \varepsilon_{2} & - i K_o g_z(\mathbf{k}) & 0
\end{pmatrix},
\eea
with the functions given below Eq. (\ref{Esk}). Here we have included the perturbations  that determine the phase diagram in Fig. \ref{Chern_Number}. In the parton mean-field theory in Sec. \ref{Sec_IV.1}, we set $\varepsilon_2=\varepsilon_3=h=0$.

According to Eqs. \eqref{Eq_Ferro_Orb_Vort} and \eqref{Eq_Anti_Oct}, the mean-field decoupling of the  Hamiltonian $H=H_s+H_I$  leads to
\begin{align}
H_\text{MF}  =&  \sum_{\mathbf{k} \in \frac{1}{6}(\text{BZ})}\begin{pmatrix} \boldsymbol{\Theta}(\mathbf{k}) \\ \boldsymbol{\Theta}(\mathbf{k} + \mathbf{Q}_1) \\ \boldsymbol{\Theta}(\mathbf{k} + \mathbf{Q}_2) \end{pmatrix}^\dagger \begin{pmatrix} 
\boldsymbol{\mathcal{H}}_s(\mathbf{k}) + \boldsymbol{\mathcal{V}} & \boldsymbol{\mathcal{V}}_{\mathbf{Q}_1} & \boldsymbol{\mathcal{V}}_{\mathbf{Q}_2} \\
\boldsymbol{\mathcal{V}}^{\dagger}_{\mathbf{Q}_1} & \boldsymbol{\mathcal{H}}_s(\mathbf{k} + \mathbf{Q}_1) + \boldsymbol{\mathcal{V}} & \mathbb{0}  \\
\boldsymbol{\mathcal{V}}^{\dagger}_{\mathbf{Q}_2} & \mathbb{0} & \boldsymbol{\mathcal{H}}_s(\mathbf{k} + \mathbf{Q}_2) + \boldsymbol{\mathcal{V}}
\end{pmatrix}
\begin{pmatrix} \boldsymbol{\Theta}(\mathbf{k}) \\ \boldsymbol{\Theta}(\mathbf{k} + \mathbf{Q}_1) \\ \boldsymbol{\Theta}(\mathbf{k} + \mathbf{Q}_2) \end{pmatrix} \nonumber \\
& - \frac{K_q'}{2} N\rho^2 \sum^{2}_{\mu = 1}\sum_{\gamma=1}^3  [f_\mu^\gamma(\varphi) ]^2 \cos(\mathbf{Q}_\mu \cdot \boldsymbol{\delta}_\gamma)   + \frac{3 K_o'}{2} N \chi^2,
\end{align}
where we have set
\begin{align}
\boldsymbol{\mathcal{V}}_{\mathbf{Q}_\mu} & = 3 K_q'
\begin{pmatrix} 
0 & 0 & i \rho^{xy *}_{\mu, A} & 0 & 0 & 0 \\
0 & 0 & 0 & i \rho^{xy}_{\mu, B} & 0 & 0 \\
- i \rho^{xy *}_{\mu, A} & 0 & 0 & 0 & - i \rho^{yz *}_{\mu, A} & 0 \\
0 & - i \rho^{xy}_{\mu, B} & 0 & 0 & 0 & - i \rho^{yz}_{\mu, B} \\
0 & 0 & i \rho^{yz *}_{\mu, A} & 0 & 0 & 0 \\
0 & 0 & 0 & i \rho^{yz}_{\mu, B} & 0 & 0
\end{pmatrix}, \\
\boldsymbol{\mathcal{V}} & = 6 K_o'
\begin{pmatrix} 
0 & 0 & 0 & 0 & - i \chi & 0 \\
0 & 0 & 0 & 0 & 0 & i \chi \\
0 & 0 & 0 & 0 & 0 & 0 \\
0 & 0 & 0 & 0 & 0 & 0 \\
i \chi & 0 & 0 & 0 & 0 & 0 \\
0 & - i \chi & 0 & 0 & 0 & 0 
\end{pmatrix}.
\end{align}
The matrices $\mathcal{V}_{\mathbf{Q}_\mu}$ are defined in terms of the functions
\begin{align}
\rho^{xy}_{\mu, A} & = \frac{\rho}{2} \left[\frac{f^1_\mu(\varphi) e^{i \mathbf{Q}_\mu \cdot \boldsymbol{\delta}_1} + f^2_\mu(\varphi)  e^{i \mathbf{Q}_\mu \cdot \boldsymbol{\delta}_2}}{2} -f^3_\mu(\varphi)  e^{i \mathbf{Q}_\mu \cdot \boldsymbol{\delta}_3} \right] , \\
\rho^{yz}_{\mu, A} & = \frac{\sqrt{3}\rho}{4} \left[ - f^1_\mu(\varphi)  e^{i \mathbf{Q}_\mu \cdot \boldsymbol{\delta}_1} + f^2_\mu(\varphi)  e^{i \mathbf{Q}_\mu \cdot \boldsymbol{\delta}_2} \right]  , \\
\rho^{xy}_{\mu, B} & = \frac{\rho}{2} \left[ \frac{f^1_\mu(\varphi)  e^{i \mathbf{Q}_\mu \cdot \mathbf{n}_1} + f^2_\mu(\varphi)  e^{i \mathbf{Q}_\mu \cdot \mathbf{n}_2}}{2} -f^3_\mu(\varphi)  \right]  , \\
\rho^{yz}_{\mu, B} & = \frac{\sqrt{3}\rho}{4} \left[ -f^1_\mu(\varphi)  e^{i \mathbf{Q}_\mu \cdot \mathbf{n}_1} +f^2_\mu(\varphi) e^{i \mathbf{Q}_\mu \cdot \mathbf{n}_2} \right]  .
\end{align}

\twocolumngrid 


\begin{thebibliography}{69}%
\makeatletter
\providecommand \@ifxundefined [1]{%
 \@ifx{#1\undefined}
}%
\providecommand \@ifnum [1]{%
 \ifnum #1\expandafter \@firstoftwo
 \else \expandafter \@secondoftwo
 \fi
}%
\providecommand \@ifx [1]{%
 \ifx #1\expandafter \@firstoftwo
 \else \expandafter \@secondoftwo
 \fi
}%
\providecommand \natexlab [1]{#1}%
\providecommand \enquote  [1]{``#1''}%
\providecommand \bibnamefont  [1]{#1}%
\providecommand \bibfnamefont [1]{#1}%
\providecommand \citenamefont [1]{#1}%
\providecommand \href@noop [0]{\@secondoftwo}%
\providecommand \href [0]{\begingroup \@sanitize@url \@href}%
\providecommand \@href[1]{\@@startlink{#1}\@@href}%
\providecommand \@@href[1]{\endgroup#1\@@endlink}%
\providecommand \@sanitize@url [0]{\catcode `\\12\catcode `\$12\catcode
  `\&12\catcode `\#12\catcode `\^12\catcode `\_12\catcode `\%12\relax}%
\providecommand \@@startlink[1]{}%
\providecommand \@@endlink[0]{}%
\providecommand \url  [0]{\begingroup\@sanitize@url \@url }%
\providecommand \@url [1]{\endgroup\@href {#1}{\urlprefix }}%
\providecommand \urlprefix  [0]{URL }%
\providecommand \Eprint [0]{\href }%
\providecommand \doibase [0]{http://dx.doi.org/}%
\providecommand \selectlanguage [0]{\@gobble}%
\providecommand \bibinfo  [0]{\@secondoftwo}%
\providecommand \bibfield  [0]{\@secondoftwo}%
\providecommand \translation [1]{[#1]}%
\providecommand \BibitemOpen [0]{}%
\providecommand \bibitemStop [0]{}%
\providecommand \bibitemNoStop [0]{.\EOS\space}%
\providecommand \EOS [0]{\spacefactor3000\relax}%
\providecommand \BibitemShut  [1]{\csname bibitem#1\endcsname}%
\let\auto@bib@innerbib\@empty
\bibitem [{\citenamefont {Balents}(2010)}]{Balents-N(2010)}%
  \BibitemOpen
  \bibfield  {author} {\bibinfo {author} {\bibfnamefont {L.}~\bibnamefont
  {Balents}},\ }\bibfield  {title} {\bibinfo {title} {{Spin liquids in
  frustrated magnets}},\ }\href {\doibase 10.1038/nature08917} {\bibfield
  {journal} {\bibinfo  {journal} {Nature}\ }\textbf {\bibinfo {volume} {464}},\
  \bibinfo {pages} {199} (\bibinfo {year} {2010})}\BibitemShut {NoStop}%
\bibitem [{\citenamefont {Broholm}\ \emph {et~al.}(2020)\citenamefont
  {Broholm}, \citenamefont {Cava}, \citenamefont {Kivelson}, \citenamefont
  {Nocera}, \citenamefont {Norman},\ and\ \citenamefont
  {Senthil}}]{Senthil-S(2020)}%
  \BibitemOpen
  \bibfield  {author} {\bibinfo {author} {\bibfnamefont {C.}~\bibnamefont
  {Broholm}}, \bibinfo {author} {\bibfnamefont {R.~J.}\ \bibnamefont {Cava}},
  \bibinfo {author} {\bibfnamefont {S.~A.}\ \bibnamefont {Kivelson}}, \bibinfo
  {author} {\bibfnamefont {D.~G.}\ \bibnamefont {Nocera}}, \bibinfo {author}
  {\bibfnamefont {M.~R.}\ \bibnamefont {Norman}}, \ and\ \bibinfo {author}
  {\bibfnamefont {T.}~\bibnamefont {Senthil}},\ }\bibfield  {title} {\bibinfo
  {title} {{Quantum spin liquids}},\ }\href {\doibase 10.1126/science.aay0668}
  {\bibfield  {journal} {\bibinfo  {journal} {Science}\ }\textbf {\bibinfo
  {volume} {367}},\ \bibinfo {pages} {eaay0668} (\bibinfo {year}
  {2020})}\BibitemShut {NoStop}%
\bibitem [{\citenamefont {Savary}\ and\ \citenamefont
  {Balents}(2017)}]{Savary-RPP(2017)}%
  \BibitemOpen
  \bibfield  {author} {\bibinfo {author} {\bibfnamefont {L.}~\bibnamefont
  {Savary}}\ and\ \bibinfo {author} {\bibfnamefont {L.}~\bibnamefont
  {Balents}},\ }\bibfield  {title} {\bibinfo {title} {{Quantum spin liquids: a
  review}},\ }\href {http://stacks.iop.org/0034-4885/80/i=1/a=016502}
  {\bibfield  {journal} {\bibinfo  {journal} {Rep. Prog. Phys.}\ }\textbf
  {\bibinfo {volume} {80}},\ \bibinfo {pages} {016502} (\bibinfo {year}
  {2017})}\BibitemShut {NoStop}%
\bibitem [{\citenamefont {Takagi}\ \emph {et~al.}(2019)\citenamefont {Takagi},
  \citenamefont {Takayama}, \citenamefont {Jackeli}, \citenamefont
  {Khaliullin},\ and\ \citenamefont {Nagler}}]{Takagi-NRP(2019)}%
  \BibitemOpen
  \bibfield  {author} {\bibinfo {author} {\bibfnamefont {H.}~\bibnamefont
  {Takagi}}, \bibinfo {author} {\bibfnamefont {T.}~\bibnamefont {Takayama}},
  \bibinfo {author} {\bibfnamefont {G.}~\bibnamefont {Jackeli}}, \bibinfo
  {author} {\bibfnamefont {G.}~\bibnamefont {Khaliullin}}, \ and\ \bibinfo
  {author} {\bibfnamefont {S.~E.}\ \bibnamefont {Nagler}},\ }\bibfield  {title}
  {\bibinfo {title} {{Concept and realization of Kitaev quantum spin
  liquids}},\ }\href {\doibase 10.1038/s42254-019-0038-2} {\bibfield  {journal}
  {\bibinfo  {journal} {Nat. Rev. Phys.}\ }\textbf {\bibinfo {volume} {1}},\
  \bibinfo {pages} {264} (\bibinfo {year} {2019})}\BibitemShut {NoStop}%
\bibitem [{\citenamefont {Trebst}\ and\ \citenamefont
  {Hickey}(2022)}]{Trebst-PR(2022)}%
  \BibitemOpen
  \bibfield  {author} {\bibinfo {author} {\bibfnamefont {S.}~\bibnamefont
  {Trebst}}\ and\ \bibinfo {author} {\bibfnamefont {C.}~\bibnamefont
  {Hickey}},\ }\bibfield  {title} {\bibinfo {title} {{Kitaev materials}},\
  }\href {\doibase https://doi.org/10.1016/j.physrep.2021.11.003} {\bibfield
  {journal} {\bibinfo  {journal} {Phys. Rep.}\ }\textbf {\bibinfo {volume}
  {950}},\ \bibinfo {pages} {1} (\bibinfo {year} {2022})}\BibitemShut {NoStop}%
\bibitem [{\citenamefont {Kitaev}(2006)}]{Kitaev-AP(2006)}%
  \BibitemOpen
  \bibfield  {author} {\bibinfo {author} {\bibfnamefont {A.}~\bibnamefont
  {Kitaev}},\ }\bibfield  {title} {\bibinfo {title} {{Anyons in an exactly
  solved model and beyond}},\ }\href {\doibase
  http://dx.doi.org/10.1016/j.aop.2005.10.005} {\bibfield  {journal} {\bibinfo
  {journal} {Ann. Phys.}\ }\textbf {\bibinfo {volume} {321}},\ \bibinfo {pages}
  {2} (\bibinfo {year} {2006})}\BibitemShut {NoStop}%
\bibitem [{\citenamefont {Jackeli}\ and\ \citenamefont
  {Khaliullin}(2009)}]{Jackeli-PRL(2009)}%
  \BibitemOpen
  \bibfield  {author} {\bibinfo {author} {\bibfnamefont {G.}~\bibnamefont
  {Jackeli}}\ and\ \bibinfo {author} {\bibfnamefont {G.}~\bibnamefont
  {Khaliullin}},\ }\bibfield  {title} {\bibinfo {title} {{Mott Insulators in
  the Strong Spin-Orbit Coupling Limit: From Heisenberg to a Quantum Compass
  and Kitaev Models}},\ }\href {\doibase 10.1103/PhysRevLett.102.017205}
  {\bibfield  {journal} {\bibinfo  {journal} {Phys. Rev. Lett.}\ }\textbf
  {\bibinfo {volume} {102}},\ \bibinfo {pages} {017205} (\bibinfo {year}
  {2009})}\BibitemShut {NoStop}%
\bibitem [{\citenamefont {Singh}\ and\ \citenamefont
  {Gegenwart}(2010)}]{Singh-PRB(2010)}%
  \BibitemOpen
  \bibfield  {author} {\bibinfo {author} {\bibfnamefont {Y.}~\bibnamefont
  {Singh}}\ and\ \bibinfo {author} {\bibfnamefont {P.}~\bibnamefont
  {Gegenwart}},\ }\bibfield  {title} {\bibinfo {title} {{Antiferromagnetic Mott
  insulating state in single crystals of the honeycomb lattice material
  ${\text{Na}}_{2}{\text{IrO}}_{3}$}},\ }\href {\doibase
  10.1103/PhysRevB.82.064412} {\bibfield  {journal} {\bibinfo  {journal} {Phys.
  Rev. B}\ }\textbf {\bibinfo {volume} {82}},\ \bibinfo {pages} {064412}
  (\bibinfo {year} {2010})}\BibitemShut {NoStop}%
\bibitem [{\citenamefont {Singh}\ \emph {et~al.}(2012)\citenamefont {Singh},
  \citenamefont {Manni}, \citenamefont {Reuther}, \citenamefont {Berlijn},
  \citenamefont {Thomale}, \citenamefont {Ku}, \citenamefont {Trebst},\ and\
  \citenamefont {Gegenwart}}]{Singh-PRL(2012)}%
  \BibitemOpen
  \bibfield  {author} {\bibinfo {author} {\bibfnamefont {Y.}~\bibnamefont
  {Singh}}, \bibinfo {author} {\bibfnamefont {S.}~\bibnamefont {Manni}},
  \bibinfo {author} {\bibfnamefont {J.}~\bibnamefont {Reuther}}, \bibinfo
  {author} {\bibfnamefont {T.}~\bibnamefont {Berlijn}}, \bibinfo {author}
  {\bibfnamefont {R.}~\bibnamefont {Thomale}}, \bibinfo {author} {\bibfnamefont
  {W.}~\bibnamefont {Ku}}, \bibinfo {author} {\bibfnamefont {S.}~\bibnamefont
  {Trebst}}, \ and\ \bibinfo {author} {\bibfnamefont {P.}~\bibnamefont
  {Gegenwart}},\ }\bibfield  {title} {\bibinfo {title} {{Relevance of the
  Heisenberg-Kitaev Model for the Honeycomb Lattice Iridates
  ${A}_{2}{\mathrm{IrO}}_{3}$}},\ }\href {\doibase
  10.1103/PhysRevLett.108.127203} {\bibfield  {journal} {\bibinfo  {journal}
  {Phys. Rev. Lett.}\ }\textbf {\bibinfo {volume} {108}},\ \bibinfo {pages}
  {127203} (\bibinfo {year} {2012})}\BibitemShut {NoStop}%
\bibitem [{\citenamefont {Cao}\ \emph {et~al.}(2013)\citenamefont {Cao},
  \citenamefont {Qi}, \citenamefont {Li}, \citenamefont {Terzic}, \citenamefont
  {Cao}, \citenamefont {Yuan}, \citenamefont {Tovar}, \citenamefont {Murthy},\
  and\ \citenamefont {Kaul}}]{Cao-PRB(2013)}%
  \BibitemOpen
  \bibfield  {author} {\bibinfo {author} {\bibfnamefont {G.}~\bibnamefont
  {Cao}}, \bibinfo {author} {\bibfnamefont {T.~F.}\ \bibnamefont {Qi}},
  \bibinfo {author} {\bibfnamefont {L.}~\bibnamefont {Li}}, \bibinfo {author}
  {\bibfnamefont {J.}~\bibnamefont {Terzic}}, \bibinfo {author} {\bibfnamefont
  {V.~S.}\ \bibnamefont {Cao}}, \bibinfo {author} {\bibfnamefont {S.~J.}\
  \bibnamefont {Yuan}}, \bibinfo {author} {\bibfnamefont {M.}~\bibnamefont
  {Tovar}}, \bibinfo {author} {\bibfnamefont {G.}~\bibnamefont {Murthy}}, \
  and\ \bibinfo {author} {\bibfnamefont {R.~K.}\ \bibnamefont {Kaul}},\
  }\bibfield  {title} {\bibinfo {title} {{Evolution of magnetism in the
  single-crystal honeycomb iridates
  $({\text{Na}}_{1\ensuremath{-}x}$Li${}_{x}{)}_{2}\text{Ir}{\text{O}}_{3}$}},\
  }\href {\doibase 10.1103/PhysRevB.88.220414} {\bibfield  {journal} {\bibinfo
  {journal} {Phys. Rev. B}\ }\textbf {\bibinfo {volume} {88}},\ \bibinfo
  {pages} {220414} (\bibinfo {year} {2013})}\BibitemShut {NoStop}%
\bibitem [{\citenamefont {Kitagawa}\ \emph {et~al.}(2018)\citenamefont
  {Kitagawa}, \citenamefont {Takayama}, \citenamefont {Matsumoto},
  \citenamefont {Kato}, \citenamefont {Takano}, \citenamefont {Kishimoto},
  \citenamefont {Bette}, \citenamefont {Dinnebier}, \citenamefont {Jackeli},\
  and\ \citenamefont {Takagi}}]{Kitagawa-Nature(2018)}%
  \BibitemOpen
  \bibfield  {author} {\bibinfo {author} {\bibfnamefont {K.}~\bibnamefont
  {Kitagawa}}, \bibinfo {author} {\bibfnamefont {T.}~\bibnamefont {Takayama}},
  \bibinfo {author} {\bibfnamefont {Y.}~\bibnamefont {Matsumoto}}, \bibinfo
  {author} {\bibfnamefont {A.}~\bibnamefont {Kato}}, \bibinfo {author}
  {\bibfnamefont {R.}~\bibnamefont {Takano}}, \bibinfo {author} {\bibfnamefont
  {Y.}~\bibnamefont {Kishimoto}}, \bibinfo {author} {\bibfnamefont
  {S.}~\bibnamefont {Bette}}, \bibinfo {author} {\bibfnamefont
  {R.}~\bibnamefont {Dinnebier}}, \bibinfo {author} {\bibfnamefont
  {G.}~\bibnamefont {Jackeli}}, \ and\ \bibinfo {author} {\bibfnamefont
  {H.}~\bibnamefont {Takagi}},\ }\bibfield  {title} {\bibinfo {title} {{A
  spin-orbital-entangled quantum liquid on a honeycomb lattice}},\ }\href
  {\doibase 10.1038/nature25482} {\bibfield  {journal} {\bibinfo  {journal}
  {Nature (London)}\ }\textbf {\bibinfo {volume} {554}},\ \bibinfo {pages}
  {341} (\bibinfo {year} {2018})}\BibitemShut {NoStop}%
\bibitem [{\citenamefont {Plumb}\ \emph {et~al.}(2014)\citenamefont {Plumb},
  \citenamefont {Clancy}, \citenamefont {Sandilands}, \citenamefont {Shankar},
  \citenamefont {Hu}, \citenamefont {Burch}, \citenamefont {Kee},\ and\
  \citenamefont {Kim}}]{Plumb-PRB(2014)}%
  \BibitemOpen
  \bibfield  {author} {\bibinfo {author} {\bibfnamefont {K.~W.}\ \bibnamefont
  {Plumb}}, \bibinfo {author} {\bibfnamefont {J.~P.}\ \bibnamefont {Clancy}},
  \bibinfo {author} {\bibfnamefont {L.~J.}\ \bibnamefont {Sandilands}},
  \bibinfo {author} {\bibfnamefont {V.~V.}\ \bibnamefont {Shankar}}, \bibinfo
  {author} {\bibfnamefont {Y.~F.}\ \bibnamefont {Hu}}, \bibinfo {author}
  {\bibfnamefont {K.~S.}\ \bibnamefont {Burch}}, \bibinfo {author}
  {\bibfnamefont {H.-Y.}\ \bibnamefont {Kee}}, \ and\ \bibinfo {author}
  {\bibfnamefont {Y.-J.}\ \bibnamefont {Kim}},\ }\bibfield  {title} {\bibinfo
  {title} {{$\ensuremath{\alpha}\ensuremath{-}{\mathrm{RuCl}}_{3}$: A
  spin-orbit assisted Mott insulator on a honeycomb lattice}},\ }\href
  {\doibase 10.1103/PhysRevB.90.041112} {\bibfield  {journal} {\bibinfo
  {journal} {Phys. Rev. B}\ }\textbf {\bibinfo {volume} {90}},\ \bibinfo
  {pages} {041112} (\bibinfo {year} {2014})}\BibitemShut {NoStop}%
\bibitem [{\citenamefont {Kasahara}\ \emph {et~al.}(2018)\citenamefont
  {Kasahara}, \citenamefont {Ohnishi}, \citenamefont {Mizukami}, \citenamefont
  {Tanaka}, \citenamefont {Ma}, \citenamefont {Sugii}, \citenamefont {Kurita},
  \citenamefont {Tanaka}, \citenamefont {Nasu}, \citenamefont {Motome},
  \citenamefont {Shibauchi},\ and\ \citenamefont
  {Matsuda}}]{Kasahara-Nature(2018)}%
  \BibitemOpen
  \bibfield  {author} {\bibinfo {author} {\bibfnamefont {Y.}~\bibnamefont
  {Kasahara}}, \bibinfo {author} {\bibfnamefont {T.}~\bibnamefont {Ohnishi}},
  \bibinfo {author} {\bibfnamefont {Y.}~\bibnamefont {Mizukami}}, \bibinfo
  {author} {\bibfnamefont {O.}~\bibnamefont {Tanaka}}, \bibinfo {author}
  {\bibfnamefont {S.}~\bibnamefont {Ma}}, \bibinfo {author} {\bibfnamefont
  {K.}~\bibnamefont {Sugii}}, \bibinfo {author} {\bibfnamefont
  {N.}~\bibnamefont {Kurita}}, \bibinfo {author} {\bibfnamefont
  {H.}~\bibnamefont {Tanaka}}, \bibinfo {author} {\bibfnamefont
  {J.}~\bibnamefont {Nasu}}, \bibinfo {author} {\bibfnamefont {Y.}~\bibnamefont
  {Motome}}, \bibinfo {author} {\bibfnamefont {T.}~\bibnamefont {Shibauchi}}, \
  and\ \bibinfo {author} {\bibfnamefont {Y.}~\bibnamefont {Matsuda}},\
  }\bibfield  {title} {\bibinfo {title} {{Majorana quantization and
  half-integer thermal quantum Hall effect in a Kitaev spin liquid}},\ }\href
  {\doibase 10.1038/s41586-018-0274-0} {\bibfield  {journal} {\bibinfo
  {journal} {Nature (London)}\ }\textbf {\bibinfo {volume} {559}},\ \bibinfo
  {pages} {227} (\bibinfo {year} {2018})}\BibitemShut {NoStop}%
\bibitem [{\citenamefont {Tanaka}\ \emph {et~al.}(2022)\citenamefont {Tanaka},
  \citenamefont {Mizukami}, \citenamefont {Harasawa}, \citenamefont
  {Hashimoto}, \citenamefont {Hwang}, \citenamefont {Kurita}, \citenamefont
  {Tanaka}, \citenamefont {Fujimoto}, \citenamefont {Matsuda}, \citenamefont
  {Moon},\ and\ \citenamefont {Shibauchi}}]{Tanaka-NP(2022)}%
  \BibitemOpen
  \bibfield  {author} {\bibinfo {author} {\bibfnamefont {O.}~\bibnamefont
  {Tanaka}}, \bibinfo {author} {\bibfnamefont {Y.}~\bibnamefont {Mizukami}},
  \bibinfo {author} {\bibfnamefont {R.}~\bibnamefont {Harasawa}}, \bibinfo
  {author} {\bibfnamefont {K.}~\bibnamefont {Hashimoto}}, \bibinfo {author}
  {\bibfnamefont {K.}~\bibnamefont {Hwang}}, \bibinfo {author} {\bibfnamefont
  {N.}~\bibnamefont {Kurita}}, \bibinfo {author} {\bibfnamefont
  {H.}~\bibnamefont {Tanaka}}, \bibinfo {author} {\bibfnamefont
  {S.}~\bibnamefont {Fujimoto}}, \bibinfo {author} {\bibfnamefont
  {Y.}~\bibnamefont {Matsuda}}, \bibinfo {author} {\bibfnamefont {E.-G.}\
  \bibnamefont {Moon}}, \ and\ \bibinfo {author} {\bibfnamefont
  {T.}~\bibnamefont {Shibauchi}},\ }\bibfield  {title} {\bibinfo {title}
  {{Thermodynamic evidence for a field-angle-dependent Majorana gap in a Kitaev
  spin liquid}},\ }\href {\doibase 10.1038/s41567-021-01488-6} {\bibfield
  {journal} {\bibinfo  {journal} {Nat. Phys.}\ }\textbf {\bibinfo {volume}
  {1}},\ \bibinfo {pages} {1} (\bibinfo {year} {2022})}\BibitemShut {NoStop}%
\bibitem [{\citenamefont {Witczak-Krempa}\ \emph {et~al.}(2014)\citenamefont
  {Witczak-Krempa}, \citenamefont {Chen}, \citenamefont {Kim},\ and\
  \citenamefont {Balents}}]{Balents-ARCMP(2014)}%
  \BibitemOpen
  \bibfield  {author} {\bibinfo {author} {\bibfnamefont {W.}~\bibnamefont
  {Witczak-Krempa}}, \bibinfo {author} {\bibfnamefont {G.}~\bibnamefont
  {Chen}}, \bibinfo {author} {\bibfnamefont {Y.~B.}\ \bibnamefont {Kim}}, \
  and\ \bibinfo {author} {\bibfnamefont {L.}~\bibnamefont {Balents}},\
  }\bibfield  {title} {\bibinfo {title} {{Correlated Quantum Phenomena in the
  Strong Spin-Orbit Regime}},\ }\href {\doibase
  10.1146/annurev-conmatphys-020911-125138} {\bibfield  {journal} {\bibinfo
  {journal} {Annu. Rev. Condens. Matter Phys.}\ }\textbf {\bibinfo {volume}
  {5}},\ \bibinfo {pages} {57} (\bibinfo {year} {2014})}\BibitemShut {NoStop}%
\bibitem [{\citenamefont {Takayama}\ \emph {et~al.}(2021)\citenamefont
  {Takayama}, \citenamefont {Chaloupka}, \citenamefont {Smerald}, \citenamefont
  {Khaliullin},\ and\ \citenamefont {Takagi}}]{Takagi-JPSJ(2021)}%
  \BibitemOpen
  \bibfield  {author} {\bibinfo {author} {\bibfnamefont {T.}~\bibnamefont
  {Takayama}}, \bibinfo {author} {\bibfnamefont {J.}~\bibnamefont {Chaloupka}},
  \bibinfo {author} {\bibfnamefont {A.}~\bibnamefont {Smerald}}, \bibinfo
  {author} {\bibfnamefont {G.}~\bibnamefont {Khaliullin}}, \ and\ \bibinfo
  {author} {\bibfnamefont {H.}~\bibnamefont {Takagi}},\ }\bibfield  {title}
  {\bibinfo {title} {{Spin-Orbit-Entangled Electronic Phases in 4$d$ and 5$d$
  Transition-Metal Compounds}},\ }\href {\doibase 10.7566/JPSJ.90.062001}
  {\bibfield  {journal} {\bibinfo  {journal} {J. Phys. Soc. Jpn.}\ }\textbf
  {\bibinfo {volume} {90}},\ \bibinfo {pages} {062001} (\bibinfo {year}
  {2021})}\BibitemShut {NoStop}%
\bibitem [{\citenamefont {Stavropoulos}\ \emph {et~al.}(2019)\citenamefont
  {Stavropoulos}, \citenamefont {Pereira},\ and\ \citenamefont
  {Kee}}]{Stavropoulos-PRL(2019)}%
  \BibitemOpen
  \bibfield  {author} {\bibinfo {author} {\bibfnamefont {P.~P.}\ \bibnamefont
  {Stavropoulos}}, \bibinfo {author} {\bibfnamefont {D.}~\bibnamefont
  {Pereira}}, \ and\ \bibinfo {author} {\bibfnamefont {H.-Y.}\ \bibnamefont
  {Kee}},\ }\bibfield  {title} {\bibinfo {title} {{Microscopic Mechanism for a
  Higher-Spin Kitaev Model}},\ }\href {\doibase 10.1103/PhysRevLett.123.037203}
  {\bibfield  {journal} {\bibinfo  {journal} {Phys. Rev. Lett.}\ }\textbf
  {\bibinfo {volume} {123}},\ \bibinfo {pages} {037203} (\bibinfo {year}
  {2019})}\BibitemShut {NoStop}%
\bibitem [{\citenamefont {Xu}\ \emph {et~al.}(2020)\citenamefont {Xu},
  \citenamefont {Feng}, \citenamefont {Kawamura}, \citenamefont {Yamaji},
  \citenamefont {Nahas}, \citenamefont {Prokhorenko}, \citenamefont {Qi},
  \citenamefont {Xiang},\ and\ \citenamefont {Bellaiche}}]{Xu-PRL(2020)}%
  \BibitemOpen
  \bibfield  {author} {\bibinfo {author} {\bibfnamefont {C.}~\bibnamefont
  {Xu}}, \bibinfo {author} {\bibfnamefont {J.}~\bibnamefont {Feng}}, \bibinfo
  {author} {\bibfnamefont {M.}~\bibnamefont {Kawamura}}, \bibinfo {author}
  {\bibfnamefont {Y.}~\bibnamefont {Yamaji}}, \bibinfo {author} {\bibfnamefont
  {Y.}~\bibnamefont {Nahas}}, \bibinfo {author} {\bibfnamefont
  {S.}~\bibnamefont {Prokhorenko}}, \bibinfo {author} {\bibfnamefont
  {Y.}~\bibnamefont {Qi}}, \bibinfo {author} {\bibfnamefont {H.}~\bibnamefont
  {Xiang}}, \ and\ \bibinfo {author} {\bibfnamefont {L.}~\bibnamefont
  {Bellaiche}},\ }\bibfield  {title} {\bibinfo {title} {{Possible Kitaev
  Quantum Spin Liquid State in 2D Materials with $S=3/2$}},\ }\href {\doibase
  10.1103/PhysRevLett.124.087205} {\bibfield  {journal} {\bibinfo  {journal}
  {Phys. Rev. Lett.}\ }\textbf {\bibinfo {volume} {124}},\ \bibinfo {pages}
  {087205} (\bibinfo {year} {2020})}\BibitemShut {NoStop}%
\bibitem [{\citenamefont {Xu}\ \emph {et~al.}(2018)\citenamefont {Xu},
  \citenamefont {Feng}, \citenamefont {Xiang},\ and\ \citenamefont
  {Bellaiche}}]{Xu-NPJ(2018)}%
  \BibitemOpen
  \bibfield  {author} {\bibinfo {author} {\bibfnamefont {C.}~\bibnamefont
  {Xu}}, \bibinfo {author} {\bibfnamefont {J.}~\bibnamefont {Feng}}, \bibinfo
  {author} {\bibfnamefont {H.}~\bibnamefont {Xiang}}, \ and\ \bibinfo {author}
  {\bibfnamefont {L.}~\bibnamefont {Bellaiche}},\ }\bibfield  {title} {\bibinfo
  {title} {{Interplay between Kitaev interaction and single ion anisotropy in
  ferromagnetic CrI$_3$ and CrGeTe$_3$ monolayers}},\ }\href {\doibase
  10.1038/s41524-018-0115-6} {\bibfield  {journal} {\bibinfo  {journal} {npj
  Comput. Mater.}\ }\textbf {\bibinfo {volume} {4}},\ \bibinfo {pages} {57}
  (\bibinfo {year} {2018})}\BibitemShut {NoStop}%
\bibitem [{\citenamefont {Stavropoulos}\ \emph {et~al.}(2021)\citenamefont
  {Stavropoulos}, \citenamefont {Liu},\ and\ \citenamefont
  {Kee}}]{Stavropoulos-PRB(2021)}%
  \BibitemOpen
  \bibfield  {author} {\bibinfo {author} {\bibfnamefont {P.~P.}\ \bibnamefont
  {Stavropoulos}}, \bibinfo {author} {\bibfnamefont {X.}~\bibnamefont {Liu}}, \
  and\ \bibinfo {author} {\bibfnamefont {H.-Y.}\ \bibnamefont {Kee}},\
  }\bibfield  {title} {\bibinfo {title} {{Magnetic anisotropy in spin-3/2 with
  heavy ligand in honeycomb Mott insulators: Application to
  ${\mathrm{CrI}}_{3}$}},\ }\href {\doibase 10.1103/PhysRevResearch.3.013216}
  {\bibfield  {journal} {\bibinfo  {journal} {Phys. Rev. Res.}\ }\textbf
  {\bibinfo {volume} {3}},\ \bibinfo {pages} {013216} (\bibinfo {year}
  {2021})}\BibitemShut {NoStop}%
\bibitem [{\citenamefont {Baskaran}\ \emph {et~al.}(2008)\citenamefont
  {Baskaran}, \citenamefont {Sen},\ and\ \citenamefont
  {Shankar}}]{Baskaran-PRB(2008)}%
  \BibitemOpen
  \bibfield  {author} {\bibinfo {author} {\bibfnamefont {G.}~\bibnamefont
  {Baskaran}}, \bibinfo {author} {\bibfnamefont {D.}~\bibnamefont {Sen}}, \
  and\ \bibinfo {author} {\bibfnamefont {R.}~\bibnamefont {Shankar}},\
  }\bibfield  {title} {\bibinfo {title} {{Spin-$S$ Kitaev model: Classical
  ground states, order from disorder, and exact correlation functions}},\
  }\href {\doibase 10.1103/PhysRevB.78.115116} {\bibfield  {journal} {\bibinfo
  {journal} {Phys. Rev. B}\ }\textbf {\bibinfo {volume} {78}},\ \bibinfo
  {pages} {115116} (\bibinfo {year} {2008})}\BibitemShut {NoStop}%
\bibitem [{\citenamefont {Koga}\ \emph {et~al.}(2018)\citenamefont {Koga},
  \citenamefont {Tomishige},\ and\ \citenamefont {Nasu}}]{Koga-JPSJ(2018)}%
  \BibitemOpen
  \bibfield  {author} {\bibinfo {author} {\bibfnamefont {A.}~\bibnamefont
  {Koga}}, \bibinfo {author} {\bibfnamefont {H.}~\bibnamefont {Tomishige}}, \
  and\ \bibinfo {author} {\bibfnamefont {J.}~\bibnamefont {Nasu}},\ }\bibfield
  {title} {\bibinfo {title} {{Ground-state and Thermodynamic Properties of an
  $S = 1$ Kitaev Model}},\ }\href {\doibase 10.7566/JPSJ.87.063703} {\bibfield
  {journal} {\bibinfo  {journal} {J. Phys. Soc. Jpn.}\ }\textbf {\bibinfo
  {volume} {87}},\ \bibinfo {pages} {063703} (\bibinfo {year}
  {2018})}\BibitemShut {NoStop}%
\bibitem [{\citenamefont {Hickey}\ \emph {et~al.}(2020)\citenamefont {Hickey},
  \citenamefont {Berke}, \citenamefont {Stavropoulos}, \citenamefont {Kee},\
  and\ \citenamefont {Trebst}}]{Hickey-PRR(2020)}%
  \BibitemOpen
  \bibfield  {author} {\bibinfo {author} {\bibfnamefont {C.}~\bibnamefont
  {Hickey}}, \bibinfo {author} {\bibfnamefont {C.}~\bibnamefont {Berke}},
  \bibinfo {author} {\bibfnamefont {P.~P.}\ \bibnamefont {Stavropoulos}},
  \bibinfo {author} {\bibfnamefont {H.-Y.}\ \bibnamefont {Kee}}, \ and\
  \bibinfo {author} {\bibfnamefont {S.}~\bibnamefont {Trebst}},\ }\bibfield
  {title} {\bibinfo {title} {{Field-driven gapless spin liquid in the spin-1
  Kitaev honeycomb model}},\ }\href {\doibase 10.1103/PhysRevResearch.2.023361}
  {\bibfield  {journal} {\bibinfo  {journal} {Phys. Rev. Res.}\ }\textbf
  {\bibinfo {volume} {2}},\ \bibinfo {pages} {023361} (\bibinfo {year}
  {2020})}\BibitemShut {NoStop}%
\bibitem [{\citenamefont {Dong}\ and\ \citenamefont
  {Sheng}(2020)}]{Dong-PRB(2020)}%
  \BibitemOpen
  \bibfield  {author} {\bibinfo {author} {\bibfnamefont {X.-Y.}\ \bibnamefont
  {Dong}}\ and\ \bibinfo {author} {\bibfnamefont {D.~N.}\ \bibnamefont
  {Sheng}},\ }\bibfield  {title} {\bibinfo {title} {{Spin-1 Kitaev-Heisenberg
  model on a honeycomb lattice}},\ }\href {\doibase
  10.1103/PhysRevB.102.121102} {\bibfield  {journal} {\bibinfo  {journal}
  {Phys. Rev. B}\ }\textbf {\bibinfo {volume} {102}},\ \bibinfo {pages}
  {121102} (\bibinfo {year} {2020})}\BibitemShut {NoStop}%
\bibitem [{\citenamefont {Zhu}\ \emph {et~al.}(2020)\citenamefont {Zhu},
  \citenamefont {Weng},\ and\ \citenamefont {Sheng}}]{Zhu-PRR(2020)}%
  \BibitemOpen
  \bibfield  {author} {\bibinfo {author} {\bibfnamefont {Z.}~\bibnamefont
  {Zhu}}, \bibinfo {author} {\bibfnamefont {Z.-Y.}\ \bibnamefont {Weng}}, \
  and\ \bibinfo {author} {\bibfnamefont {D.~N.}\ \bibnamefont {Sheng}},\
  }\bibfield  {title} {\bibinfo {title} {{Magnetic field induced spin liquids
  in $S=1$ Kitaev honeycomb model}},\ }\href {\doibase
  10.1103/PhysRevResearch.2.022047} {\bibfield  {journal} {\bibinfo  {journal}
  {Phys. Rev. Res.}\ }\textbf {\bibinfo {volume} {2}},\ \bibinfo {pages}
  {022047} (\bibinfo {year} {2020})}\BibitemShut {NoStop}%
\bibitem [{\citenamefont {Khait}\ \emph {et~al.}(2021)\citenamefont {Khait},
  \citenamefont {Stavropoulos}, \citenamefont {Kee},\ and\ \citenamefont
  {Kim}}]{Khait-PRR(2021)}%
  \BibitemOpen
  \bibfield  {author} {\bibinfo {author} {\bibfnamefont {I.}~\bibnamefont
  {Khait}}, \bibinfo {author} {\bibfnamefont {P.~P.}\ \bibnamefont
  {Stavropoulos}}, \bibinfo {author} {\bibfnamefont {H.-Y.}\ \bibnamefont
  {Kee}}, \ and\ \bibinfo {author} {\bibfnamefont {Y.~B.}\ \bibnamefont
  {Kim}},\ }\bibfield  {title} {\bibinfo {title} {{Characterizing spin-one
  Kitaev quantum spin liquids}},\ }\href {\doibase
  10.1103/PhysRevResearch.3.013160} {\bibfield  {journal} {\bibinfo  {journal}
  {Phys. Rev. Res.}\ }\textbf {\bibinfo {volume} {3}},\ \bibinfo {pages}
  {013160} (\bibinfo {year} {2021})}\BibitemShut {NoStop}%
\bibitem [{\citenamefont {Jin}\ \emph {et~al.}(2022)\citenamefont {Jin},
  \citenamefont {Natori}, \citenamefont {Pollmann},\ and\ \citenamefont
  {Knolle}}]{Jin-NC(2022)}%
  \BibitemOpen
  \bibfield  {author} {\bibinfo {author} {\bibfnamefont {H.-K.}\ \bibnamefont
  {Jin}}, \bibinfo {author} {\bibfnamefont {W.~M.~H.}\ \bibnamefont {Natori}},
  \bibinfo {author} {\bibfnamefont {F.}~\bibnamefont {Pollmann}}, \ and\
  \bibinfo {author} {\bibfnamefont {J.}~\bibnamefont {Knolle}},\ }\bibfield
  {title} {\bibinfo {title} {{Unveiling the $S=3/2$ Kitaev honeycomb spin
  liquids}},\ }\href {\doibase 10.1038/s41467-022-31503-0} {\bibfield
  {journal} {\bibinfo  {journal} {Nat. Commun.}\ }\textbf {\bibinfo {volume}
  {13}},\ \bibinfo {pages} {3813} (\bibinfo {year} {2022})}\BibitemShut
  {NoStop}%
\bibitem [{\citenamefont {Chen}\ \emph {et~al.}(2010)\citenamefont {Chen},
  \citenamefont {Pereira},\ and\ \citenamefont {Balents}}]{Balents-PRB(2011)}%
  \BibitemOpen
  \bibfield  {author} {\bibinfo {author} {\bibfnamefont {G.}~\bibnamefont
  {Chen}}, \bibinfo {author} {\bibfnamefont {R.}~\bibnamefont {Pereira}}, \
  and\ \bibinfo {author} {\bibfnamefont {L.}~\bibnamefont {Balents}},\
  }\bibfield  {title} {\bibinfo {title} {{Exotic phases induced by strong
  spin-orbit coupling in ordered double perovskites}},\ }\href {\doibase
  10.1103/PhysRevB.82.174440} {\bibfield  {journal} {\bibinfo  {journal} {Phys.
  Rev. B}\ }\textbf {\bibinfo {volume} {82}},\ \bibinfo {pages} {174440}
  (\bibinfo {year} {2010})}\BibitemShut {NoStop}%
\bibitem [{\citenamefont {Natori}\ \emph {et~al.}(2016)\citenamefont {Natori},
  \citenamefont {Andrade}, \citenamefont {Miranda},\ and\ \citenamefont
  {Pereira}}]{Pereira-PRL(2016)}%
  \BibitemOpen
  \bibfield  {author} {\bibinfo {author} {\bibfnamefont {W.~M.~H.}\
  \bibnamefont {Natori}}, \bibinfo {author} {\bibfnamefont {E.~C.}\
  \bibnamefont {Andrade}}, \bibinfo {author} {\bibfnamefont {E.}~\bibnamefont
  {Miranda}}, \ and\ \bibinfo {author} {\bibfnamefont {R.~G.}\ \bibnamefont
  {Pereira}},\ }\bibfield  {title} {\bibinfo {title} {{Chiral Spin-Orbital
  Liquids with Nodal Lines}},\ }\href {\doibase 10.1103/PhysRevLett.117.017204}
  {\bibfield  {journal} {\bibinfo  {journal} {Phys. Rev. Lett.}\ }\textbf
  {\bibinfo {volume} {117}},\ \bibinfo {pages} {017204} (\bibinfo {year}
  {2016})}\BibitemShut {NoStop}%
\bibitem [{\citenamefont {Natori}\ \emph {et~al.}(2017)\citenamefont {Natori},
  \citenamefont {Daghofer},\ and\ \citenamefont {Pereira}}]{Natori-PRB(2017)}%
  \BibitemOpen
  \bibfield  {author} {\bibinfo {author} {\bibfnamefont {W.~M.~H.}\
  \bibnamefont {Natori}}, \bibinfo {author} {\bibfnamefont {M.}~\bibnamefont
  {Daghofer}}, \ and\ \bibinfo {author} {\bibfnamefont {R.~G.}\ \bibnamefont
  {Pereira}},\ }\bibfield  {title} {\bibinfo {title} {{Dynamics of a
  $j=\frac{3}{2}$ quantum spin liquid}},\ }\href {\doibase
  10.1103/PhysRevB.96.125109} {\bibfield  {journal} {\bibinfo  {journal} {Phys.
  Rev. B}\ }\textbf {\bibinfo {volume} {96}},\ \bibinfo {pages} {125109}
  (\bibinfo {year} {2017})}\BibitemShut {NoStop}%
\bibitem [{\citenamefont {Romh\'anyi}\ \emph {et~al.}(2017)\citenamefont
  {Romh\'anyi}, \citenamefont {Balents},\ and\ \citenamefont
  {Jackeli}}]{Jackeli-PRL(2017)}%
  \BibitemOpen
  \bibfield  {author} {\bibinfo {author} {\bibfnamefont {J.}~\bibnamefont
  {Romh\'anyi}}, \bibinfo {author} {\bibfnamefont {L.}~\bibnamefont {Balents}},
  \ and\ \bibinfo {author} {\bibfnamefont {G.}~\bibnamefont {Jackeli}},\
  }\bibfield  {title} {\bibinfo {title} {{Spin-Orbit Dimers and Noncollinear
  Phases in ${d}^{1}$ Cubic Double Perovskites}},\ }\href {\doibase
  10.1103/PhysRevLett.118.217202} {\bibfield  {journal} {\bibinfo  {journal}
  {Phys. Rev. Lett.}\ }\textbf {\bibinfo {volume} {118}},\ \bibinfo {pages}
  {217202} (\bibinfo {year} {2017})}\BibitemShut {NoStop}%
\bibitem [{\citenamefont {Natori}\ \emph {et~al.}(2018)\citenamefont {Natori},
  \citenamefont {Andrade},\ and\ \citenamefont {Pereira}}]{Natori-PRB(2018)}%
  \BibitemOpen
  \bibfield  {author} {\bibinfo {author} {\bibfnamefont {W.~M.~H.}\
  \bibnamefont {Natori}}, \bibinfo {author} {\bibfnamefont {E.~C.}\
  \bibnamefont {Andrade}}, \ and\ \bibinfo {author} {\bibfnamefont {R.~G.}\
  \bibnamefont {Pereira}},\ }\bibfield  {title} {\bibinfo {title}
  {{SU(4)-symmetric spin-orbital liquids on the hyperhoneycomb lattice}},\
  }\href {\doibase 10.1103/PhysRevB.98.195113} {\bibfield  {journal} {\bibinfo
  {journal} {Phys. Rev. B}\ }\textbf {\bibinfo {volume} {98}},\ \bibinfo
  {pages} {195113} (\bibinfo {year} {2018})}\BibitemShut {NoStop}%
\bibitem [{\citenamefont {Yamada}\ \emph {et~al.}(2018)\citenamefont {Yamada},
  \citenamefont {Oshikawa},\ and\ \citenamefont {Jackeli}}]{Yamada-PRL(2018)}%
  \BibitemOpen
  \bibfield  {author} {\bibinfo {author} {\bibfnamefont {M.~G.}\ \bibnamefont
  {Yamada}}, \bibinfo {author} {\bibfnamefont {M.}~\bibnamefont {Oshikawa}}, \
  and\ \bibinfo {author} {\bibfnamefont {G.}~\bibnamefont {Jackeli}},\
  }\bibfield  {title} {\bibinfo {title} {{Emergent $\mathrm{SU}(4)$ Symmetry in
  $\ensuremath{\alpha}\text{\ensuremath{-}}{\mathrm{ZrCl}}_{3}$ and Crystalline
  Spin-Orbital Liquids}},\ }\href {\doibase 10.1103/PhysRevLett.121.097201}
  {\bibfield  {journal} {\bibinfo  {journal} {Phys. Rev. Lett.}\ }\textbf
  {\bibinfo {volume} {121}},\ \bibinfo {pages} {097201} (\bibinfo {year}
  {2018})}\BibitemShut {NoStop}%
\bibitem [{\citenamefont {Ishikawa}\ \emph {et~al.}(2019)\citenamefont
  {Ishikawa}, \citenamefont {Takayama}, \citenamefont {Kremer}, \citenamefont
  {Nuss}, \citenamefont {Dinnebier}, \citenamefont {Kitagawa}, \citenamefont
  {Ishii},\ and\ \citenamefont {Takagi}}]{Ishikawa-PRB(2019)}%
  \BibitemOpen
  \bibfield  {author} {\bibinfo {author} {\bibfnamefont {H.}~\bibnamefont
  {Ishikawa}}, \bibinfo {author} {\bibfnamefont {T.}~\bibnamefont {Takayama}},
  \bibinfo {author} {\bibfnamefont {R.~K.}\ \bibnamefont {Kremer}}, \bibinfo
  {author} {\bibfnamefont {J.}~\bibnamefont {Nuss}}, \bibinfo {author}
  {\bibfnamefont {R.}~\bibnamefont {Dinnebier}}, \bibinfo {author}
  {\bibfnamefont {K.}~\bibnamefont {Kitagawa}}, \bibinfo {author}
  {\bibfnamefont {K.}~\bibnamefont {Ishii}}, \ and\ \bibinfo {author}
  {\bibfnamefont {H.}~\bibnamefont {Takagi}},\ }\bibfield  {title} {\bibinfo
  {title} {{Ordering of hidden multipoles in spin-orbit entangled $5{d}^{1}$ Ta
  chlorides}},\ }\href {\doibase 10.1103/PhysRevB.100.045142} {\bibfield
  {journal} {\bibinfo  {journal} {Phys. Rev. B}\ }\textbf {\bibinfo {volume}
  {100}},\ \bibinfo {pages} {045142} (\bibinfo {year} {2019})}\BibitemShut
  {NoStop}%
\bibitem [{\citenamefont {de~Farias}\ \emph {et~al.}(2020)\citenamefont
  {de~Farias}, \citenamefont {de~Carvalho}, \citenamefont {Miranda},\ and\
  \citenamefont {Pereira}}]{Pereira-PRB(2020)}%
  \BibitemOpen
  \bibfield  {author} {\bibinfo {author} {\bibfnamefont {C.~S.}\ \bibnamefont
  {de~Farias}}, \bibinfo {author} {\bibfnamefont {V.~S.}\ \bibnamefont
  {de~Carvalho}}, \bibinfo {author} {\bibfnamefont {E.}~\bibnamefont
  {Miranda}}, \ and\ \bibinfo {author} {\bibfnamefont {R.~G.}\ \bibnamefont
  {Pereira}},\ }\bibfield  {title} {\bibinfo {title} {{Quadrupolar spin liquid,
  octupolar Kondo coupling, and odd-frequency superconductivity in an exactly
  solvable model}},\ }\href {\doibase 10.1103/PhysRevB.102.075110} {\bibfield
  {journal} {\bibinfo  {journal} {Phys. Rev. B}\ }\textbf {\bibinfo {volume}
  {102}},\ \bibinfo {pages} {075110} (\bibinfo {year} {2020})}\BibitemShut
  {NoStop}%
\bibitem [{\citenamefont {Chen}\ and\ \citenamefont
  {Wu}(2021)}]{Chen-arXiv(2021)}%
  \BibitemOpen
  \bibfield  {author} {\bibinfo {author} {\bibfnamefont {G.}~\bibnamefont
  {Chen}}\ and\ \bibinfo {author} {\bibfnamefont {C.}~\bibnamefont {Wu}},\
  }\bibfield  {title} {\bibinfo {title} {{Mott insulators with large local
  Hilbert spaces in quantum materials and ultracold atoms}},\ }\href
  {https://arxiv.org/abs/2112.02630} {\bibfield  {journal} {\bibinfo  {journal}
  {arXiv:2112.02630}\ } (\bibinfo {year} {2021})}\BibitemShut {NoStop}%
\bibitem [{\citenamefont {Natori}\ \emph {et~al.}(2023)\citenamefont {Natori},
  \citenamefont {Jin},\ and\ \citenamefont {Knolle}}]{Natori-arXiv(2023)}%
  \BibitemOpen
  \bibfield  {author} {\bibinfo {author} {\bibfnamefont {W.~M.~H.}\
  \bibnamefont {Natori}}, \bibinfo {author} {\bibfnamefont {H.-K.}\
  \bibnamefont {Jin}}, \ and\ \bibinfo {author} {\bibfnamefont
  {J.}~\bibnamefont {Knolle}},\ }\bibfield  {title} {\bibinfo {title} {{Quantum
  liquids of the $S=3/2$ Kitaev honeycomb and related Kugel-Khomskii models}},\
  }\href {https://ui.adsabs.harvard.edu/abs/2023arXiv230413378N} {\bibfield
  {journal} {\bibinfo  {journal} {arXiv:2304.13378}\ } (\bibinfo {year}
  {2023})}\BibitemShut {NoStop}%
\bibitem [{\citenamefont {Li}\ and\ \citenamefont
  {Chen}(2017)}]{Chen-PRB(2017)}%
  \BibitemOpen
  \bibfield  {author} {\bibinfo {author} {\bibfnamefont {Y.-D.}\ \bibnamefont
  {Li}}\ and\ \bibinfo {author} {\bibfnamefont {G.}~\bibnamefont {Chen}},\
  }\bibfield  {title} {\bibinfo {title} {{Symmetry enriched U(1) topological
  orders for dipole-octupole doublets on a pyrochlore lattice}},\ }\href
  {\doibase 10.1103/PhysRevB.95.041106} {\bibfield  {journal} {\bibinfo
  {journal} {Phys. Rev. B}\ }\textbf {\bibinfo {volume} {95}},\ \bibinfo
  {pages} {041106} (\bibinfo {year} {2017})}\BibitemShut {NoStop}%
\bibitem [{\citenamefont {Sibille}\ \emph {et~al.}(2020)\citenamefont
  {Sibille}, \citenamefont {Gauthier}, \citenamefont {Lhotel}, \citenamefont
  {Por\'ee}, \citenamefont {Pomjakushin}, \citenamefont {Ewings}, \citenamefont
  {Perring}, \citenamefont {Ollivier}, \citenamefont {Wildes}, \citenamefont
  {Ritter}, \citenamefont {Hansen}, \citenamefont {Keen}, \citenamefont
  {Nilsen}, \citenamefont {Keller}, \citenamefont {Petit},\ and\ \citenamefont
  {Fennell}}]{Sibille-NP(2020)}%
  \BibitemOpen
  \bibfield  {author} {\bibinfo {author} {\bibfnamefont {R.}~\bibnamefont
  {Sibille}}, \bibinfo {author} {\bibfnamefont {N.}~\bibnamefont {Gauthier}},
  \bibinfo {author} {\bibfnamefont {E.}~\bibnamefont {Lhotel}}, \bibinfo
  {author} {\bibfnamefont {V.}~\bibnamefont {Por\'ee}}, \bibinfo {author}
  {\bibfnamefont {V.}~\bibnamefont {Pomjakushin}}, \bibinfo {author}
  {\bibfnamefont {R.~A.}\ \bibnamefont {Ewings}}, \bibinfo {author}
  {\bibfnamefont {T.~G.}\ \bibnamefont {Perring}}, \bibinfo {author}
  {\bibfnamefont {J.}~\bibnamefont {Ollivier}}, \bibinfo {author}
  {\bibfnamefont {A.}~\bibnamefont {Wildes}}, \bibinfo {author} {\bibfnamefont
  {C.}~\bibnamefont {Ritter}}, \bibinfo {author} {\bibfnamefont {T.~C.}\
  \bibnamefont {Hansen}}, \bibinfo {author} {\bibfnamefont {D.~A.}\
  \bibnamefont {Keen}}, \bibinfo {author} {\bibfnamefont {G.~J.}\ \bibnamefont
  {Nilsen}}, \bibinfo {author} {\bibfnamefont {L.}~\bibnamefont {Keller}},
  \bibinfo {author} {\bibfnamefont {S.}~\bibnamefont {Petit}}, \ and\ \bibinfo
  {author} {\bibfnamefont {T.}~\bibnamefont {Fennell}},\ }\bibfield  {title}
  {\bibinfo {title} {{A quantum liquid of magnetic octupoles on the pyrochlore
  lattice}},\ }\href {\doibase 10.1038/s41567-020-0827-7} {\bibfield  {journal}
  {\bibinfo  {journal} {Nat. Phys.}\ }\textbf {\bibinfo {volume} {16}},\
  \bibinfo {pages} {546} (\bibinfo {year} {2020})}\BibitemShut {NoStop}%
\bibitem [{\citenamefont {Rayyan}\ \emph {et~al.}(2023)\citenamefont {Rayyan},
  \citenamefont {Churchill},\ and\ \citenamefont {Kee}}]{Rayyan-PRB(2023)}%
  \BibitemOpen
  \bibfield  {author} {\bibinfo {author} {\bibfnamefont {A.}~\bibnamefont
  {Rayyan}}, \bibinfo {author} {\bibfnamefont {D.}~\bibnamefont {Churchill}}, \
  and\ \bibinfo {author} {\bibfnamefont {H.-Y.}\ \bibnamefont {Kee}},\
  }\bibfield  {title} {\bibinfo {title} {{Field-induced Kitaev multipolar
  liquid in spin-orbit coupled ${d}^{2}$ honeycomb Mott insulators}},\ }\href
  {\doibase 10.1103/PhysRevB.107.L020408} {\bibfield  {journal} {\bibinfo
  {journal} {Phys. Rev. B}\ }\textbf {\bibinfo {volume} {107}},\ \bibinfo
  {pages} {L020408} (\bibinfo {year} {2023})}\BibitemShut {NoStop}%
\bibitem [{\citenamefont {Khomskii}\ and\ \citenamefont
  {Streltsov}(2021)}]{Khomskii-CR(2021)}%
  \BibitemOpen
  \bibfield  {author} {\bibinfo {author} {\bibfnamefont {D.~I.}\ \bibnamefont
  {Khomskii}}\ and\ \bibinfo {author} {\bibfnamefont {S.~V.}\ \bibnamefont
  {Streltsov}},\ }\bibfield  {title} {\bibinfo {title} {Orbital effects in
  solids: Basics, recent progress, and opportunities},\ }\href {\doibase
  10.1021/acs.chemrev.0c00579} {\bibfield  {journal} {\bibinfo  {journal}
  {Chem. Rev.}\ }\textbf {\bibinfo {volume} {121}},\ \bibinfo {pages} {2992}
  (\bibinfo {year} {2021})}\BibitemShut {NoStop}%
\bibitem [{\citenamefont {Streltsov}\ and\ \citenamefont
  {Khomskii}(2020)}]{Streltsov-PRX(2020)}%
  \BibitemOpen
  \bibfield  {author} {\bibinfo {author} {\bibfnamefont {S.~V.}\ \bibnamefont
  {Streltsov}}\ and\ \bibinfo {author} {\bibfnamefont {D.~I.}\ \bibnamefont
  {Khomskii}},\ }\bibfield  {title} {\bibinfo {title} {{Jahn-Teller Effect and
  Spin-Orbit Coupling: Friends or Foes?}}\ }\href {\doibase
  10.1103/PhysRevX.10.031043} {\bibfield  {journal} {\bibinfo  {journal} {Phys.
  Rev. X}\ }\textbf {\bibinfo {volume} {10}},\ \bibinfo {pages} {031043}
  (\bibinfo {year} {2020})}\BibitemShut {NoStop}%
\bibitem [{\citenamefont {Rau}\ \emph {et~al.}(2014)\citenamefont {Rau},
  \citenamefont {Lee},\ and\ \citenamefont {Kee}}]{Rau-PRL(2014)}%
  \BibitemOpen
  \bibfield  {author} {\bibinfo {author} {\bibfnamefont {J.~G.}\ \bibnamefont
  {Rau}}, \bibinfo {author} {\bibfnamefont {E.~K.-H.}\ \bibnamefont {Lee}}, \
  and\ \bibinfo {author} {\bibfnamefont {H.-Y.}\ \bibnamefont {Kee}},\
  }\bibfield  {title} {\bibinfo {title} {{Generic Spin Model for the Honeycomb
  Iridates beyond the Kitaev Limit}},\ }\href {\doibase
  10.1103/PhysRevLett.112.077204} {\bibfield  {journal} {\bibinfo  {journal}
  {Phys. Rev. Lett.}\ }\textbf {\bibinfo {volume} {112}},\ \bibinfo {pages}
  {077204} (\bibinfo {year} {2014})}\BibitemShut {NoStop}%
\bibitem [{\citenamefont {Baker}(1971)}]{Baker-RPP(1971)}%
  \BibitemOpen
  \bibfield  {author} {\bibinfo {author} {\bibfnamefont {J.~M.}\ \bibnamefont
  {Baker}},\ }\bibfield  {title} {\bibinfo {title} {Interactions between ions
  with orbital angular momentum in insulators},\ }\href {\doibase
  10.1088/0034-4885/34/1/303} {\bibfield  {journal} {\bibinfo  {journal} {Rep.
  Prog. Phys.}\ }\textbf {\bibinfo {volume} {34}},\ \bibinfo {pages} {109}
  (\bibinfo {year} {1971})}\BibitemShut {NoStop}%
\bibitem [{\citenamefont {Gehring}\ and\ \citenamefont
  {Gehring}(1975)}]{Gehring-RPP(1975)}%
  \BibitemOpen
  \bibfield  {author} {\bibinfo {author} {\bibfnamefont {G.~A.}\ \bibnamefont
  {Gehring}}\ and\ \bibinfo {author} {\bibfnamefont {K.~A.}\ \bibnamefont
  {Gehring}},\ }\bibfield  {title} {\bibinfo {title} {{Co-operative Jahn-Teller
  effects}},\ }\href {\doibase 10.1088/0034-4885/38/1/001} {\bibfield
  {journal} {\bibinfo  {journal} {Rep. Prog. Phys.}\ }\textbf {\bibinfo
  {volume} {38}},\ \bibinfo {pages} {1} (\bibinfo {year} {1975})}\BibitemShut
  {NoStop}%
\bibitem [{\citenamefont {Voleti}\ \emph {et~al.}(2021)\citenamefont {Voleti},
  \citenamefont {Haldar},\ and\ \citenamefont
  {Paramekanti}}]{Voleti-PRB(2021)}%
  \BibitemOpen
  \bibfield  {author} {\bibinfo {author} {\bibfnamefont {S.}~\bibnamefont
  {Voleti}}, \bibinfo {author} {\bibfnamefont {A.}~\bibnamefont {Haldar}}, \
  and\ \bibinfo {author} {\bibfnamefont {A.}~\bibnamefont {Paramekanti}},\
  }\bibfield  {title} {\bibinfo {title} {{Octupolar order and Ising quantum
  criticality tuned by strain and dimensionality: Application to $d$-orbital
  Mott insulators}},\ }\href {\doibase 10.1103/PhysRevB.104.174431} {\bibfield
  {journal} {\bibinfo  {journal} {Phys. Rev. B}\ }\textbf {\bibinfo {volume}
  {104}},\ \bibinfo {pages} {174431} (\bibinfo {year} {2021})}\BibitemShut
  {NoStop}%
\bibitem [{\citenamefont {Yadav}\ \emph {et~al.}(2018)\citenamefont {Yadav},
  \citenamefont {Rachel}, \citenamefont {Hozoi}, \citenamefont {van~den
  Brink},\ and\ \citenamefont {Jackeli}}]{Yadav-PRB(2018)}%
  \BibitemOpen
  \bibfield  {author} {\bibinfo {author} {\bibfnamefont {R.}~\bibnamefont
  {Yadav}}, \bibinfo {author} {\bibfnamefont {S.}~\bibnamefont {Rachel}},
  \bibinfo {author} {\bibfnamefont {L.}~\bibnamefont {Hozoi}}, \bibinfo
  {author} {\bibfnamefont {J.}~\bibnamefont {van~den Brink}}, \ and\ \bibinfo
  {author} {\bibfnamefont {G.}~\bibnamefont {Jackeli}},\ }\bibfield  {title}
  {\bibinfo {title} {{Strain- and pressure-tuned magnetic interactions in
  honeycomb Kitaev materials}},\ }\href {\doibase 10.1103/PhysRevB.98.121107}
  {\bibfield  {journal} {\bibinfo  {journal} {Phys. Rev. B}\ }\textbf {\bibinfo
  {volume} {98}},\ \bibinfo {pages} {121107} (\bibinfo {year}
  {2018})}\BibitemShut {NoStop}%
\bibitem [{\citenamefont {Santini}\ \emph {et~al.}(2009)\citenamefont
  {Santini}, \citenamefont {Carretta}, \citenamefont {Amoretti}, \citenamefont
  {Caciuffo}, \citenamefont {Magnani},\ and\ \citenamefont
  {Lander}}]{Santini-RMP(2009)}%
  \BibitemOpen
  \bibfield  {author} {\bibinfo {author} {\bibfnamefont {P.}~\bibnamefont
  {Santini}}, \bibinfo {author} {\bibfnamefont {S.}~\bibnamefont {Carretta}},
  \bibinfo {author} {\bibfnamefont {G.}~\bibnamefont {Amoretti}}, \bibinfo
  {author} {\bibfnamefont {R.}~\bibnamefont {Caciuffo}}, \bibinfo {author}
  {\bibfnamefont {N.}~\bibnamefont {Magnani}}, \ and\ \bibinfo {author}
  {\bibfnamefont {G.~H.}\ \bibnamefont {Lander}},\ }\bibfield  {title}
  {\bibinfo {title} {Multipolar interactions in $f$-electron systems: The
  paradigm of actinide dioxides},\ }\href {\doibase 10.1103/RevModPhys.81.807}
  {\bibfield  {journal} {\bibinfo  {journal} {Rev. Mod. Phys.}\ }\textbf
  {\bibinfo {volume} {81}},\ \bibinfo {pages} {807} (\bibinfo {year}
  {2009})}\BibitemShut {NoStop}%
\bibitem [{\citenamefont {Iwahara}\ \emph {et~al.}(2018)\citenamefont
  {Iwahara}, \citenamefont {Vieru},\ and\ \citenamefont
  {Chibotaru}}]{Iwahara-PRB(2018)}%
  \BibitemOpen
  \bibfield  {author} {\bibinfo {author} {\bibfnamefont {N.}~\bibnamefont
  {Iwahara}}, \bibinfo {author} {\bibfnamefont {V.}~\bibnamefont {Vieru}}, \
  and\ \bibinfo {author} {\bibfnamefont {L.~F.}\ \bibnamefont {Chibotaru}},\
  }\bibfield  {title} {\bibinfo {title} {Spin-orbital-lattice entangled states
  in cubic ${d}^{1}$ double perovskites},\ }\href {\doibase
  10.1103/PhysRevB.98.075138} {\bibfield  {journal} {\bibinfo  {journal} {Phys.
  Rev. B}\ }\textbf {\bibinfo {volume} {98}},\ \bibinfo {pages} {075138}
  (\bibinfo {year} {2018})}\BibitemShut {NoStop}%
\bibitem [{\citenamefont {Wu}(2008)}]{Wu-PRL(2008)}%
  \BibitemOpen
  \bibfield  {author} {\bibinfo {author} {\bibfnamefont {C.}~\bibnamefont
  {Wu}},\ }\bibfield  {title} {\bibinfo {title} {{Orbital Ordering and
  Frustration of $p$-Band Mott Insulators}},\ }\href {\doibase
  10.1103/PhysRevLett.100.200406} {\bibfield  {journal} {\bibinfo  {journal}
  {Phys. Rev. Lett.}\ }\textbf {\bibinfo {volume} {100}},\ \bibinfo {pages}
  {200406} (\bibinfo {year} {2008})}\BibitemShut {NoStop}%
\bibitem [{\citenamefont {Khaliullin}\ \emph {et~al.}(2021)\citenamefont
  {Khaliullin}, \citenamefont {Churchill}, \citenamefont {Stavropoulos},\ and\
  \citenamefont {Kee}}]{Khaliulin-PRR(2021)}%
  \BibitemOpen
  \bibfield  {author} {\bibinfo {author} {\bibfnamefont {G.}~\bibnamefont
  {Khaliullin}}, \bibinfo {author} {\bibfnamefont {D.}~\bibnamefont
  {Churchill}}, \bibinfo {author} {\bibfnamefont {P.~P.}\ \bibnamefont
  {Stavropoulos}}, \ and\ \bibinfo {author} {\bibfnamefont {H.-Y.}\
  \bibnamefont {Kee}},\ }\bibfield  {title} {\bibinfo {title} {{Exchange
  interactions, Jahn-Teller coupling, and multipole orders in pseudospin
  one-half $5{d}^{2}$ Mott insulators}},\ }\href {\doibase
  10.1103/PhysRevResearch.3.033163} {\bibfield  {journal} {\bibinfo  {journal}
  {Phys. Rev. Research}\ }\textbf {\bibinfo {volume} {3}},\ \bibinfo {pages}
  {033163} (\bibinfo {year} {2021})}\BibitemShut {NoStop}%
\bibitem [{\citenamefont {Kubo}\ and\ \citenamefont
  {Kuramoto}(2003)}]{Kubo-JPSJ(2003)}%
  \BibitemOpen
  \bibfield  {author} {\bibinfo {author} {\bibfnamefont {K.}~\bibnamefont
  {Kubo}}\ and\ \bibinfo {author} {\bibfnamefont {Y.}~\bibnamefont
  {Kuramoto}},\ }\bibfield  {title} {\bibinfo {title} {{Lattice Distortion and
  Octupole Ordering Model in Ce$_x$La$_{1 - x}$B$_6$}},\ }\href {\doibase
  10.1143/JPSJ.72.1859} {\bibfield  {journal} {\bibinfo  {journal} {J. Phys.
  Soc. Jpn.}\ }\textbf {\bibinfo {volume} {72}},\ \bibinfo {pages} {1859}
  (\bibinfo {year} {2003})}\BibitemShut {NoStop}%
\bibitem [{\citenamefont {Kubo}\ and\ \citenamefont
  {Kuramoto}(2004)}]{Kubo-JPSJ(2004)}%
  \BibitemOpen
  \bibfield  {author} {\bibinfo {author} {\bibfnamefont {K.}~\bibnamefont
  {Kubo}}\ and\ \bibinfo {author} {\bibfnamefont {Y.}~\bibnamefont
  {Kuramoto}},\ }\bibfield  {title} {\bibinfo {title} {{Octupole Ordering Model
  for the Phase IV of Ce$_x$La$_{1 - x}$B$_6$}},\ }\href {\doibase
  10.1143/JPSJ.73.216} {\bibfield  {journal} {\bibinfo  {journal} {J. Phys.
  Soc. Jpn.}\ }\textbf {\bibinfo {volume} {73}},\ \bibinfo {pages} {216}
  (\bibinfo {year} {2004})}\BibitemShut {NoStop}%
\bibitem [{\citenamefont {Paramekanti}\ \emph {et~al.}(2020)\citenamefont
  {Paramekanti}, \citenamefont {Maharaj},\ and\ \citenamefont
  {Gaulin}}]{Paramekanti-PRB(2020)}%
  \BibitemOpen
  \bibfield  {author} {\bibinfo {author} {\bibfnamefont {A.}~\bibnamefont
  {Paramekanti}}, \bibinfo {author} {\bibfnamefont {D.~D.}\ \bibnamefont
  {Maharaj}}, \ and\ \bibinfo {author} {\bibfnamefont {B.~D.}\ \bibnamefont
  {Gaulin}},\ }\bibfield  {title} {\bibinfo {title} {{Octupolar order in
  $d$-orbital Mott insulators}},\ }\href {\doibase 10.1103/PhysRevB.101.054439}
  {\bibfield  {journal} {\bibinfo  {journal} {Phys. Rev. B}\ }\textbf {\bibinfo
  {volume} {101}},\ \bibinfo {pages} {054439} (\bibinfo {year}
  {2020})}\BibitemShut {NoStop}%
\bibitem [{\citenamefont {Churchill}\ and\ \citenamefont
  {Kee}(2022)}]{Churchill-PRB(2022)}%
  \BibitemOpen
  \bibfield  {author} {\bibinfo {author} {\bibfnamefont {D.}~\bibnamefont
  {Churchill}}\ and\ \bibinfo {author} {\bibfnamefont {H.-Y.}\ \bibnamefont
  {Kee}},\ }\bibfield  {title} {\bibinfo {title} {Competing multipolar orders
  in a face-centered cubic lattice: Application to the osmium double
  perovskites},\ }\href {\doibase 10.1103/PhysRevB.105.014438} {\bibfield
  {journal} {\bibinfo  {journal} {Phys. Rev. B}\ }\textbf {\bibinfo {volume}
  {105}},\ \bibinfo {pages} {014438} (\bibinfo {year} {2022})}\BibitemShut
  {NoStop}%
\bibitem [{\citenamefont {Shiina}\ \emph {et~al.}(1997)\citenamefont {Shiina},
  \citenamefont {Shiba},\ and\ \citenamefont {Thalmeier}}]{Shiina-JPSJ(1997)}%
  \BibitemOpen
  \bibfield  {author} {\bibinfo {author} {\bibfnamefont {R.}~\bibnamefont
  {Shiina}}, \bibinfo {author} {\bibfnamefont {H.}~\bibnamefont {Shiba}}, \
  and\ \bibinfo {author} {\bibfnamefont {P.}~\bibnamefont {Thalmeier}},\
  }\bibfield  {title} {\bibinfo {title} {{Magnetic-Field Effects on Quadrupolar
  Ordering in a $\Gamma_{8}$-Quartet System CeB$_{6}$}},\ }\href {\doibase
  10.1143/JPSJ.66.1741} {\bibfield  {journal} {\bibinfo  {journal} {J. Phys.
  Soc. Jpn.}\ }\textbf {\bibinfo {volume} {66}},\ \bibinfo {pages} {1741}
  (\bibinfo {year} {1997})}\BibitemShut {NoStop}%
\bibitem [{\citenamefont {Knolle}\ \emph {et~al.}(2014)\citenamefont {Knolle},
  \citenamefont {Kovrizhin}, \citenamefont {Chalker},\ and\ \citenamefont
  {Moessner}}]{Chalker-PRL(2014)}%
  \BibitemOpen
  \bibfield  {author} {\bibinfo {author} {\bibfnamefont {J.}~\bibnamefont
  {Knolle}}, \bibinfo {author} {\bibfnamefont {D.~L.}\ \bibnamefont
  {Kovrizhin}}, \bibinfo {author} {\bibfnamefont {J.~T.}\ \bibnamefont
  {Chalker}}, \ and\ \bibinfo {author} {\bibfnamefont {R.}~\bibnamefont
  {Moessner}},\ }\bibfield  {title} {\bibinfo {title} {{Dynamics of a
  Two-Dimensional Quantum Spin Liquid: Signatures of Emergent Majorana Fermions
  and Fluxes}},\ }\href {\doibase 10.1103/PhysRevLett.112.207203} {\bibfield
  {journal} {\bibinfo  {journal} {Phys. Rev. Lett.}\ }\textbf {\bibinfo
  {volume} {112}},\ \bibinfo {pages} {207203} (\bibinfo {year}
  {2014})}\BibitemShut {NoStop}%
\bibitem [{\citenamefont {Fukui}\ \emph {et~al.}(2005)\citenamefont {Fukui},
  \citenamefont {Hatsugai},\ and\ \citenamefont {Suzuki}}]{Fukui-JPSJ(2005)}%
  \BibitemOpen
  \bibfield  {author} {\bibinfo {author} {\bibfnamefont {T.}~\bibnamefont
  {Fukui}}, \bibinfo {author} {\bibfnamefont {Y.}~\bibnamefont {Hatsugai}}, \
  and\ \bibinfo {author} {\bibfnamefont {H.}~\bibnamefont {Suzuki}},\
  }\bibfield  {title} {\bibinfo {title} {{Chern Numbers in Discretized
  Brillouin Zone: Efficient Method of Computing (Spin) Hall Conductances}},\
  }\href {\doibase 10.1143/JPSJ.74.1674} {\bibfield  {journal} {\bibinfo
  {journal} {J. Phys Soc. Jpn.}\ }\textbf {\bibinfo {volume} {74}},\ \bibinfo
  {pages} {1674} (\bibinfo {year} {2005})}\BibitemShut {NoStop}%
\bibitem [{\citenamefont {Zhang}\ \emph {et~al.}(2020)\citenamefont {Zhang},
  \citenamefont {Batista},\ and\ \citenamefont {Hal\'asz}}]{Batista-PRR(2020)}%
  \BibitemOpen
  \bibfield  {author} {\bibinfo {author} {\bibfnamefont {S.-S.}\ \bibnamefont
  {Zhang}}, \bibinfo {author} {\bibfnamefont {C.~D.}\ \bibnamefont {Batista}},
  \ and\ \bibinfo {author} {\bibfnamefont {G.~B.}\ \bibnamefont {Hal\'asz}},\
  }\bibfield  {title} {\bibinfo {title} {{Toward Kitaev's sixteenfold way in a
  honeycomb lattice model}},\ }\href {\doibase
  10.1103/PhysRevResearch.2.023334} {\bibfield  {journal} {\bibinfo  {journal}
  {Phys. Rev. Res.}\ }\textbf {\bibinfo {volume} {2}},\ \bibinfo {pages}
  {023334} (\bibinfo {year} {2020})}\BibitemShut {NoStop}%
\bibitem [{\citenamefont {Chulliparambil}\ \emph {et~al.}(2020)\citenamefont
  {Chulliparambil}, \citenamefont {Seifert}, \citenamefont {Vojta},
  \citenamefont {Janssen},\ and\ \citenamefont
  {Tu}}]{Chulliparambil-PRB(2020)}%
  \BibitemOpen
  \bibfield  {author} {\bibinfo {author} {\bibfnamefont {S.}~\bibnamefont
  {Chulliparambil}}, \bibinfo {author} {\bibfnamefont {U.~F.~P.}\ \bibnamefont
  {Seifert}}, \bibinfo {author} {\bibfnamefont {M.}~\bibnamefont {Vojta}},
  \bibinfo {author} {\bibfnamefont {L.}~\bibnamefont {Janssen}}, \ and\
  \bibinfo {author} {\bibfnamefont {H.-H.}\ \bibnamefont {Tu}},\ }\bibfield
  {title} {\bibinfo {title} {{Microscopic models for Kitaev's sixteenfold way
  of anyon theories}},\ }\href {\doibase 10.1103/PhysRevB.102.201111}
  {\bibfield  {journal} {\bibinfo  {journal} {Phys. Rev. B}\ }\textbf {\bibinfo
  {volume} {102}},\ \bibinfo {pages} {201111(R)} (\bibinfo {year}
  {2020})}\BibitemShut {NoStop}%
\bibitem [{\citenamefont {Fernandes}\ \emph {et~al.}(2019)\citenamefont
  {Fernandes}, \citenamefont {Orth},\ and\ \citenamefont
  {Schmalian}}]{Fernandes-ARCMP(2019)}%
  \BibitemOpen
  \bibfield  {author} {\bibinfo {author} {\bibfnamefont {R.~M.}\ \bibnamefont
  {Fernandes}}, \bibinfo {author} {\bibfnamefont {P.~P.}\ \bibnamefont {Orth}},
  \ and\ \bibinfo {author} {\bibfnamefont {J.}~\bibnamefont {Schmalian}},\
  }\bibfield  {title} {\bibinfo {title} {{Intertwined Vestigial Order in
  Quantum Materials: Nematicity and Beyond}},\ }\href
  {https://doi.org/10.1146/annurev-conmatphys-031218-013200} {\bibfield
  {journal} {\bibinfo  {journal} {Annu. Rev. Condens. Matter Phys.}\ }\textbf
  {\bibinfo {volume} {10}},\ \bibinfo {pages} {133} (\bibinfo {year}
  {2019})}\BibitemShut {NoStop}%
\bibitem [{\citenamefont {Murthy}\ \emph {et~al.}(1997)\citenamefont {Murthy},
  \citenamefont {Arovas},\ and\ \citenamefont {Auerbach}}]{Murthy-PRB(1997)}%
  \BibitemOpen
  \bibfield  {author} {\bibinfo {author} {\bibfnamefont {G.}~\bibnamefont
  {Murthy}}, \bibinfo {author} {\bibfnamefont {D.}~\bibnamefont {Arovas}}, \
  and\ \bibinfo {author} {\bibfnamefont {A.}~\bibnamefont {Auerbach}},\
  }\bibfield  {title} {\bibinfo {title} {{Superfluids and supersolids on
  frustrated two-dimensional lattices}},\ }\href {\doibase
  10.1103/PhysRevB.55.3104} {\bibfield  {journal} {\bibinfo  {journal} {Phys.
  Rev. B}\ }\textbf {\bibinfo {volume} {55}},\ \bibinfo {pages} {3104}
  (\bibinfo {year} {1997})}\BibitemShut {NoStop}%
\bibitem [{\citenamefont {Zhao}\ and\ \citenamefont
  {Liu}(2008)}]{Liu-PRL(2008)}%
  \BibitemOpen
  \bibfield  {author} {\bibinfo {author} {\bibfnamefont {E.}~\bibnamefont
  {Zhao}}\ and\ \bibinfo {author} {\bibfnamefont {W.~V.}\ \bibnamefont {Liu}},\
  }\bibfield  {title} {\bibinfo {title} {{Orbital Order in Mott Insulators of
  Spinless $p$-Band Fermions}},\ }\href {\doibase
  10.1103/PhysRevLett.100.160403} {\bibfield  {journal} {\bibinfo  {journal}
  {Phys. Rev. Lett.}\ }\textbf {\bibinfo {volume} {100}},\ \bibinfo {pages}
  {160403} (\bibinfo {year} {2008})}\BibitemShut {NoStop}%
\bibitem [{\citenamefont {van~den Brink}\ \emph {et~al.}(1998)\citenamefont
  {van~den Brink}, \citenamefont {Stekelenburg}, \citenamefont {Khomskii},
  \citenamefont {Sawatzky},\ and\ \citenamefont {Kugel}}]{Brink-PRB(1998)}%
  \BibitemOpen
  \bibfield  {author} {\bibinfo {author} {\bibfnamefont {J.}~\bibnamefont
  {van~den Brink}}, \bibinfo {author} {\bibfnamefont {W.}~\bibnamefont
  {Stekelenburg}}, \bibinfo {author} {\bibfnamefont {D.~I.}\ \bibnamefont
  {Khomskii}}, \bibinfo {author} {\bibfnamefont {G.~A.}\ \bibnamefont
  {Sawatzky}}, \ and\ \bibinfo {author} {\bibfnamefont {K.~I.}\ \bibnamefont
  {Kugel}},\ }\bibfield  {title} {\bibinfo {title} {{Elementary excitations in
  the coupled spin-orbital model}},\ }\href {\doibase
  10.1103/PhysRevB.58.10276} {\bibfield  {journal} {\bibinfo  {journal} {Phys.
  Rev. B}\ }\textbf {\bibinfo {volume} {58}},\ \bibinfo {pages} {10276}
  (\bibinfo {year} {1998})}\BibitemShut {NoStop}%
\bibitem [{\citenamefont {Hou}\ \emph {et~al.}(2007)\citenamefont {Hou},
  \citenamefont {Chamon},\ and\ \citenamefont {Mudry}}]{Hou-PRL(2007)}%
  \BibitemOpen
  \bibfield  {author} {\bibinfo {author} {\bibfnamefont {C.-Y.}\ \bibnamefont
  {Hou}}, \bibinfo {author} {\bibfnamefont {C.}~\bibnamefont {Chamon}}, \ and\
  \bibinfo {author} {\bibfnamefont {C.}~\bibnamefont {Mudry}},\ }\bibfield
  {title} {\bibinfo {title} {{Electron Fractionalization in Two-Dimensional
  Graphenelike Structures}},\ }\href {\doibase 10.1103/PhysRevLett.98.186809}
  {\bibfield  {journal} {\bibinfo  {journal} {Phys. Rev. Lett.}\ }\textbf
  {\bibinfo {volume} {98}},\ \bibinfo {pages} {186809} (\bibinfo {year}
  {2007})}\BibitemShut {NoStop}%
\bibitem [{\citenamefont {{A. C. Qu and P. Nigge and S. Link and G. Levy and M.
  Michiardi and P. L. Spandar and T. Matth\'e and M. Schneider and S.
  Zhdanovich and U. Starke and C. Guti\'errez and A.
  Damascelli}}(2022)}]{Qu-SciAdv(2022)}%
  \BibitemOpen
  \bibfield  {author} {\bibinfo {author} {\bibnamefont {{A. C. Qu and P. Nigge
  and S. Link and G. Levy and M. Michiardi and P. L. Spandar and T. Matth\'e
  and M. Schneider and S. Zhdanovich and U. Starke and C. Guti\'errez and A.
  Damascelli}}},\ }\bibfield  {title} {\bibinfo {title} {Ubiquitous
  defect-induced density wave instability in monolayer graphene},\ }\href
  {\doibase 10.1126/sciadv.abm5180} {\bibfield  {journal} {\bibinfo  {journal}
  {Sci. Adv.}\ }\textbf {\bibinfo {volume} {8}},\ \bibinfo {pages} {eabm5180}
  (\bibinfo {year} {2022})}\BibitemShut {NoStop}%
\bibitem [{\citenamefont {Jiang}\ \emph {et~al.}(2020)\citenamefont {Jiang},
  \citenamefont {Liang}, \citenamefont {Chen}, \citenamefont {Qi},
  \citenamefont {Li},\ and\ \citenamefont {Wang}}]{Jiang-PRL(2020)}%
  \BibitemOpen
  \bibfield  {author} {\bibinfo {author} {\bibfnamefont {M.-H.}\ \bibnamefont
  {Jiang}}, \bibinfo {author} {\bibfnamefont {S.}~\bibnamefont {Liang}},
  \bibinfo {author} {\bibfnamefont {W.}~\bibnamefont {Chen}}, \bibinfo {author}
  {\bibfnamefont {Y.}~\bibnamefont {Qi}}, \bibinfo {author} {\bibfnamefont
  {J.-X.}\ \bibnamefont {Li}}, \ and\ \bibinfo {author} {\bibfnamefont {Q.-H.}\
  \bibnamefont {Wang}},\ }\bibfield  {title} {\bibinfo {title} {{Tuning
  Topological Orders by a Conical Magnetic Field in the Kitaev Model}},\ }\href
  {\doibase 10.1103/PhysRevLett.125.177203} {\bibfield  {journal} {\bibinfo
  {journal} {Phys. Rev. Lett.}\ }\textbf {\bibinfo {volume} {125}},\ \bibinfo
  {pages} {177203} (\bibinfo {year} {2020})}\BibitemShut {NoStop}%
\bibitem [{\citenamefont {Zhang}\ \emph {et~al.}(2022)\citenamefont {Zhang},
  \citenamefont {Hal\'asz},\ and\ \citenamefont {Batista}}]{Zhang-NC(2022)}%
  \BibitemOpen
  \bibfield  {author} {\bibinfo {author} {\bibfnamefont {S.-S.}\ \bibnamefont
  {Zhang}}, \bibinfo {author} {\bibfnamefont {G.~B.}\ \bibnamefont {Hal\'asz}},
  \ and\ \bibinfo {author} {\bibfnamefont {C.~D.}\ \bibnamefont {Batista}},\
  }\bibfield  {title} {\bibinfo {title} {{Theory of the Kitaev model in a [111]
  magnetic field}},\ }\href {\doibase 10.1038/s41467-022-28014-3} {\bibfield
  {journal} {\bibinfo  {journal} {Nat. Commun.}\ }\textbf {\bibinfo {volume}
  {13}},\ \bibinfo {pages} {399} (\bibinfo {year} {2022})}\BibitemShut
  {NoStop}%
\bibitem [{\citenamefont {Lieb}(1994)}]{Lieb-PRL(1994)}%
  \BibitemOpen
  \bibfield  {author} {\bibinfo {author} {\bibfnamefont {E.~H.}\ \bibnamefont
  {Lieb}},\ }\bibfield  {title} {\bibinfo {title} {{Flux Phase of the
  Half-Filled Band}},\ }\href {\doibase 10.1103/PhysRevLett.73.2158} {\bibfield
   {journal} {\bibinfo  {journal} {Phys. Rev. Lett.}\ }\textbf {\bibinfo
  {volume} {73}},\ \bibinfo {pages} {2158} (\bibinfo {year}
  {1994})}\BibitemShut {NoStop}%
\end{thebibliography}

%

\onecolumngrid

\end{document}